\documentclass[prd,amsmath,amssymb,aps,twocolumn]{revtex4-1}

\usepackage{subcaption}
\usepackage[dvipsnames]{xcolor}
\usepackage{graphicx}
\usepackage[colorlinks=true,citecolor=blue]{hyperref}
\usepackage{soul}
\setstcolor{Green}
\usepackage{cancel}
\DeclareMathAlphabet{\mathcalligra}{T1}{calligra}{m}{n}
\usepackage[utf8]{inputenc}

\newcommand{\subDS}{\textrm{\tiny{DS}}}
\newcommand{\subADS}{\textrm{\tiny{AdS}}}

\newcommand{\be}{\begin{equation} }
	\newcommand{\ee}{\end{equation}}
\newcommand{\bes}{\begin{equation*} }
	\newcommand{\ees}{\end{equation*}}
\newcommand{\bea}{\begin{eqnarray} }
	\newcommand{\eea}{\end{eqnarray}}
\newcommand{\beas}{\begin{eqnarray*} }
	\newcommand{\eeas}{\end{eqnarray*}}
\newcommand{\ba}{\begin{align} }
	\newcommand{\ea}{\end{align} }
\newcommand{\bas}{\begin{align*} }
	\newcommand{\eas}{\end{align*} }

\newcommand{\gb}{\mathcalligra{\mathbf{g}}}
\usepackage{caption}

\begin{document}
	
	\title{Semi-Classical Backreaction on Asymptotically Anti-de Sitter Black Holes}
	
	\author{Peter Taylor}
	\email{peter.taylor@dcu.ie}
	\affiliation{
		Center for Astrophysics and Relativity, School of Mathematical Sciences, Dublin City University, Glasnevin, Dublin 9, Ireland}
	
	\author{Cormac Breen}
	\email{cormac.breen@tudublin.ie}
	\affiliation{School of Mathematical Sciences, Technological University Dublin, Kevin Street, Dublin 8, Ireland}

	\date{\today}
	\begin{abstract}
		We consider a quantum scalar field on the classical background of an asymptotically anti-de Sitter black hole and the backreaction the field's stress-energy tensor induces on the black hole geometry. The backreaction is computed by solving the reduced-order semi-classical Einstein field equations sourced by simple analytical approximations for the renormalized expectation value of the scalar field stress-energy tensor. When the field is massless and conformally coupled, we adopt Page's approximation to the renormalised stress energy tensor while for massive fields, we adopt a modified version of the DeWitt-Schwinger approximation. The latter approximation must be modified so that it possesses the correct renormalization freedom required to ensure the semi-classical equations are consistent. Equipped with these approximations, the reduced-order field equations are easily integrated and the first-order (in $\hbar$) corrections to the metric are obtained. We also compute the corrections to the black hole event horizon, surface gravity and minimum temperature as well as corrections to the photon sphere and quadratic curvature invariants. We pay particular attention to the temperature profiles of the semi-classical black holes compared with their classical counterparts, pointing out some interesting qualitative features produced by the backreaction. These results ought to provide reasonable approximations to the first-order (one-loop) quantum backreaction on the geometry of asymptotically anti-de Sitter black holes when the exact numerical stress-energy tensor sources the semi-classical equations.

	\end{abstract}
	\maketitle
	\section{Introduction}
	\label{sec:into}
	Quantum Field Theory in black hole spacetimes has been an important area of research since Hawking's discovery \cite{Hawking:1975} that the gravitational field of a black hole stimulates the emission of a low-energy quantum radiation. This insight has led to some deep connections between gravity, quantum mechanics and thermodynamics (see, e.g., Ref.~\cite{Hawking:1976}) and has, in some sense, been one of the arenas in which a search for a quantum theory of gravity has been played out. While there has been significant progress in our understanding of these fundamental ideas over the decades, there still remains a number of outstanding problems in the general semi-classical framework that underpinned Hawking's original discovery, both conceptual and computational challenges. For example, are the semi-classical Einstein equations (\ref{eq:semiclassicaleqns}) the equations that emerge from a full quantum theory of gravity in an appropriate semi-classical limit? What about existence and uniqueness of solutions to these equations? On the pragmatic front, assuming solutions exist, how do we compute them numerically and what are their regimes of validity? There are, of course, partial answers to aspects of these questions in the literature (see, e.g., Refs.~\cite{Horowitz:1980, HorowitzWald:1978, Singh:1989, PazSinha:1991, Padmanabhan:1990, Simon:1990, ParkerSimon:1993, FlanaganWald:1996} for some relevant discussion), but much of the endeavour has had a narrow focus on how to compute the source term in the semi-classical equations, the quantum expectation value of the stress-energy tensor \cite{CandelasHoward:84, CandelasHoward:1984, Anderson:1990, AHS:1995, FlachiTanaka:08, OttewillTaylor:10, breenottewill:2012a, breenottewill:2012b, taylorbreen:2016, taylorbreen:2017, amoslevi:2016, LeviOriKerr:2017}. Moreover, usually the background spacetime is fixed and possesses symmetries. This is understandable since computing the source term is a necessary step and already proved to be very challenging since it required a regularization procedure to render the quantity of interest well-defined (see Sec.~\ref{sec:semi-classical} for details). While the conceptual framework for this regularization prescription has been long understood, the so-called point-split regularization \cite{DeWitt:1975, Christensen:1976, Christensen:1978, FullingWald:1978, FullingWald:1981, BrownOttewill:1986}, only recently have efficient numerical schemes been developed for its implementation in black hole spacetimes \cite{taylorbreen:2017,taylorbreen:2016,amoslevi:2016}. Armed with efficient numerical schemes for computing regularized quantum expectation values, a natural progression now is to turn towards solving the semi-classical equations, which remains a formidable challenge. Given the labour and investment involved in solving the semi-classical equations, it is prudent first to attempt to gain some qualitative insight on the solutions from approximate analytical means, if possible. 
	
	In this paper, we are interested in quantum effects on asymptotically anti-de Sitter (AdS) black holes and in particular, the quantum backreaction on such spacetimes via the (reduced order) semi-classical equations. Self-consistent backreactions on these spacetimes are important as a matter of principle since the end state of such a system is not expected to be complete evaporation but rather the black hole in stable thermal equilibrium. The reason is that the AdS potential grows without bound as $r\to\infty$ and this effectively provides a reflective barrier resulting in an evaporating black hole eventually reaching a point where it is absorbing as much radiation as it emits. This raises an issue of information loss in gravitational collapse \cite{KlemmVanzo98} that is different from the context often discussed where the black hole ostensibly completely evaporates. In any case, while this context provides a motivating factor for studying these systems, here we make no attempt to address the dynamical problem involving gravitational collapse, but instead we seek static solutions to the semi-classical equations perturbed about the Schwarzschild-AdS black hole. Rather than solve the problem exactly, which as mentioned above is a computationally intensive endeavour, we adopt a simplified approach which replaces the renormalized expectation of the stress-energy tensor (RSET) with a simple closed-form approximation. While this strategy for computing the backreaction on a black hole is not a novel one \cite{York:1985, York:1993, AndersonYork:1994, Massar:1995, LoustoSanchez:1988, THA:2000, Fabbri:2006}, oftentimes approximations are employed which are either (i) only valid for a very restrictive set of field parameters, (ii) violate the properties required to make the semi-classical equations consistent, (iii) not particularly accurate, or (iv) pathological in some region of the spacetime. In order to cover a wide range of field parameters, we adopt a two-pronged approach for our approximation to the RSET, we use the Page approximation \cite{Page1982} for a massless, conformally coupled field and a modified version of the DeWitt-Schwinger approximation for massive, arbitrarily coupled fields. It is essential that the latter approximation, as it is usually presented in the literature, be modified in order to accommodate the correct renormalization freedom which is needed to make the semi-classical equations consistent. For the range of parameters studied, we find an approximate stress-energy tensor that is regular on the entire exterior, possesses the correct renormalization freedom and is reasonably accurate.
	
	With these approximations sourcing the semi-classical equations and further assuming the perturbation induced by the field's stress-energy is static and spherically symmetric, the equations reduce to two simple ODEs which are straightforwardly integrated. The result is a multi-dimensional space of black hole solutions parametrized by the black hole mass, the AdS lengthscale and the quantum field parameters. Varying these parameters gives a rich phenomenology to explore. We focus on the temperature profiles in the space of solutions, that is, we consider the temperature as a function of black hole radius assuming the AdS lengthscale is fixed. We plot the semi-classical and classical temperature profiles for a range of parameters to glean some qualitative predictions about the nature of the semi-classical corrections. Moreover, we used these profiles to examine for what values of these parameters does the semi-classical approximation breakdown. We also examine the semi-classical corrections to the unstable null circular orbit (the photon sphere) and the quadratic curvature invariants.
	
	The layout of the paper is as follows: In Sec.~\ref{sec:semi-classical}, we review the regularization and renormalization prescription needed to make sense of the semi-classical equations, as well as a reduction of order technique that rids the theory of runaway solutions. In Sec.~\ref{sec:approximations}, we introduce our approximation scheme for the stress-energy tensor and compare its accuracy against exact results obtained numerically. In Sec.~\ref{sec:backreaction}, we compute the backreaction due to this approximate source term via the reduced-order semi-classical equations. Finally, we discuss the results in Sec.~\ref{sec:conclusions}.
	
	\section{The Semi-Classical Equations}
	\label{sec:semi-classical}
	\subsection{Stress-Tensor Regularization and Renormalization}
	In this section, we review and discuss the renormalization prescription that is essential to making sense of the semi-classical equations. The starting point in the semi-classical approximation is to replace the stress-energy tensor that appears on the right-hand side of the classical Einstein field equations with the expectation value of the stress-energy tensor operator for some quantum fields in a unit-norm quantum state $|A\rangle$,
	\begin{align}
		G_{ab}+\Lambda\,g_{ab}=8\pi\,G\,\langle A|\hat{T}_{ab}|A\rangle,
	\end{align}
	in units where $c=\hbar=1$. As is well-known, the right-hand side of these equations is meaningless since the stress-energy tensor operator is quadratic in an operator-valued distribution and hence the expectation value is ill-defined. In order to make sense of the theory, one requires a map $\langle A|\hat{T}_{ab}|A\rangle\to \langle A|\hat{T}_{ab}|A\rangle_{\textrm{reg}} $ which renders the expectation value finite which is compensated by an infinite renormalization by some finite set of parameters in the theory. In order to see what an appropriate map might look like explicitly, let us simplify by assuming the only non-gravitational field present is a scalar field, then we define the point-split stress-energy tensor as
	\begin{align}
		\langle A |\hat{T}_{ab}|A\rangle =-i\lim_{x'\to x}\hat{\tau}_{ab}G_{A}(x,x')
	\end{align}
	where
	\begin{align}
		G_{A}(x,x')\equiv \langle A|\mathcal{T}\hat{\phi}(x)\hat{\phi}(x')|A\rangle
	\end{align}
	is the Feynmann Green function for the scalar field in the state $|A\rangle$. The operator $\mathcal{T}$ appearing in the expression above is the time-ordering operator ensuring that the expression above satisfies causality. The operator $\hat{\tau}_{ab}$ is any differential operator that gives back the known expression for the stress-energy tensor in the coincidence limit $x'\to x$, for example, we take
	\begin{align}
		\hat{\tau}_{ab}\equiv (1-2\xi)g_{b'}{}^{b}\nabla^{a}\nabla^{b'}+(2\xi-\tfrac{1}{2})g^{ab}g_{c'}{}^{c}\nabla_{c}\nabla^{c'}\nonumber\\
		-2\xi g^{ac}\nabla^{c}\nabla^{b}+2\xi g^{ab}\Box+\xi G^{ab}-\tfrac{1}{2}m^{2}g^{ab},
	\end{align}
	where $m$ and $\xi$ are the scalar field mass and coupling to the background curvature, respectively, $\nabla_{a}$ is the covariant derivative with respect to the metric $g_{ab}$, $\Box=\nabla_{a}\nabla^{a}$ is the d'Alembertian operator and $g_{b'}{}^{b}$ is the bivector of parallel transport. Since $G_{A}(x,x')$ is a Green function, it satisfies the inhomogeneous wave equation
	\begin{align}
		(\Box-\xi\,R-m^{2})G_{A}(x,x')=-\delta(x,x')
	\end{align}
	where the right-hand side is the covariant Dirac delta distribution. We restrict to a class of quantum states that satisfy the so-called Hadamard condition (see, e.g., Ref.~\cite{WaldBook:1994}), that is, states for which the Green function has the following short-distance behaviour \cite{DecaniniFolacci:2008}
	\begin{align}
		G_{A}(x,x')=&\frac{i\Gamma_{d}}{2}\Bigg\{\frac{U(x,x')}{(\sigma(x,x')+i\epsilon)^{d/2-1}}\nonumber\\
		&+V(x,x')\log\left(\frac{2\sigma(x,x')}{\ell^{2}}+i\epsilon\right)+W_{A}(x,x')\Bigg\}
	\end{align}
	where $U$, $V$ and $W_{A}$ are symmetric biscalars and $\sigma$ is Synge's world function corresponding to half the square of the geodetic distance between two points (assuming there's a unique geodesic connecting them). The parameter $\ell$ here is an arbitrary lenghtscale needed to make the argument of the log term dimensionless. In odd dimensions, this term is irrelevant since $V\equiv 0$ for odd $d$. The coefficient appearing in this expression is given by $\Gamma_{d}\equiv \Gamma(d/2-1)/(2\pi)^{d/2}$. The term involving $U(x,x')$ above is called the direct part of the Hadamard form while the term involving $V(x,x')$ is known as the tail of the Hadamard form. Both of these terms contain all the short-distance (or ultraviolet) divergences. They are constructed only from the geometry through the metric and its derivatives. The remaining term $W_{A}(x,x')$ depends on the quantum state and cannot be determined by a local expansion.
	
	Now in order to obtain a map that regularizes the quantum stress-energy tensor, we postulate the existence of a two-point function $G_{\textrm{S}}(x,x')$ whose properties we will discuss shortly. Now we can express the stress-energy tensor as
	\begin{align}
		\langle A|\hat{T}_{ab}|A\rangle=-i\lim_{x'\to x}\hat{\tau}_{ab}\left(G_{A}(x,x')-G_{\textrm{S}}(x,x')\right)\nonumber\\
		+i\lim_{x'\to x}\hat{\tau}_{ab}G_{\textrm{S}}(x,x').
	\end{align}
	For this to be a useful procedure demands that the first limit above be well-defined. This implies that the second limit will diverge but we demand that this term can be absorbed into some other terms on the left-hand side of the field equations by an infinite renormalization of some parameters in the theory. This gives a set of properties that we can demand of $G_{\textrm{S}}(x,x')$ (see Ref.~\cite{HarteFlanaganTaylor} for a similar axiomatic construction of the regularized retarded Green function in the classical self-force problem):
	\begin{enumerate}
		\item $G_{\textrm{S}}(x,x')$ is a parametrix for the Klein-Gordon wave operator.
		\item $G_{\textrm{S}}(x,x')$ is symmetric.
		\item $G_{\textrm{S}}(x,x')$ is locally constructed from the geometry.
	\end{enumerate}
	Existense of such a $G_{\textrm{S}}(x,x')$ satisfying these properties is straightforward, for example, taking $G_{\textrm{S}}(x,x')$ to be
	\begin{align}
		\label{eq:GS}
		G_{\textrm{S}}(x,x')=&\frac{i\Gamma_{d}}{2}\Bigg\{\frac{U(x,x')}{(\sigma(x,x')+i\epsilon)^{d/2-1}}\nonumber\\
		&+V(x,x')\log\left(\frac{2\sigma(x,x')}{\ell^{2}}+i\epsilon\right)+W(x,x')\Bigg\},
	\end{align}
	where $W(x,x')$ is any regular symmetric biscalar constructed only from the geometry. The simplest choice is the trivial one $W(x,x')\equiv 0$. Making this choice, we have
	\begin{align}
		\langle A|\hat{T}_{ab}|A\rangle_{\textrm{reg}}&\equiv -i\lim_{x'\to x}\hat{\tau}_{ab}\left(G_{A}(x,x')-G_{\textrm{S}}(x,x')\right)\nonumber\\
		&=\frac{\Gamma_{d}}{2}\lim_{x'\to x}\hat{\tau}_{ab}\left(W_{A}(x,x')\right),
	\end{align}
	which is manifestly finite. Further defining
	\begin{align}
		\Theta_{ab}\equiv i\,\hat{\tau}_{ab}G_{\textrm{S}}(x,x')
	\end{align}
	and introducing the notation $\left[\cdot\right]$ to denote the coincidence limit $x'\to x$, we have
	\begin{align}
		\langle A|\hat{T}_{ab}|A\rangle=\langle A|\hat{T}_{ab}|A\rangle_{\textrm{reg}}+\left[\Theta_{ab}\right].
	\end{align}
	At this stage a couple of comments are in order. First, $\left[\Theta_{ab}\right]$ is clearly ill-defined. We return to this below. Second, what we have defined as $\langle A|\hat{T}_{ab}|A\rangle_{\textrm{reg}}$ is not conserved in even spacetime dimensions. In fact, for $d$ even, it can be shown that
	\begin{align}
		\nabla_{b} \langle A|\hat{T}^{ab}|A\rangle_{\textrm{reg}}=\frac{\Gamma_{d}}{2}\left[\hat{\tau}^{ab}{}_{;b}(W_{A})\right]=-\frac{d\,\Gamma_{d}}{4}g^{ab}\nabla_{b}v_{1},
	\end{align}
	where $v_{1}(x)$ is the coincidence limit of the Hadamard biscalar $V_{1}(x,x')$ \cite{BrownOttewill1985, DecaniniFolacci:2008}. This suggests the following mapping
	\begin{align}
		\label{eq:RegMap}
		\langle A|\hat{T}^{ab}|A\rangle\to \frac{\Gamma_{d}}{2}\left(\left[\hat{\tau}_{ab}W_{A}(x,x')\right]+\frac{d}{2}g_{ab}v_{1}\right)+\tilde{\Theta}_{ab}
	\end{align}
	where we now identify the first term in parenthesis as the regularized conserved quantum stress-energy tensor and the second term is
	\begin{align}
		\tilde{\Theta}_{ab}=\left[\Theta_{ab}\right]-\frac{d\,\Gamma_{d}}{4}g_{ab}\,v_{1}.
	\end{align}
	To see how to deal with these terms, let us first implement the mapping (\ref{eq:RegMap}) in the semi-classical equations
	\begin{align}
		G^{-1}G_{ab}+\tilde{\Lambda}g_{ab}=&4\pi\,\Gamma_{d}\left(\left[\hat{\tau}_{ab}W_{A}(x,x')\right]+\frac{d}{2}g_{ab}v_{1}\right)\nonumber\\
		&+8\pi\,\tilde{\Theta}_{ab}
	\end{align}
	where we have divided across by $G$ and defined $\tilde{\Lambda}=\Lambda/G$ for later convenience. Explicit expressions for $\tilde{\Theta}_{ab}$ are derivable for each $d$ \cite{DecaniniFolacci:2008}. First note that $\tilde{\Theta}_{ab}$ has dimensions $\left[\textrm{length}\right]^{-d}$ and can only depend on the local geometry. For $d=3$, an expansion of $\Theta_{ab}$ about $x$ before we consider coincidence limits gives terms of the form $m^{3}g_{ab}$ and $m^{2}G_{ab}$. However, the coefficients of these terms are singular in the coincidence limits. Nevertheless, we reabsorb these terms into an infinite renormalization of $G^{-1}$ and $\tilde{\Lambda}$ on the left-hand side of the semi-classical equations. Hence for $d=3$, the semi-classical equations are
	\begin{align}
		G_{\textrm{ren}}^{-1}G_{ab}+\tilde{\Lambda}_{\textrm{ren}}g_{ab}=4\pi\,\Gamma_{3}\left[\hat{\tau}_{ab}W_{A}(x,x')\right].
	\end{align}
	Things are particularly simple for massless fields in any odd dimensions where no parameters are renormalized; one simply replaces the naive and ill-defined quantum stress-energy tensor with the regularized one $\left[\hat{\tau}_{ab}W_{A}\right]$.
	
	For the case of present interest, $d=4$, the situation is considerably more complicated. The terms appearing in the expansion of $\tilde{\Theta}_{ab}$ are $m^{4}g_{ab}$ and $m^{2}G_{ab}$, as well as two independent quadratic curvature tensors $A_{ab}$ and $B_{ab}$. Explicit expressions for $A_{ab}$ and $B_{ab}$ are given below, but importantly they are each separately conserved. The mass dependant terms again renormalize $G^{-1}$ and $\tilde{\Lambda}$. However, there are no terms on the left-hand side of the field equations to absorb the higher curvature terms. Hence, we need to consider the quadratic field equations in the semi-classical approximation, the semi-classical equations being
	\begin{align}
		\label{eq:semiclassicaleqns}
		G^{-1}_{\textrm{ren}}G_{ab}+\tilde{\Lambda}_{\textrm{ren}}g_{ab}+\alpha_{\textrm{ren}}A_{ab}+\beta_{\textrm{ren}}B_{ab}\nonumber\\
		=4\pi\Gamma_{4}\left(\left[\hat{\tau}_{ab}W_{A}\right]+2g_{ab}v_{1}\right).
	\end{align}
	These equations are the starting point of the semi-classical development. We note that oftentimes in the literature, the quadratic terms on the left-hand side are omitted. Without these, however, the solutions are dependent on the particular choice of parametrix used to regularize the right-hand side. These terms are necessary therefore to ensure that different choices in the parametrix, i.e., different choices for the regularization prescription, are degenerate with different choices of renormalization of the constants appearing in front of the geometrical tensors on the left-hand side.

	\subsection{Reduction Of Order}
	Solving the equations (\ref{eq:semiclassicaleqns}) exactly is problematic for a number of reasons. First, it is not even clear whether these equations are well-posed. They are fourth-order non-linear partial differential equations which are known to possess runaway solutions \cite{Simon:1990, FlanaganWald:1996}. Moreover, because the source-term depends on the quantum state, one needs to have information everywhere on a past Cauchy surface to determine the metric anywhere in the future, unlike the classical Cauchy development.
	
	One possible resolution to the problem of well-posedness is the so-called reduction-of-order prescription. This scheme is imported from electrodynamics \cite{LandauLifschitz} where the self-interaction of the electron field with itself gives rise to a third-order time derivative in the equations of motion. This is resolved by expanding the solution around a background solution which reduces the order of the equation when terms above first order (in an appropriate small parameter) are ignored. To see how this idea might be applied in the semi-classical context, it is convenient to work again in units where $G=1$, but reinstating $\hbar$ as a small parameter, whence the equations are (dropping the subscript on renormalized parameters for typographical convenience)
	\begin{align}
		G_{ab}+\Lambda\,g_{ab}+\alpha\,\hbar\,A_{ab}+\beta\,\hbar\,B_{ab}=8\pi\hbar\,\langle A|\hat{T}_{ab}[g]|A\rangle_{\textrm{reg}},
	\end{align}
	where here it is convenient to adopt the notation
	\begin{align}
		\langle A|\hat{T}_{ab}[g]|A\rangle_{\textrm{reg}}\equiv \frac{\Gamma_{4}}{2}\left([\hat{\tau}_{ab}W_{A}]+2g_{ab}v_{1}\right).
	\end{align}
	Now we assume a perturbative expansion of the metric and quantum state of the form \cite{FlanaganWald:1996}
	\begin{align}
		g_{ab}&=\gb_{ab}+\hbar\,h_{ab}+\mathcal{O}(\hbar^{2}),\nonumber\\
		|A\rangle &=|A_{\gb}\rangle +\mathcal{O}(\hbar),
	\end{align}
	where, for simplicity, we are assuming that $|A_{\gb}\rangle$ is a quantum state for a Klein-Gordon scalar field on the fixed background metric $\gb_{ab}$. Then to zeroth order in $\hbar$, $\gb_{ab}$ is a vacuum solution to the classical Einstein field equations. The next order gives $\mathcal{O}(\hbar)$ quantum corrections to this classical solution satisfying
	\begin{align}
		\label{eq:perturbationeqn}
		G_{ab}^{(1)}[h]+\Lambda\,h_{ab}+\alpha\,A_{ab}^{(0)}[\gb]+\beta\,B_{ab}^{(0)}[\gb]\nonumber\\
		=8\pi \langle A_{\gb}|\hat{T}_{ab}[\gb]|A_{\gb}\rangle,
	\end{align}
	where
	\begin{align}
		\label{eq:A0B0}
		A_{ab}^{(0)}[\gb]=&\frac{1}{\sqrt{-\gb}}\frac{\delta}{\delta \gb^{ab}}\int d^{4}x\sqrt{-\gb}\,C_{cdef}C^{cdef}\nonumber\\
		=&-2\Box R_{ab}+\tfrac{2}{3}\nabla_{a}\nabla_{b}R+\tfrac{1}{3}\gb_{ab}\Box R-\tfrac{1}{3}\gb_{ab}R^{2}\nonumber\\
		&+\tfrac{4}{3}R\,R_{ab}+(R_{cd}R^{cd})\gb_{ab}-4 R_{acbd}R^{cd},\nonumber\\
		B_{ab}^{(0)}[\gb]=&\frac{1}{\sqrt{-\gb}}\frac{\delta}{\delta \gb^{ab}}\int d^{4}x\sqrt{-\gb}\,R^{2}\nonumber\\
		=&2\nabla_{a}\nabla_{b}R-2\gb_{ab}\Box R+\tfrac{1}{2}R^{2}\gb_{ab}-2 R\,R_{ab},
	\end{align}
	where all curvature terms and covariant derivatives here are with respect to the background metric $\gb_{ab}$. The term $G_{ab}^{(1)}[h]$ is the linearized Einstein tensor which is explicitly given by
	\begin{align}
		\label{eq:linearized}
		-2G_{ab}^{(1)}[h]=\Box \bar{h}_{ab}+\gb_{ab}\nabla^{c}\nabla^{d}\bar{h}_{cd}-2\nabla^{c}\nabla_{(a}\bar{h}_{b)c}\nonumber\\
		-\gb_{ab}R^{cd}\bar{h}_{cd}+R\,\bar{h}_{ab}
	\end{align}
	where again covariant derivatives and curvature terms are with respect to $\gb_{ab}$ and $\bar{h}_{ab}=h_{ab}-\tfrac{1}{2}\gb_{ab}\gb^{cd}h_{cd}$ is the trace-reversed metric perturbation. Finally, the source term in (\ref{eq:perturbationeqn}) is the conserved quantum stress-energy tensor for a scalar field on the fixed background $\gb_{ab}$ in the quantum state $|A_{\gb}\rangle$.
	
	The crucial point is that the perturbation satisfies an equation with time derivatives of second order. It is in this sense that the perturbation scheme has reduced the order of the equations and presumably removed unphysical runaway solutions. Still these are extremely difficult equations to solve, even numerically. Indeed, even computing the source term is computationally challenging since an efficient regularization prescription is required. While much effort has been devoted to such regularization schemes in recent years \cite{LeviOri:2015, LeviOri:2016, amoslevi:2016,  taylorbreen:2016, taylorbreen:2017}, there has been less progress on solving exactly the reduced-order semi-classical equations in contexts of physical interest.

	\section{Approximating the Stress-Energy Tensor for Anti-de Sitter Black Holes}
	\label{sec:approximations}
	In this section, we discuss the analytical approximations for the expectation of the stress-energy tensor for a quantum scalar field that we will use as the source term when solving the reduced-order semi-classical equations. We review the existing approximations in the literature, as well as introducing a modified version of the well-known DeWitt-Schwinger approximation for massive fields. First, however, we briefly review the geometry and thermodynamics of the Schwarzschild-AdS black hole.
	
	\subsection{Schwarzschild-AdS black holes}
	\label{sec:AdSBH}
	The Schwarzschild-AdS black hole spacetime is a static, spherically-symmetric solution to the vacuum Einstein equations with negative cosmological constant $\Lambda<0$. In Schwarzschild-like coordinates, this solution has a line-element of the form 
	\begin{eqnarray}
		\label{eq:metric}
		ds^{2}=\gb_{ab}dx^{a}dx^{b}=-f(r)dt^{2}+dr^{2}/f(r)+r^{2}d\Omega^{2}_{2},\nonumber\\
	\end{eqnarray}
	where $d\Omega^{2}_{2}$ is the line element for $\mathbb{S}^{2}$ and
	\begin{eqnarray}
		\label{eq:fSchwarzschild_AdS}
		f(r)=1-\frac{2 M}{r}+\frac{r^2}{L^2},
	\end{eqnarray}
	in units where $G=c=1$. The parameter $M$ is the conserved mass and $L=\sqrt{-3/\Lambda}$ is the AdS curvature lenghtscale. The Ricci tensor and scalar for Schwarzschild-AdS are, respectively,  given by
	\begin{eqnarray}
		\label{eq:Ricci}
		R_{ab}=-\frac{3}{L^2}\gb_{ab},~~~~~~~R=-\frac{12}{L^2}.
	\end{eqnarray}
	These coordinates are singular at $r=r_{+}$ where $r_{+}$ is the only real root of $f$, which has a rather complicated expression in terms of $M$ and $L$.
	
	The hypersurface defined by $r=r_{+}$ is the black hole event horizon. The surface gravity at the event horizon is
	\begin{equation}
		\label{eq:kappa}
		\kappa=\frac{ M}{r_{+}^2}+\frac{r_{+} }{L^2}.
	\end{equation}
	
	We will find it natural to parametrize the black hole not in terms of $(M, L)$, but rather in terms of $(r_{+}, L)$ since this gives a clearer interpretation of the two lengthscales at play. Moreover, we will also find it useful for the numerical calculations described later in this paper to introduce a new dimensionless radial coordinate $\zeta$ for which the event horizon radius takes the same value irrespective of the parameters $r_{+}$ and $L$, defined by
	\begin{equation}
		\zeta = \frac{2 r}{r_{+}} - 1.
		\label{eq:zeta}
	\end{equation}
	In this new coordinate, the metric function assumes the form
	\begin{equation}
		{\overline {f}}(\zeta ) =f(r)= \frac{\zeta -1}{ \zeta + 1 } h(\zeta),\qquad h(\zeta)\equiv 1+\frac{r_{+}^{2}}{4 L^{2}}(\zeta^{2}+4\zeta+7).
		\label{eq:fzeta}
	\end{equation}
	The event horizon $r=r_{+}$ is then located at $\zeta =1$ for all $r_{+}$ and $L$, and the curvature singularity at $r=0$ is at $\zeta=-1$. The function $h(\zeta)$ has no real roots. The asymptotically flat Schwarzschild limit is given by $L\to\infty$ whence $h(\zeta)=1$. In terms of this parametrization $(r_{+}, L)$, the surface gravity of the black hole is
	\begin{align}
		\label{eq:kappaR}
		\kappa=\frac{1}{2}f'(r_{+})=\frac{(3 r_{+}^{2}+L^{2})}{2 r_{+}L^{2}}.
	\end{align}
	
	Like asymptotically flat black holes, there is a natural temperature $T$ associated with Schwarzschild-AdS. To see this, we note that the Euclidean section of the spacetime obtained by Wick rotating $t\to i\,\tau$ would possess an essential conical singularity at the horizon unless we identify the Euclidean time coordinate with periodicity $\beta=T^{-1}=2\pi/\kappa$. Unlike asymptotically flat black holes, there exists a minimum temperature which can be seen by differentiating Eq.~(\ref{eq:kappaR}) with respect to $r_{+}$ and solving for the critical points. The only stationary point occurs at $r_{+}^{\textrm{min}}=L/\sqrt{3}$ and the corresponding minimum temperature is $T_{\textrm{min}}=(2\pi)^{-1}(\sqrt{3}/L)$.
	
	Thermal states in Schwarzschild-AdS are in thermal equilibrium with a locally measured temperature
	\begin{equation}
		T_{\mathrm{loc}}=T f^{-1/2}(r).
	\end{equation}
	Due to the confining nature of the AdS potential as $r\to\infty$, the locally measured temperature decreases indefinitely as the boundary is approached, unlike the asymptotically flat case. This implies that the total energy of thermal radiation in Schwarzschild-AdS is finite. This provides part of the motivation for considering thermal states in asymptotically AdS black holes. In contrast, in Ref.~ \cite{York:1985}, the backreaction on a Schwarzschild black hole due to a quantum scalar field in a thermal state was considered, but the approximation used diverged at infinity and it was necessary to place the black hole in a box to avoid these issues. In asymptotically AdS spacetimes, no such measures are needed since the AdS potential effectively provides a confining box.

	\subsection{Review of Existing Approximations}
	There exists several approximations for the vacuum polarization and stress-energy tensor in the literature. The most common approximations are for conformally invariant fields. These approximations usually rely on computing closed-form expressions for $\langle A|\hat{T}^{a}{}_{b}|A\rangle$ in a simplified (usually ultrastatic) spacetime conformally related to the spacetime of interest. This technique was first adopted by Page \cite{Page1982} to approximate the stress-energy tensor for conformal scalar fields in the Hartle-Hawking state in Schwarzschild, a result which was extended to the Boulware vacuum state by Frolov and Zel'nikov \cite{FZ1985a}. It is Page's approximation that was used by York \cite{York:1985} to compute the backreaction on a Schwarzschild black hole.
	
	Brown and Ottewill \cite{BrownOttewill1985} generalized Page's approximation to include a more general class of conformal transformations than those considered by Page which allowed them to construct new approximations for the stress-energy tensor. They also generalized to other conformal fields, deriving approximations for neutrino and electromagnetic field contributions to the stress-energy tensor, though it was later shown that the approximation for electromagnetic fields was a poor one near a Schwarzschild black hole \cite{JensenOttewill1989}. This shortcoming was rectified by Frolov and Zel'nikov in a more general approximation scheme that was valid for any static spacetime. Their method has a free parameter which can be chosen so that the stress-energy tensor matches exact values on the horizon of a Schwarzschild black hole. However, their approximation for the stress-energy tensor of a scalar field in the Hartle-Hawking state of a Reissner-Nordstr{\"o}m black hole is singular on the horizon, while the exact tensor is finite there.
	
	Thus far, we have mentioned only approximations for conformal fields. For fields with large mass, an oft-used approximation to the stress energy tensor is the DeWitt-Schwinger approximation (see, for example,\cite{Decanini_2007}). Based on heat kernel methods for expanding the singular field, it is a purely local approximation independent of the quantum state, unlike the other approximations we discussed. For an arbitrary spacetime, the DeWitt-Schwinger approximation takes on a rather complicated form, however for Einstein spacetimes satisfying $R_{ab}=\Lambda\,g_{ab}$ for some constant $\Lambda$, the expressions simplifies considerably yielding:
	\begin{align}
		\label{eq:T_DSGen}
		(96 \pi^{2}& m^{2}) \langle \hat{T}_{ab}\rangle_\subDS=-\tfrac{1}{15}(\xi-\tfrac{3}{14})R^{pqrs}R_{pqrs;(ab)}\nonumber\\
		& -\tfrac{1}{15}(\xi-\tfrac{13}{56})R^{pqrs}{}_{;a}R_{pqrs;b}-\tfrac{1}{70}R^{pqr}{}_{a;s}R_{pqrb}{}^{;s}\nonumber\\
		& +\tfrac{2}{315}R^{pqrs}R_{pqta}R_{rs}{}^{t}{}_{b}+\tfrac{4}{63}R^{prqs}R^{t}{}_{pqa}R_{trsb}\nonumber\\
		& -\tfrac{2}{315}R^{pqr}{}_{s}R_{pqrt}R^{s}{}_{a}{}^{t}{}_{b}+\tfrac{4}{15}\Lambda\,(\xi-\tfrac{1}{4})R^{pqr}{}_{a}R_{pqrb}\nonumber\\
		&  +g_{ab}\Big(\tfrac{1}{15}(\xi-\tfrac{19}{112})R_{pqrs;t}R^{pqrs;t}\nonumber\\
		&  +\tfrac{1}{10}(\xi-\tfrac{1}{6})\Lambda R_{pqrs}R^{pqrs}\nonumber\\
		& -\tfrac{4}{15}(\xi-\tfrac{41}{252})R_{pqrs}R^{p}{}_{u}{}^{q}{}_{v}R^{rusv}\nonumber\\
		&  -\tfrac{1}{15}(\xi-\tfrac{47}{252})R_{pqrs}R^{pquv}R^{rs}{}_{uv}\nonumber\\
		& +\Lambda^{3}[16(\xi-\tfrac{1}{6})^{3}-\tfrac{2}{15}(\xi-\tfrac{1}{6})+\tfrac{2}{945}]\Big).
	\end{align}
	Evaluating this tensor for the particular case of the Schwarzschild-AdS spacetime, the non-zero components are explicitly given by
	\begin{widetext}
		\begin{eqnarray}
			\label{eq:T_DS}
			\langle \hat{T}^{r}_{~r} \rangle_\subDS=\frac{1}{10080 \pi ^2 m^2 L^6 r^9}\bigg[14 L^6 M^3 (216 \xi -47)-21 L^4 M^2 r \left(3 L^2 (32 \xi -7)+(168 \xi -37) r^2\right)\nonumber\\+\left(185-45360 \xi ^3+22680 \xi ^2-3654 \xi\right) r^9\bigg]\nonumber\\
			\langle \hat{T}^{t}_{~t} \rangle_\subDS=\frac{1}{10080 \pi ^2 m^2 L^6 r^9}\bigg[2 L^6 M^3 (1237-5544 \xi )+3  L^4 M^2 r \left(15 L^2 (112 \xi -25)+(1176 \xi -263) r^2\right)\nonumber\\+\left(185-45360 \xi ^3+22680 \xi ^2-3654 \xi\right) r^9\bigg]\nonumber\\
			\langle \hat{T}^{\theta}_{~\theta} \rangle_\subDS=\frac{1}{10080 \pi ^2 m^2 L^6 r^9}\bigg[2  L^6 M^3 (1543-7056 \xi )+3  L^4 M^2 r \left(63 L^2 (32 \xi -7)+(1176 \xi -257) r^2\right)\nonumber\\
			+\left(185 -45360 \xi ^3+22680 \xi ^2-3654 \xi\right) r^9\bigg].\nonumber\\
		\end{eqnarray}
	\end{widetext}

	\subsection{Approximations to $\langle\hat{T}^{a}_{~b}\rangle$ for the Schwarzchild AdS spacetime. }
	\label{sec:newapprox}
	We now turn to the task of developing appropriate approximations for the expectation value of the stress-energy tensor for a scalar field in Schwarzschild-AdS that will enable us to accurately compute the backreaction on this family of black holes. We will discuss two distinct and separate approximations, one which is based on the DeWitt-Schwinger approximation for massive fields and the other which is based on Page's approximation for conformal fields. Before discussing these approximations, however, we first discuss the general properties we wish our approximation to satisfy.
	
	While, our approximate stress-energy tensor operators are not exact, in order for the semi-classical equations to make sense mathematically and for the physics to be reasonable, we insist on the following:
	\begin{enumerate}
		\item The approximate RSET should be exactly conserved.
		\item The approximate RSET represents a one-parameter family of tensors parametrized by an arbitrary lengthscale $\ell$ and transforms appropriately under a change in this lengthscale (see Sec.~\ref{sec:backreaction}). The exception to this is for conformal fields ($m=0$, $\xi=1/6$) whence the ambiguity is fixed by insisting we retrieve the standard trace anomaly $\langle \hat{T}^{a}{}_{a}\rangle=v_{1}(x)/(4\pi^{2})$ \cite{Wald1978}.
		\item Our approximate RSET should asymptote to the exact tensor as $r\to\infty$.
		\item Our approximate RSET should be regular on the black hole horizon in a freely falling frame and respect the symmetries of the background spacetime. 
	\end{enumerate}
	For the second of these properties, the renormalization ambiguity is particularly simple when the background spacetime is an Einstein spacetime, since the geometrical tensors $A_{ab}^{(0)}$ and $B_{ab}^{(0)}$ that appear on the left-hand side of the semi-classical equations (\textit{c.f.} Eq.~(\ref{eq:perturbationeqn})-(\ref{eq:A0B0})) vanish identically. Hence changes in the arbitrary lengthscale of our approximate RSET must correspond to a change in terms proportional to the metric and are degenerate with different renormalizations of the cosmological constant $\Lambda_{\textrm{ren}}$ on the left-hand side of the semi-classical equations. Imposing property 4 implies a natural association with the field in the Hartle-Hawking state. For the Page stress-energy tensor, the thermal nature of the approximation is explicit in its construction and can be unambiguously associated with the Hartle-Hawking state when applied to a stationary black hole spacetime. However, the DeWitt-Schwinger approximation is a purely local one independent of the quantum state. On the other hand, of the quantum states usually considered on a black hole spacetime, the Boulware state diverges on the future horizon, the Unruh state does not exist since the AdS potential barrier prevents a flux of radiation out to infinity \cite{KlemmVanzo98}, whereas the Hartle-Hawking state is regular on both the past and future horizon and the RSET in this state has the same symmetries as the background spacetime. Hence there is still a sense in which one would expect the backreaction sourced by this approximation to best approximate the exact backreaction for the field in the Hartle-Hawking state.

	\subsubsection{Modified DeWitt-Schwinger Approximation}
	In this subsection, we develop an approximation for large mass fields which takes, as its starting point, the DeWitt-Schwinger approximation (\ref{eq:T_DS}). This expression is not appropriate as is since it does not satisfy properties 2 and 3 above. In order to guarantee that all properties are satisfied, we must exploit the freedom we have to add to this any geometrical second-rank conserved tensor. In this case, the only freedom we have is to add a term proportional to the metric which ought to vanish for conformal fields (see Sec.~\ref{sec:semi-classical}). Hence, we take as an ansatz
	\begin{align}
		\langle \hat{T}^{a}{}_{b}\rangle^{\ell}_{\subDS}= \langle \hat{T}^{a}{}_{b}\rangle_{\subDS}+\frac{3}{128 \pi^{2}}\delta^{a}{}_{b}\Bigg\{C_{1}\,m_{\xi}^{4}+C_{2}m^{2}m_{\xi}^{2}\nonumber\\
		+C_{3}\frac{m^{2}}{L^{2}}+\frac{C_{4}}{L^{4}}\Bigg\},
	\end{align}
	where $C_{1},...,C_{4}$ are dimensionless parameters and we have parametrized the dependence on $\xi$ in terms of an effective mass
	\begin{align}
		m_{\xi}^{2}=m^{2}-\frac{12}{L^{2}}(\xi-\tfrac{1}{6}),
	\end{align}
	which is defined in such a way that it vanishes for conformally invariant fields. Note that the Klein-Gordon equation satisfied by the classical scalar field has stable solutions only for fields satisfying the so-called Breitenlohner-Freedman bound $m_{\xi}^{2}\,L^{2}\ge -1/4$ \cite{Holzegel2011, Vasy}. Now the constants $C_{1},...,C_{4}$ are determined by demanding that our approximation reproduces the correct asymptotic value as $r\to\infty$, i.e., we recover the exact known stress energy tensor in a global vacuum state for pure AdS spacetime \cite{winstanleykent:2014}
	\begin{align}
		\label{eq:T_adS}
		\langle \hat{T}^{a}{}_{b}\rangle_{\subADS}=\frac{3}{128 \pi^{2}}\delta^{a}{}_{b}\Bigg(m^{2}m_{\xi}^{2}\Bigg\{\frac{4}{3}\ln\left(\frac{2 L}{\ell}\right)\nonumber\\
		-\frac{4}{3}\psi\left(\sqrt{L^{2}m_{\xi}^{2}+\tfrac{1}{4}}+\tfrac{1}{2}\right)
		-\frac{1}{3}\Bigg\}\nonumber\\
		+\frac{1}{3}m_{\xi}^{4}+\frac{2 m^{2}}{9 L^{2}}-\frac{2}{45 L^{4}}\Bigg).
	\end{align}
	We remark that this is not precisely the same form for the tensor given in Ref.~\cite{winstanleykent:2014} but the two are easily seen to be equivalent through a particular choice of arbitrary lengthscale $\ell$. We have chosen a representation with a particularly simple flat spacetime limit,
	\begin{align}
		\label{eq:stressenergyflatlimit}
		\langle \hat{T}^{a}{}_{b}\rangle_{\subADS}\sim \delta^{a}{}_{b} \frac{ m^{4}}{32\pi^{2}}\ln\left(\frac{2}{\ell\,m}\right),\qquad L\to\infty.
	\end{align}
	We also note here that (\ref{eq:T_adS}) is the RSET for a scalar field in the global vacuum state satisfying Dirichlet boundary conditions. However, we wish to associate our approximation with the Hartle-Hawking state satisfying Dirichlet boundary conditions which is a mixed thermal state. While simple closed-form expressions are not known for thermal states in pure AdS, the asymptotic behaviour is the same as those for the vacuum state. Since we are only requiring that our approximation has the correct asymptotics as $r\to\infty$, it matters not whether we take the limit of the thermal state RSET or vacuum state RSET on pure AdS, so we may as well take the latter. A further complication is that the correct comparison for the asymptotics is not between Eq.~(\ref{eq:T_adS}) and Eq.~(\ref{eq:T_DS}) but rather between a large-$m$ expansion of Eq.~(\ref{eq:T_adS}) and Eq.~(\ref{eq:T_DS}). From the asymptotic expansion of the digamma function, $\psi(z)=\ln z+O(1/z)$, we see that
	\begin{align}
		\label{eq:TAdSLargem}
		\langle \hat{T}^{a}{}_{b}\rangle_{\subADS}=\frac{\delta^{a}{}_{b}}{128 \pi^{2}}\Bigg(-4 m^{2}m_{\xi}^{2}\ln(\ell\,m/2)
		+12(\xi-\tfrac{1}{6})\frac{m^{2}}{L^{2}}\nonumber\\
		+\frac{4(185-45360 \xi^{3}+22680\xi^{2}-3654\xi)}{315 m^{2}L^{6}}+O(m^{-4})\Bigg).
	\end{align}
	Comparing with (\ref{eq:T_DS}), we recognise the last term here as the dominant contribution from the DeWitt- Schwinger approximation in the large $r$ limit. Hence, obtaining the correct asymptotics necessitates that $C_{1}=C_{4}=0$ and
	\begin{align}
		C_{2}=-\tfrac{4}{3}\ln(\ell m/2),\qquad C_{3}=4(\xi-\tfrac{1}{6}),
	\end{align}
	whence we can express our modified DeWitt-Schwinger approximation as
	\begin{align}
		\label{eq:TModifiedDS}
		\langle \hat{T}^{a}{}_{b}\rangle^{\ell}_{\subDS}= \langle \hat{T}^{a}{}_{b}\rangle_{\subDS}-\frac{1}{32 \pi^{2}}\delta^{a}{}_{b}\Bigg\{m^{2}m_{\xi}^{2}\ln(\ell \,m/2)\nonumber\\
		-3(\xi-\tfrac{1}{6})\frac{m^{2}}{L^{2}}\Bigg\}.
	\end{align}
	This expression now satisfies all the properties that we required, at least up to $O(m^{-4})$ in a large mass expansion.
	
	In order to arrive at this form of our approximation, we relied on the well-known ambiguity in the renormalization of the stress-energy tensor, i.e., the renormalization prescription provides an equivalence class of RSETs parametrized by $\ell$ rather than a particular tensor. This fact makes discussing the accuracy of an approximation somewhat confusing. For example, the DeWitt-Schwinger approximation as it is usually presented possesses no renormalization ambiguity and is claimed to be a good approximation to the exact RSET. However, the exact RSET is computed for a particular choice of lengthscale in, say, the Hadamard regularization procedure but the different choice of this lengthscale can produce very different numerical results. It is not completely obvious which $\ell$ in the family of exact RSETs we should be comparing with any given approximation. Instead, we say there exists a mapping between the family of approximate RSETs and exact RSETs for which each member in the latter family of tensors can be well approximated by a member in the former family of tensors. There is a sense in which it doesn't matter since the RSET itself is not the object of physical interest but rather the geometry of the backreaction, and this does not distinguish between different renormalization lengthscales. Hence, the most pragmatic approach to solving the backreaction equations is to source the semi-classical equations by $\langle \hat{T}^{a}{}_{b}\rangle_{\subDS}$ and absorb the terms with an explicit $\ell$-dependence into a renormalization of the cosmological constant on the left-hand side of the semi-classical equations. Nevertheless, it is useful for assessing the accuracy of an approximation to keep the $\ell$-dependence in the RSET.
	
	\subsubsection{Approximation for Conformal Fields}
	While the above approximation is valid for large field mass, it is clearly pathological for massless fields. We require a separate approximation for this case. It is difficult to develop a massless approximation for general couplings. However, for conformal coupling $\xi=1/6$, one can leverage the invariance of the scalar field theory under conformal rescalings to enable us to work on the conformally related ultrastatic spacetime where good approximations for the thermal propagator can be obtained. This is the essence of the Page approximation \cite{Page1982}, which uses as its basis the Gaussian path-integral approximation for the thermal propagator \cite{BekensteinParker1981} on the ultrastatic spacetime to obtain an approximation for the RSET in the Hartle-Hawking state on static Einstein spacetimes. For a line-element of the general form (\ref{eq:metric}), the result is
	\begin{align}
		\label{eq:PageGen}
		\langle \hat{T}^{a}{}_{b}\rangle_{\textrm{P}}=\frac{1}{5760 \pi^{2}f^{2}}\Bigg\{4\kappa^{4}(\delta^{a}{}_{b}-4\delta^{a}{}_{0}\delta^{0}{}_{b})\nonumber\\
		+8 \tilde{C}^{ca}{}_{db}(3 f^{-1}\,f^{1/2}_{;c}f^{1/2}{}^{;d}-\tilde{R}^{d}{}_{c})\nonumber\\
		+\tfrac{2}{3}\tilde{B}^{a}{}_{b}-\tilde{A}^{a}{}_{b}-\tfrac{2}{3}\Lambda^{2}f^{2}\delta^{a}{}_{b}\Bigg\}
	\end{align}
	where the tildes here indicate that these tensors are evaluated on the conformally related ultrastatic spacetime $\tilde{g}_{ab}=f^{-1}g_{ab}$, $\kappa=2\pi \,T$ with $T$ the temperature, $\tilde{C}^{ca}{}_{db}$ is the Weyl tensor, and the tensors $\tilde{A}^{a}{}_{b}$ and $\tilde{B}^{a}{}_{b}$ are given by Eq.~(\ref{eq:A0B0}) evaluated on this ultrastatic spacetime. More explicitly, for a conformal scalar field in the Hartle-Hawking state on Schwarzschild-AdS spacetime where $f$ is given by (\ref{eq:fSchwarzschild_AdS}), the Page approximation yields
	\begin{widetext}
		\begin{eqnarray}
			\label{eq:T_P}
			\langle \hat{T}^{r}_{~r} \rangle_{\textrm{P}}&=&\frac{1}{2880 \pi ^2 L^8 r^8 f(r)^2} \bigg[2 L^8 \left(15 M^4-12 M^3 r+2 M^2 r^2+\kappa ^4 r^8\right)+4 L^6 M r^3 \left(6 M^2-9 M r+2 r^2\right)\nonumber\\&+&L^4 r^6 \left(-72 M^2+36 M r-5 r^2\right)+6 L^2 r^9 (2 M-r)-3
			r^{12}\bigg]\nonumber\\
			\langle \hat{T}^{t}_{~t} \rangle_{\textrm{P}}&=&\frac{1}{960 \pi ^2 L^8 r^8 f(r)^2} \bigg[L^8 \left(66 M^4-72 M^3 r+20 M^2 r^2-2 \kappa ^4 r^8\right)+4 L^6 M^2 r^3 (13 r-22 M)\nonumber\\&+&L^4 r^6 \left(48 M^2-4 M r+r^2\right)+2 L^2 r^9 (6 M-r)-r^{12}\bigg]\nonumber\\
			\langle \hat{T}^{\theta}_{~\theta} \rangle_{\textrm{P}}&=&\frac{1}{2880 \pi ^2 L^8 r^8 f(r)^2} \bigg[ L^8 \left(-18 M^4+24 M^3 r-8 M^2 r^2+2 \kappa ^4 r^8\right)+4 L^6 M r^3 \left(6 M^2-3 M r-r^2\right)\nonumber\\&+&L^4 r^6 \left(-36 M^2+12 M r-5 r^2\right)-6 L^2 r^{10}-3 r^{12}\bigg].\nonumber\\
		\end{eqnarray}
	\end{widetext}
	For conformal fields, there is no renormalization ambiguity and one can speak unambiguously about the accuracy of the approximation. For Schwarzschild spacetime, Eq.~(\ref{eq:PageGen}) is known to be a very good approximation for scalar fields. However, the approximation does not seem to perform as well for Schwarzschild-AdS. The reason is likely that, as already mentioned, Page's approximation uses the Gaussian path-integral approximation for the thermal propagator counting only the contributions coming from the shortest geodesic connecting points $x$ and $x'$. The contributions from other geodesics are ignored. The error in ignoring the other terms is compounded in Schwarzschild-AdS since the geodesic structure is more complicated owing to the unbounded potential \cite{Cruz_2005}. For example, there are families of bound null geodesics in the Schwarzschild-AdS spacetime that do not exist in the Schwarzschild spacetime. Simply put, ignoring all but the shortest geodesic connecting two points means there is more to ignore in Schwarzschild-AdS geometry relative to the Schwarzschild geometry. Notwithstanding the fact that this approximation is less accurate relative to its accuracy in Schwarzschild spacetime, we still adopt this approximation for conformal scalar fields in our backreaction scheme in the following section. In part because we are not aware of a better approximation for massless fields but also we note that adopting this approximation gives consistent equations (up to first order in our perturbation scheme) which satisfy all the criteria we set out above. Moreover, the approximation should still give reasonable order-of-magnitude estimates for the semi-classical effects on the background geometry.

	\subsection{Numerical Comparison of Approximations}
	\label{sec:assess}
	In order to draw conclusions on the accuracy of the Page and modifed DeWitt-Schwinger approximations, we require full numerical results for the expectation value of the stress-energy tensor of a quantum scalar field for comparison. To obtain these we employ the method of \cite{breenottewill:2012b}. As is well documented, obtaining numerical results for the quantum stress-energy tensor in a given spacetime is a long and arduous process and we refer the reader to \cite{breenottewill:2012b} for a detailed description of the approach taken in this paper. For now, it suffices to say that we used this method to calculate the non-zero components of the expectation of stress-energy tensor of a quantum scalar field in the Hartle-Hawking state on a 'tortoise-like' grid of 30 points in the exterior region of a Schwarzschild-AdS black hole, including the exact horizon value. For clarity of presentation we perform the numerical calculations in terms of the dimensionless radial coordinate $\zeta$ described in Sec (\ref{sec:AdSBH}) as it fixes the location of the event horizon to be unity for each parameter set. In the plots contained in this section we focus on a conformally and a minimally coupled field with  $r_{+}=L=10$. As we wish to compare both the Page approximation and the modified DeWitt-Schwinger approximation we include plots for both conformal and massive fields. Ideally in order to assess the accuracy of the modified DeWitt-Schwinger approximation, we would include comparison plots for increasing values of $m$ until a reasonable level of accuracy is observed.  However the error in the numerical scheme increases with increasing $m$, rendering comparisons unreliable for $m r_{+}>2$. Indeed even for $m r_+=2$ the errors in the numerical calculation are such that the results plotted in this section are only useful for comparison purposes and would not be of the accuracy required to, for example, solve the semi-classical field equations numerically. Inspection of the comparison plots for $m r_+=2$ demonstrate that the largest error in the approximation occurs at the event horizon and therefore a comparison on the horizon itself, using the exact results obtained in \cite{breenottewill:2012a}, should provide a good indication of the accuracy of the  modified DeWitt-Schwinger approximation in the entire exterior region for values of $m r_+>2$. The difficulty in obtaining highly accurate numerical results reinforces the need for approximations.
	
	To obtain plots for non-conformal fields, we must fix the renormalization ambiguity. We set the renormalisation lengthscale $\ell$ contained in the modified DeWitt-Schwinger approximation Eq.(\ref{eq:TModifiedDS}) to be
	\begin{align}
		\label{eq:lDS}
		\ell=
		\frac{2}{m}\exp\left\{\frac{6\xi-1}{2 m^{2}L^{2}}\right\}.
	\end{align}
	With this particular choice, $\langle \hat{T}^{a}{}_{b}\rangle^{\ell}_{\subDS}$ remains bounded in the large $m$ limit, in keeping with the large $m$ behaviour of the standard DeWitt-Swinger approximation.
	
	We must also fix a value for the arbitrary length scale contained in the numerical calculations of the exact $\langle \hat{T}^{a}_{~b} \rangle$, which we choose to set as
	\begin{align}
		\ell_{\textrm{numeric}}=\sqrt{2}\exp\left\{3/4-\gamma\right\} \ell 
	\end{align}
	where $\gamma$ is Euler's constant and $\ell$ is given by Eq. (\ref{eq:lDS}). Making this choice for the arbitrary length scales for both the approximate and exact $\langle \hat{T}^{a}_{~b} \rangle$ leads to agreement on the AdS boundary to leading order in $m$. Here and henceforth, we suppress the dependence of the numerical RSET on the Hartle-Hawking state and we suppress any reference to whether the approximate RSET is the Page approximation or the DeWitt-Schwinger approximation; the particular stress-energy tensor being considered will be clear from the context.

	\begin{figure*}
		\begin{subfigure}{.5\textwidth}
			\includegraphics[width=.8\linewidth]{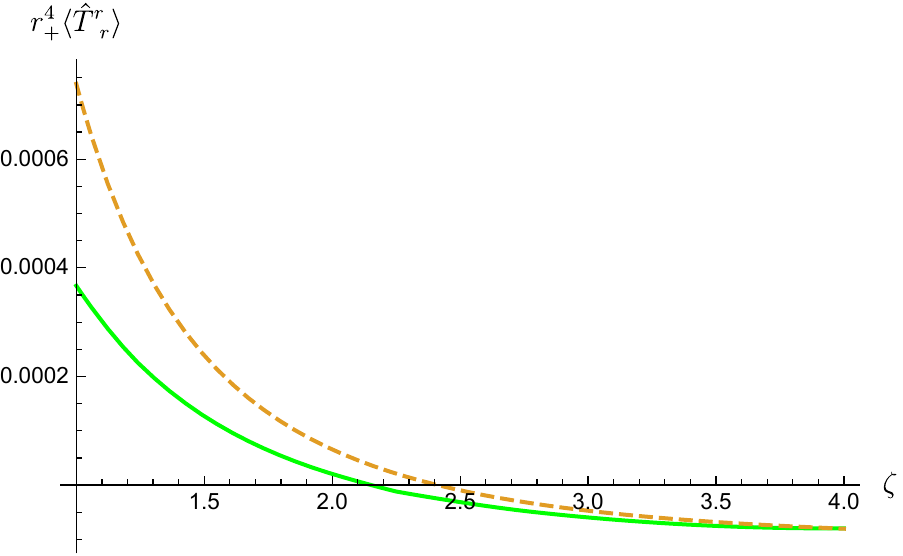}
			\caption{$r_+=L=10,m=0,\xi=1/6$ }
		\end{subfigure}%
		\begin{subfigure}{.5\textwidth}
			\includegraphics[width=.8\linewidth]{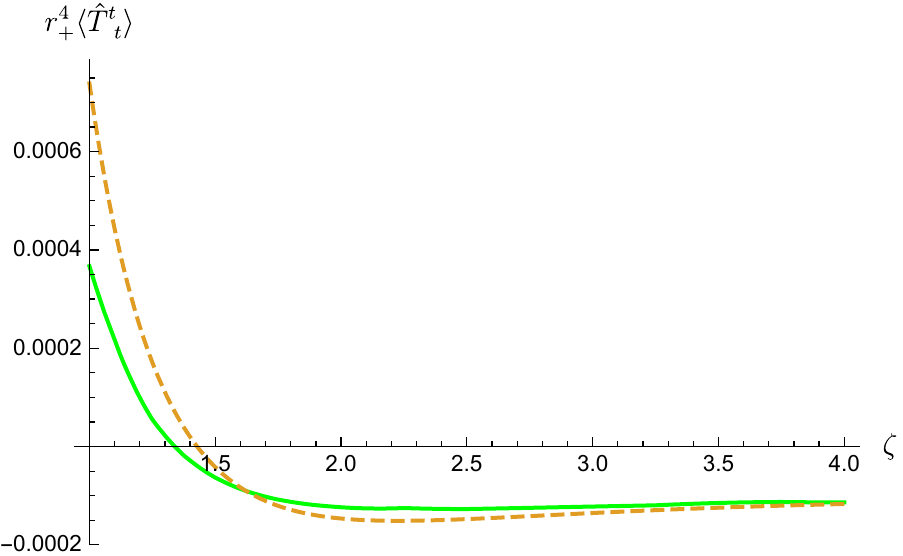}
			\caption{$r_+=L=10,m=0,\xi=1/6$}
		\end{subfigure}
		\caption{\label{fig:tabapproxxi16P} Plots of the $\langle \hat{T}^{r}_{~r} \rangle$ and $\langle \hat{T}^{t}_{~t} \rangle$ components in green alongside the Page approximation (dashed) for a conformal field in the Schwarzschild-AdS spacetime with $r_+=L=10$ }
	\end{figure*}
	\begin{figure*}
		\begin{subfigure}{.5\textwidth}
			\includegraphics[width=.8\linewidth]{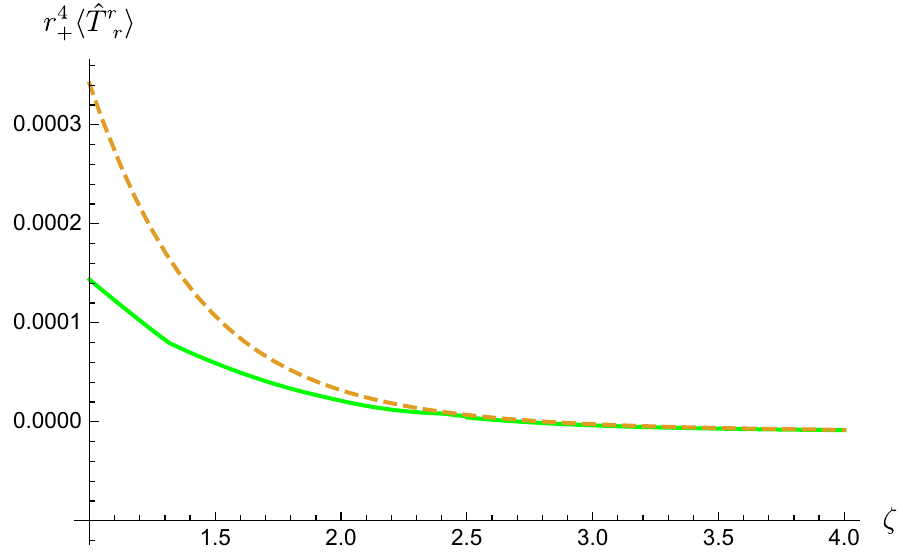}
			\caption{$r_+=L=10,m r_+=2,\xi=1/6$}
		\end{subfigure}%
		\begin{subfigure}{.5\textwidth}
			\includegraphics[width=.8\linewidth]{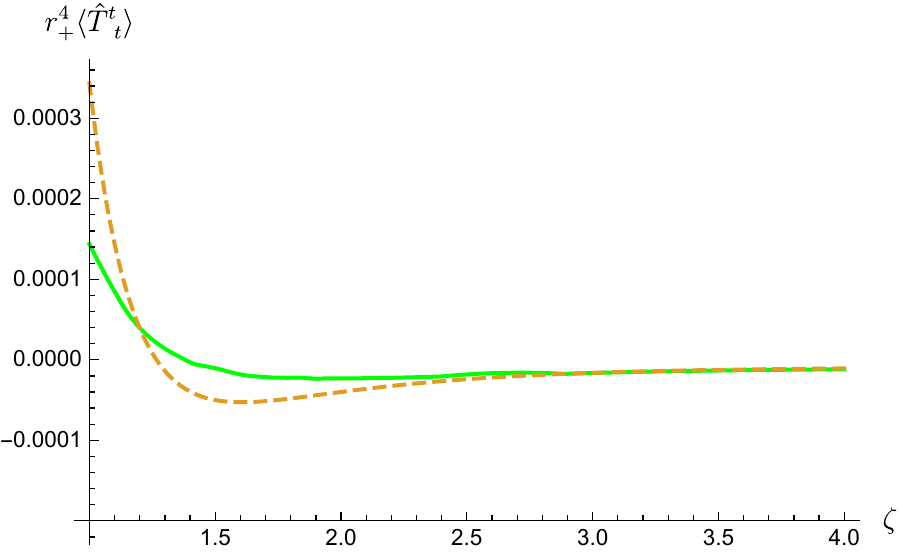}
			\caption{$r_+=L=10,m r_+=2,\xi=1/6$}
		\end{subfigure}
		\begin{subfigure}{.5\textwidth}
			\includegraphics[width=.8\linewidth]{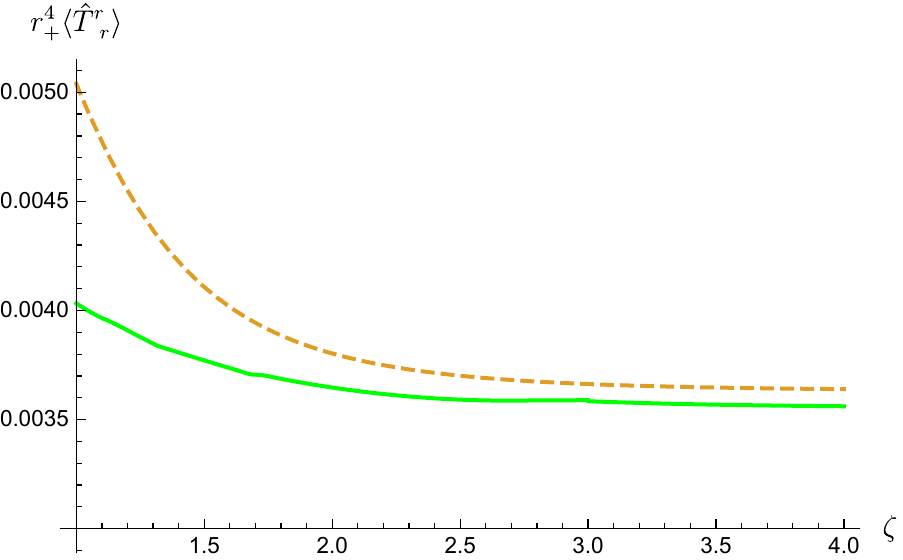}
			\caption{$r_+=L=10,m r_+=2 ,\xi=0$}
		\end{subfigure}%
		\begin{subfigure}{.5\textwidth}
			\includegraphics[width=.8\linewidth]{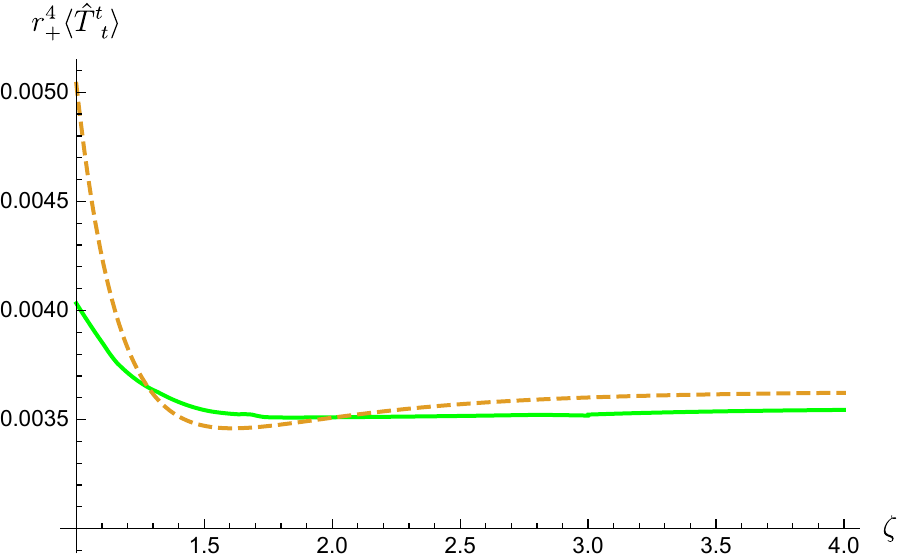}
			\caption{$r_+=L=10,mr_+=2 ,\xi=0$}
		\end{subfigure}
		\caption{\label{fig:tabapproxex} Plots of the $\langle \hat{T}^{r}_{~r} \rangle$ and $\langle \hat{T}^{t}_{~t} \rangle$ components in green alongside the modified DeWitt-Schwinger (dashed) for both a conformally and minimally coupled scalar field in Schwarzschild-AdS spacetime with $r_+=L=10$ and for $m r_+=2 $. }
	\end{figure*}

	\begin{figure*}
		\begin{subfigure}{.5\textwidth}
			\includegraphics[width=.8\linewidth]{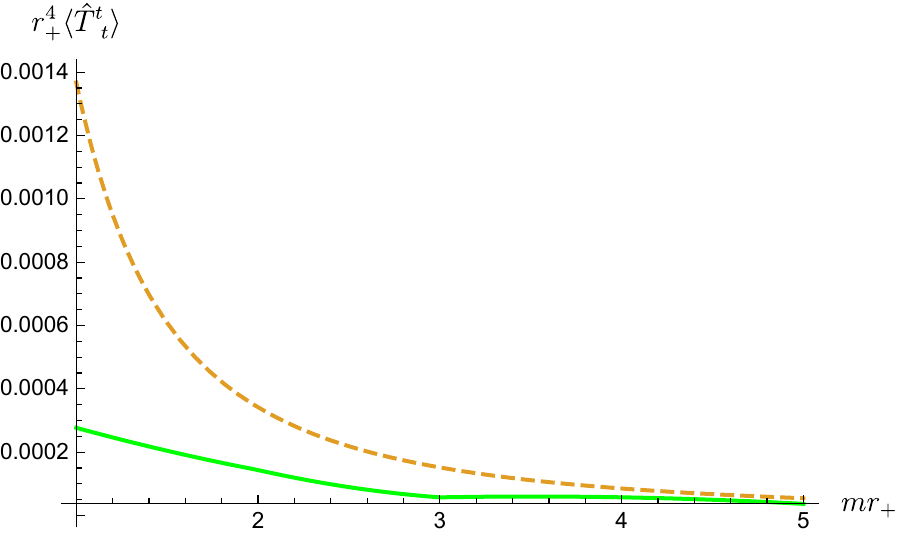}
			\caption{$r_+=L=10,\xi=1/6$}
		\end{subfigure}%
		\begin{subfigure}{.5\textwidth}
			\includegraphics[width=.8\linewidth]{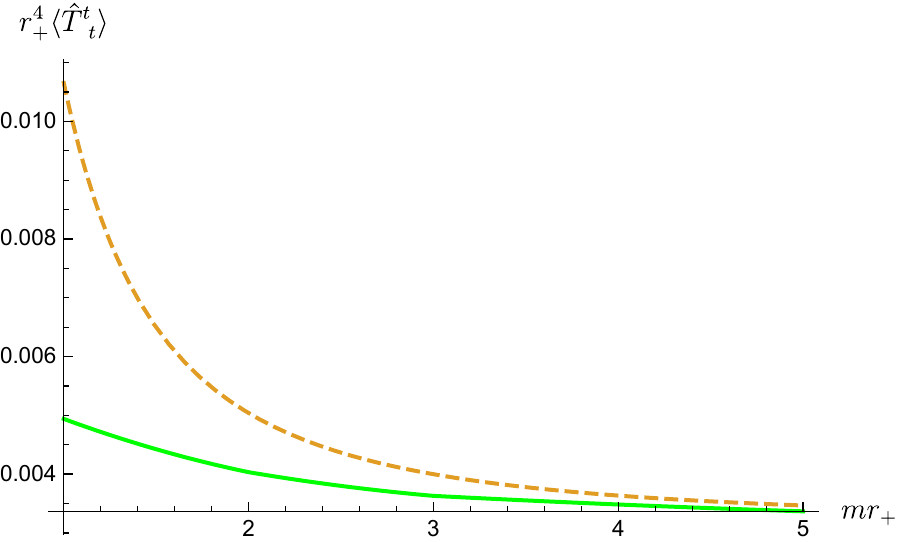}
			\caption{$r_+=L=10,\xi=0$}
		\end{subfigure}
		\caption{\label{fig:tabapproxhor} Plots of the $\langle \hat{T}^{t}_{~t} \rangle$ (=$\langle \hat{T}^{r}_{~r} \rangle$) component in green alongside the modified DeWitt-Schwinger (dashed). Here we plot event horizon values as a function of $m r_+$ for both a minimally and conformally coupled scalar field in Schwarzschild-AdS spacetime with $r_+=L=10$. }
	\end{figure*}
	
	In Figs. \ref{fig:tabapproxxi16P} and \ref{fig:tabapproxex}  we compare the full numerical results for the $\langle \hat{T}^{r}_{~r} \rangle$ and $\langle \hat{T}^{t}_{~t} \rangle$ components, calculated in terms of the variable $\zeta$ using the method of \cite{breenottewill:2012b}, with both the Page approximation (for conformal fields only) and the modified DeWitt-Schwinger approximation for both a conformally and minimally coupled scalar field in the Hartle-Hawking state with $r_{+}=L=10$ and  $m=0$ and $m=2/r_+$. In Fig. \ref{fig:tabapproxhor}, we plot the event horizon values of the $\langle \hat{T}^{t}_{~t} \rangle$ (=$\langle \hat{T}^{r}_{~r} \rangle$) component alongside its modified DeWitt-Schwinger approximation, with $r_{+}=L=10$, for both a conformally and minimally coupled scalar field as a function of increasing $m$. We choose to compare the $\langle \hat{T}^{r}_{~r} \rangle$ and $\langle \hat{T}^{t}_{~t} \rangle$ components only since, as will be seen in Section \ref{sec:backreaction}, they appear in the source terms of the semi-classical field equations. 
	
	In Fig. \ref{fig:tabapproxxi16P} for the conformal field we see that Page's approximation fails to be an accurate approximation in the vicinity of the event horizon but is accurate for intermediary values of $\zeta$ and is exact in the limit of $\zeta \to \infty$. On general grounds, we expect that the DeWitt-Schwinger approximation should be reasonably accurate when the Compton wavelength of the scalar field is much smaller than the black hole, which in Planck units implies that $m\,r_{+}\gg 1$. It is often the case however, that the approximation works even for modest values, for example, in Ref.~\cite{Anderson:1990} it was found that for the Reissner-Nordstr\"{o}m  spacetime, the DeWitt-Schwinger approximation is reliable in the vicinity of the event horizon provided that $m r_{+} \gtrsim 4$. For the Schwarzschild-AdS spacetime, we see in Figs. \ref{fig:tabapproxex} and \ref{fig:tabapproxhor} that the modified DeWitt-Schwinger approximation is not a very accurate approximation near the black hole for small values of $m$ but as expected the accuracy of this approximation, at least at the event horizon, increases with $m$. In the case of the Reissner-Nordstr\"{o}m  spacetime, the approximation fails to be accurate for large values of $r$ where the non-local temperature-dependent contribution to the quantum stress tensor dominates \cite{Anderson:1990}. However, in the Schwarzschild-AdS spacetime considered in this paper, we see that the modified DeWitt-Schwinger approximation becomes increasingly accurate as the distance from the black hole increases, at least with an appropriate choice of $\ell$. This is a consequence of the locally-measured temperature decreasing indefinitely as the AdS boundary is approached.
	
	\section{The Back Reaction}
	\label{sec:backreaction}
	\subsection{The Field Equations}

	In this section, we will use our approximation for the expectation of the stress-energy tensor for a quantum scalar field to solve the back reaction equations on the Schwarzschild-AdS spacetime to first order in $\hbar$. 
	
	We seek solutions of Eq.~(\ref{eq:perturbationeqn}) with the source term $\langle A_{\gb}|\hat{T}_{ab}[\gb]|A_{\gb}\rangle$ given by Eq.~(\ref{eq:TModifiedDS}) for massive fields and Eq.~(\ref{eq:T_P}) for conformal fields. As discussed in the previous section, both $A_{ab}^{(0)}[\gb]$ and $B_{ab}^{(0)}[\gb]$ in Eq.~(\ref{eq:perturbationeqn}) vanish for the Schwarzschild-AdS spacetime, by virtue of Eq.~(\ref{eq:Ricci}). In order to solve Eq.~(\ref{eq:perturbationeqn}), one requires a functional form for the perturbation of the classical metric, $h_{ab}$. Following the method of \cite{York:1985}, we are able to reduce Eq.~(\ref{eq:perturbationeqn}) to a system of two ODEs, which are easily integrated in terms of elementary functions. While we assume in this study that the backreaction and the background geometry are static, we refer the reader to \cite{Hayward:2006,Frolov:1981,Stephens:1994,Ashtekar:2005,Haggard:2015,Yokokura:2013,Yokokura:2016,Yokokura:2017,Yokokura:2020,Ho:2016} for interesting work on the backreaction of evaporating black holes.
	
	Before proceeding we note that the approximate RSET for both massive fields and conformally invariant fields have a contribution that is proportional to the background metric. Moreover, for massive fields, this contribution contains the arbitrary lengthscale $\ell$ which encodes the renormalization ambiguity. As we have stressed above, this ambiguity cannot be physical and is degenerate with ambiguities in the renormalization of coefficients on the left-hand side of the semi-classical equation. It is therefore more transparent at this point to absorb the terms in the RSET that are proportional to the metric into a renormalization of the cosmological constant. In doing this, we render the equations and hence solutions explicitly independent of the lengthscale $\ell$. In other words, we now consider a perturbation about a Schwarzschild-AdS black hole with a renormalized cosmological constant $\hat{\Lambda}=-3/\hat{L}^{2}$, where $\hat{\Lambda}$ has absorbed terms in the RSET proportional to the metric. For the DeWitt-Schwinger approximation for massive fields, the renormalization is explicitly given by
	\begin{align}
		\hat{\Lambda}=\Lambda-8\pi\,\hbar\,T_{\textrm{AdS}}
	\end{align}
	where we have identified the coefficient of the metric in the RSET as
	\begin{align}
		T_{\subADS}=\frac{1}{2520 \pi^{2}}\Bigg\{\frac{1}{m^{2}\hat{L}^{6}}(185-45360\xi^{3}+22680\xi^{2}-3654\xi)\nonumber\\
		-\frac{315 m^{2}}{\hat{L}^{2}}\left(m_{\xi}^{2}\hat{L}^{2}\ln(\ell\,m/2)-3(\xi-\tfrac{1}{6})\right)\Bigg\}.
	\end{align}
	Then the perturbation $h_{ab}$ satisfies a simplified equation
	\begin{align}
		\label{eq:pertren}
		G_{ab}^{(1)}[h]+\hat{\Lambda}\,h_{ab}=8\pi\,\langle \hat{T}_{ab}\rangle
	\end{align}
	where $\langle \hat{T}_{ab}\rangle$ is given by Eq.~(\ref{eq:T_DS}) with the constant terms omitted and with the bare cosmological lengthscale $L$ replaced by the renormalized lengthscale $\hat{L}$, namely,
	\begin{widetext}
		\begin{align}
			\label{eq:T_DS_ren}
			\langle \hat{T}^{r}_{~r} \rangle &=\frac{1}{10080 \pi ^2 m^2 \hat{L}^6 r^9}\bigg[14 \hat{L}^6 M^3 (216 \xi -47)-21 \hat{L}^4 M^2 r \left(3 \hat{L}^2 (32 \xi -7)+(168 \xi -37) r^2\right)\bigg]\nonumber\\
			\langle \hat{T}^{t}_{~t} \rangle &=\frac{1}{10080 \pi ^2 m^2 \hat{L}^6 r^9}\bigg[2 \hat{L}^6 M^3 (1237-5544 \xi )+3  \hat{L}^4 M^2 r \left(15 \hat{L}^2 (112 \xi -25)+(1176 \xi -263) r^2\right)\bigg]\nonumber\\
			\langle \hat{T}^{\theta}_{~\theta} \rangle &=\frac{1}{10080 \pi ^2 m^2 \hat{L}^6 r^9}\bigg[2  \hat{L}^6 M^3 (1543-7056 \xi )+3  \hat{L}^4 M^2 r \left(63 \hat{L}^2 (32 \xi -7)+(1176 \xi -257) r^2\right)\bigg].
		\end{align}
	\end{widetext}
	Similarly, for conformal scalar fields, the last term in Page's approximation given by Eq.~(\ref{eq:PageGen}) is proportional to the metric and can be renormalized in the same way with
	\begin{align}
		\hat{\Lambda}=\Lambda+\frac{\hbar}{120 \pi \hat{L}^{4}}.
	\end{align}
	Then the perturbation satisfies (\ref{eq:pertren}) with a simplified source term given by
	\begin{widetext}
		\begin{align}
			\label{eq:T_P_ren}
			\langle \hat{T}^{r}_{~r} \rangle&=\frac{1}{1440 \pi ^2 f(r)^2} \bigg[ \frac{ \left(15 M^4-12 M^3 r+2 M^2 r^2+\kappa ^4 r^8\right)}{r^{8}}+\frac{2  M  \left(6 M^2-9 M r+2 r^2\right)}{\hat{L}^{2}r^{5}}- \frac{\left(30 M^2-12 M r+ r^2\right)}{\hat{L}^{4}r^{2}}\bigg]\nonumber\\
			\langle \hat{T}^{t}_{~t} \rangle&=\frac{1}{480 \pi ^2 f(r)^2} \bigg[\frac{ \left(33 M^4-36 M^3 r+10 M^2 r^2- \kappa ^4 r^8\right)}{r^{8}}+\frac{2 M^2(13 r-22 M)}{\hat{L}^{2}r^{5}}+\frac{\left(26 M^2-4 M r+r^2\right)}{\hat{L}^{4}r^{2}}+\frac{4 M\,r}{\hat{L}^{6}} \bigg]\nonumber\\
			\langle \hat{T}^{\theta}_{~\theta} \rangle&=\frac{1}{1440 \pi ^2 f(r)^2} \bigg[\frac{ \left(-9 M^4+12 M^3 r-4 M^2 r^2+ \kappa ^4 r^8\right)}{r^{8}}+\frac{2  M  \left(6 M^2-3 M r-r^2\right)}{\hat{L}^{2}r^{5}}+\frac{ \left(-12 M^2- r^2\right)}{\hat{L}^{4}r^{2}}-\frac{6 M\,r}{\hat{L}^{6}}\bigg].\nonumber\\
		\end{align}
	\end{widetext}

	We can now proceed to solve Eq.~(\ref{eq:pertren}) with these simplified source terms. We begin by transforming coordinates from the Schwarzschild-like coordinates given in Eq.~(\ref{eq:metric}) into the equivalent of the advanced-time Eddington-Finkelstein coordinates, given by the following transformation:
	\begin{eqnarray}
		\label{eqn:coord}
		dv= dt+\frac{1}{f(r)}dr,
	\end{eqnarray}
	where $f(r)=1-2 M/r+r^{2}/\hat{L}^{2}$.	In this coordinate transformation, the Schwarzschild-AdS line element takes the form:
	\begin{equation}
		ds^2=\gb_{ab}dx^adx^b=-f(r) ~dv^2+2 dv ~ dr +r^{2}d\Omega^{2}_{2}.
	\end{equation}
	and the relevant components of the quantum stress-energy tensor transform as
	\begin{eqnarray}
		&&\langle \hat{T}^{v}_{~v} \rangle=\langle \hat{T}^{t}_{~t} \rangle\nonumber\\
		&&\langle \hat{T}^{v}_{~r} \rangle=\frac{1}{f(r)}\left(\langle \hat{T}^{r}_{~r} \rangle-\langle \hat{T}^{t}_{~t} \rangle\right).
	\end{eqnarray}
	We note here that $\langle \hat{T}^{v}_{~r} \rangle$ remains finite as the event horizon is approached ($f(r) \to 0$).
	We seek solutions of the reduced-order semi-classical equations that respect the symmetries of the background metric $\gb_{ab}$, in other words we seek a perturbed metric $g_{ab}$ that is both static and spherically symmetric.
	
	In the new coordinate system, we may satisfy this condition by taking $g_{ab}$ to have the form:
	\begin{eqnarray}
		\label{eq:fullmetric}
		&&g_{v v}=-\Psi(r)^2\left(1-\frac{2 \mathfrak{M}(r)}{r}+\frac{r^2}{\hat{L}^2}\right)\nonumber\\
		&&g_{v r}=g_{rv}= \Psi(r)\nonumber\\
		&&g_{\theta \theta}=\gb_{\theta\theta}~~~~~~~g_{\phi \phi}=\gb_{\phi \phi}
	\end{eqnarray}
	where $\Psi(r)$ is assumed to be non-vanishing in the region of interest. In the spirit of the reduction-of-order approach outlined in Section (\ref{sec:semi-classical}), we assume a perturbation expansions of the functions $\Psi(r)$ and $\mathfrak{M}(r)$ of the form:
	\begin{eqnarray}
		&&\Psi(r)=1+ \hbar\, \rho(r) +O(\hbar^2)\nonumber\\
		&&\mathfrak{M}(r)=M(1+ \hbar \,\mu(r))+O(\hbar^2).
	\end{eqnarray}
	Inserting these expansions into the expressions contained in Eq.~(\ref{eq:fullmetric}) and neglecting terms that are $O(\hbar^2)$, yields the following form of the first order metric perturbation,
	\begin{eqnarray}
		\label{eq:hab}
		&&h_{v v}=-2 f(r) \rho(r)+\frac{2 M}{r} \mu(r) \nonumber\\
		&&h_{v r}=h_{rv}= \rho(r)\nonumber\\
		&&h_{\theta \theta}=h_{\phi \phi}=0.
	\end{eqnarray}
	Inserting $h_{ab}$ into Eq. (\ref{eq:linearized}) yields to first order in $\hbar$:
	\begin{eqnarray}
		G^{(1)r}_{~~~~r}+\hat{\Lambda}\, h^{r}_{~r}&&=\frac{2}{r^2}\bigg[r f(r) \rho'(r)- M \mu'(r)\bigg]\nonumber\\
		G^{(1)v}_{~~~~r}+\hat{\Lambda}\, h^{v}_{~r}&&=\frac{2 }{r}\rho'(r) \nonumber\\
		G^{(1)v}_{~~~~v}+\hat{\Lambda}\, h^{v}_{~v}&&=-\frac{2 M}{r^2}\mu'(r).
	\end{eqnarray}
	
	All other components vanish except $G^{(1)\theta}_{~~~~\theta}$ and $G^{(1)\phi}_{~~~~\phi}$, which are equal and which we can ignore as the field equations involving these terms are satisfied identically through the contracted Bianchi identity.
	Substituting these in turn into  Eq. (\ref{eq:pertren}) yields the following set of ODEs
	\begin{align}
		\label{eq:ODEs}
		&\frac{2 M}{r^2} \mu'(r)= -8 \pi  \langle \hat{T}^{t}_{~t}\rangle\nonumber\\
		&\frac{2}{r} \rho'(r)= 8 \pi \langle \hat{T}^{v}_{~r}\rangle=\frac{8 \pi}{f(r)}\left(\langle \hat{T}^{r}_{~r} \rangle-\langle \hat{T}^{t}_{~t} \rangle\right) \nonumber\\
		&\frac{2}{r^2}\bigg[r f(r) \rho'(r)-M \mu'(r)\bigg]= 8 \pi  \langle \hat{T}^{r}_{~r}\rangle,
	\end{align}
	where the components of $\langle \hat{T}^{a}{}_{b}\rangle$ are given by Eqs.~(\ref{eq:T_DS_ren}) and (\ref{eq:T_P_ren}) for massive and conformally invariant scalar fields, respectively. Only two of these equations above are independent, clearly the first two equations imply the third. 
	
	\subsection{Backreaction for Massive Fields}
	\label{sec:backreactionmassive}
	For massive fields, Eqs.~(\ref{eq:ODEs}) are straightforward to integrate with (\ref{eq:T_DS_ren}) as the source-term yielding
	
	\begin{align}
		\label{eq:soln}
		\mu(r)&=\frac{1}{2520\pi \,m^{2}}\Bigg\{\frac{M^{2}}{3r^{6}}(1237-5544\xi)+\frac{9 M}{r^{5}}(112\xi-25)\nonumber\\
		&+\frac{M}{\hat{L}^{2}r^{3}}(1176\xi-263)\Bigg\}+C_{0}\nonumber\\
		\rho(r)&=\frac{M^2 (392 \xi -87)}{840 \pi\,m^{2}  r^6}+K_0
	\end{align}	
	where the parameters $K_0$ and $C_0$ are dimensional constants of integration with $[C_{0}]=[K_{0}]=[\textrm{length}]^{-2}$. Both $\mu(r)$ and $\rho(r)$ are evidently regular at the event horizon of the background spacetime. It is useful for later convenience to express these solutions in terms of $(r_{+}, \hat{L})$ rather than $(M, \hat{L})$ using the fact that $2 M=r_{+}(1+r_{+}^{2}/\hat{L}^{2})$. Making this substitution and redefining the integration constants so that $\mu(r_{+})=C_{0}$ and $\rho(r_{+})=K_{0}$ gives
	\begin{align}
		\label{eq:solnDS}
		\mu(r)=&-\frac{(\hat{L}^{2}+r_{+}^{2})^{2}}{30240\pi \hat{L}^{4}r_{+}^{4}m^{2}}\Bigg\{(1237-5544 \xi)\left(1-\frac{r_{+}^{6}}{r^{6}}\right)\nonumber\\
		&+\frac{6 r_{+}^{2}(1176\xi-263)}{(\hat{L}^{2}+r_{+}^{2})}\left(1-\frac{r_{+}^{3}}{r^{3}}\right)\nonumber\\
		&+\frac{54 \hat{L}^{2}(112\xi-25)}{(\hat{L}^{2}+r_{+}^{2})}\left(1-\frac{r_{+}^{5}}{r^{5}}\right)\Bigg\}+C_{0},\nonumber\\
		\rho(r)=&-\frac{(\hat{L}^{2}+r_{+}^{2})^{2}(392\xi-87)}{3360\pi \hat{L}^{4}r_{+}^{4}m^{2}}\left(1-\frac{r_{+}^{6}}{r^{6}}\right)+K_{0}.
	\end{align}

	It is useful to adopt a notation,
	\begin{align}
		\mu(r)=\mu_{0}(r)+C_{0},
	\end{align}
	where $\mu_{0}(r_{+})=0$ since with this choice, we can re-express the semi-classical spacetime in Schwarzschild-like coordinates using the coordinate transformation
	\begin{align}
		dv=dt+\frac{dr}{\Psi(r)f(r)},
	\end{align}
	whence
	\begin{align}
		\label{eq:perturbedmetric1}
		ds^{2}=-\hat{f}(r)\left(1+2\hbar \,\rho(r)-2 \hbar\,\psi(r)\right)dt^{2}\nonumber\\
		+\frac{\left(1+2\hbar\,\psi(r)\right)}{\hat{f}(r)}dr^{2}+r^{2}d\Omega^{2},
	\end{align}
	where we have defined
	\begin{align}
		\label{eq:hatmetricfunctions}
		\hat{f}(r)=1-\frac{2 \hat{M}}{r}+\frac{r^{2}}{\hat{L}^{2}}\nonumber\\
		\psi(r)=\frac{\hat{M}}{r}\frac{\mu_{0}(r)}{\hat{f}(r)}.
	\end{align}
	In arriving at this particular form for the back-reacted metric, we have absorbed the constant $C_{0}$ into a quantum dressed mass $\hat{M}=M(1+\hbar\,C_{0})$ so that $C_{0}$ plays no further role. We have also freely replaced $M$ with $\hat{M}$ in $O(\hbar)$ terms to arrive at this expression, which ignores terms of order $O(\hbar^{2}).$ This metric is now explicitly regular at $r=r_{+}$, the horizon of the background spacetime. If we further define
	\begin{align}
		\rho(r)=\rho_{0}(r)+K_{0},
	\end{align}
	so that $\rho_{0}(r_{+})=0$, and rescaling the time coordinate by
	\begin{align}
		t\to \frac{t}{1+\hbar\,K_{0}}
	\end{align}
	allows us to absorb the constant $K_{0}$ into the scaling symmetry, at least up to the approprite order in $\hbar$. This implies that the constant $K_{0}$ is equivalent to the freedom we have in normalizing the Killing vector associated with the time-translation invariance. Since this Killing vector is also the generator of the horizon, a choice of normalization forms part of the definition of the surface gravity. We will revisit the consequences of this in the next section. For now, we take our semi-classical black hole to be
	\begin{align}
		\label{eq:perturbedmetric2}
		ds^{2}=-\hat{f}(r)\left(1+2\hbar \,\rho_{0}(r)-2 \hbar\,\psi(r)\right)dt^{2}\nonumber\\
		+\frac{\left(1+2\hbar\,\psi(r)\right)}{\hat{f}(r)}dr^{2}+r^{2}d\Omega^{2},
	\end{align}
	with $\hat{f}(r)$ and $\psi(r)$ given by (\ref{eq:hatmetricfunctions}). The only difference between the expressions (\ref{eq:perturbedmetric1}) and (\ref{eq:perturbedmetric2}) is that $\rho$ is replaced with $\rho_{0}$. Of course, the time coordinate is different too but this is suppressed in the notation. 
	
	Now the solution as presented is still not expressed in a form appropriate for making any physical observation since it contains both the bare and dressed mass of the black hole, the former occurring implicitly in the background horizon radius $r_{+}$ which occurs frequently in $\mu_{0}(r)$ and $\rho_{0}(r)$. The bare mass is not accessible to experiment or observation, only the quantum dressed mass and hence we need a means to replace the dependence on the background horizon radius $r_{+}$ with the perturbed horizon radius. To obtain the location of the perturbed event horizon, we consider the expansion of an outgoing future directed null vector field $l^{a}$, given in $(v,r,\theta,\phi)$ coordinates by
	\begin{equation}
		l^a=\left[1,\frac{1}{2}\Psi(r)\left(1-\frac{2 \mathfrak{M}(r)}{r}+\frac{r^2}{\hat{L}^2}\right) ,0,0\right].
	\end{equation}
	The radially ingoing null vector field is
	\begin{equation}
		\beta^{a}=\left[0, -\frac{1}{\Psi(r)},0,0\right]
	\end{equation}
	with the normalisation of $\beta_a l^a=-1$. The event horizon is then defined by the vanishing of the expansion of $l^a$, which is given by:
	\begin{eqnarray}\Theta=\nabla_a l^{a}+ \beta^al^b \nabla_b l_a=\frac{\Psi(r)}{r}\left(1-\frac{2 \mathfrak{M}(r)}{r}+\frac{r^2}{\hat{L}^2}\right)\nonumber\\
	\end{eqnarray}
	where covariant differentiation here is with respect to the full metric $g_{ab}$ in the $(v,r,\theta,\phi)$ coordinate system. In terms of $\hat{M}$ and expanded up to and including terms of first order in $\hbar$, the expansion is
	\begin{align}
		\Theta=\frac{1}{r}\hat{f}(r)\Big[1+\hbar\,\rho_{0}(r)-2\hbar\,\psi(r)\Big].
	\end{align}
	Since the $\mathcal{O}(\hbar)$ term is assumed to be sufficiently small in the perturbative scheme employed throughout, the term in square brackets above is nonzero and hence the expansion vanishes at a root of $\hat{f}(r)$. In other words, the semi-classical black hole event horizon, which we label by $\hat{r}_{+}$, satisfies
	\begin{align}
		\label{eq:rhatEq}
		1-\frac{2 \hat{M}}{\hat{r}_{+}}+\frac{\hat{r}_{+}^2}{\hat{L}^2}=0.
	\end{align}
	It is straightforward to express $\hat{r}_{+}$ in terms of the background horizon radius by recalling that $\hat{M}=M(1+\hbar\,C_{0})$ and adopting the expansion
	\begin{eqnarray}
		\label{eq:EHCorrection}
		\hat{r}_{+}&= r_{+}+ \hbar\, \delta r_{+},
	\end{eqnarray}
	in (\ref{eq:rhatEq}). Equating order-by-order in $\hbar$ yields
	\begin{align}
		\label{eq:Deltarp}
		\delta r_{+}=\frac{2 M \,\hat{L}^{2}\mu(r_{+})}{3r_{+}^{2}+\hat{L}^{2}}= \frac{ M \,C_{0}}{\kappa\, r_{+}},
	\end{align}
	where the last equality follows from Eq.~(\ref{eq:kappaR}). We can see clearly now that the constant $C_0$ effectively sets the location of the event horizon. We can also invert this expansion which allows us to express $r_{+}$ in terms of $\hat{r}_{+}$, that is,
	\begin{align}
		r_{+}[\hat{r}_{+}]=\hat{r}_{+}-\hbar\,\delta r_{+}[\hat{r}_{+}]+\mathcal{O}(\hbar^{2}),
	\end{align}
	where $\delta r_{+}[\hat{r}_{+}]$ is given by Eq.~(\ref{eq:Deltarp}) with $r_{+}$ replaced by $\hat{r}_{+}$. The explicit expression for $\delta r_{+}$ is actually irrelevant for our purposes, since we only require terms to first order in $\hbar$ which implies we can replace $r_{+}$ with $\hat{r}_{+}$ in any $\mathcal{O}(\hbar)$ terms. In particular, we define
	
	\begin{align}
		\hat{\mu}_{0}(r)=&-\frac{(\hat{L}^{2}+\hat{r}_{+}^{2})^{2}}{30240\pi \hat{L}^{4}\hat{r}_{+}^{4}m^{2}}\Bigg\{(1237-5544 \xi)\left(1-\frac{\hat{r}_{+}^{6}}{r^{6}}\right)\nonumber\\
		&+\frac{6 \hat{r}_{+}^{2}(1176\xi-263)}{(\hat{L}^{2}+\hat{r}_{+}^{2})}\left(1-\frac{\hat{r}_{+}^{3}}{r^{3}}\right)\nonumber\\
		&+\frac{54 \hat{L}^{2}(112\xi-25)}{(\hat{L}^{2}+\hat{r}_{+}^{2})}\left(1-\frac{\hat{r}_{+}^{5}}{r^{5}}\right)\Bigg\},\nonumber\\
		\hat{\rho}_{0}(r)=&-\frac{(\hat{L}^{2}+\hat{r}_{+}^{2})^{2}(392\xi-87)}{3360\pi \hat{L}^{4}\hat{r}_{+}^{4}m^{2}}\left(1-\frac{\hat{r}_{+}^{6}}{r^{6}}\right),
	\end{align}
	as well as
	\begin{align}
		\label{eq:psihat}
		\hat{\psi}(r)=\frac{\hat{M}\hat{\mu}_{0}(r)}{r\,\hat{f}(r)},
	\end{align}
	then the solution in terms of dressed mass $\hat{M}$ and the scalar hair (the field mass and coupling strength) is
	\begin{align}
		\label{eq:dressedmetric}
		ds^{2}=-\hat{f}(r)\left(1+2\hbar \,\hat{\rho}_{0}(r)-2 \hbar\,\hat{\psi}(r)\right)dt^{2}\nonumber\\
		+\frac{\left(1+2\hbar\,\hat{\psi}(r)\right)}{\hat{f}(r)}dr^{2}+r^{2}d\Omega^{2}.
	\end{align}
	
	Finally, we can express this solution in a form that is explicitly regular at $r=\hat{r}_{+}$ by converting back to Eddington-Finkelstein coordinates and keeping only terms up to the appropriate order, yielding
	\begin{align}
		\label{eq:dressedmetricEF}
		ds^{2}=-\hat{f}(r)\left(1+2\hbar\,\hat{\rho}_{0}(r)-2\hbar\,\hat{\psi}(r)\right)dv^{2}\nonumber\\
		+2(1+\hbar\,\hat{\rho}_{0}(r))\,dv\,dr+r^{2}d\Omega^{2}.
	\end{align}
	In order for the perturbative scheme to make sense, we require, at the very least, that $ |h_{ab}|/|\gb_{ab}|$ is bounded throughout the exterior. In York's exposition of the backreaction on the Schwarzschild spacetime \cite{York:1985}, which employs Page's approximation to the stress-energy tensor \cite{Page1982}, the perturbation grows with $r$ even though the background is asymptotically flat; hence, the approximation breaks down at large $r$. To circumvent this issue, York solves for the backreaction in a finite cavity around the black hole and then matches this to an asymptotically flat Schwarzschild solution with appropriate boundary conditions across the matching surface. In our case, the asymptotic behaviour of the solutions implies
	\begin{align}
		\frac{|h_{vv}|}{|\gb_{vv}|}=&\left|\frac{(392 \xi -87) \left(\hat{L}^2+\hat{r}_+^2\right)^2}{1680 \pi  \hat{L}^4 m^2 \hat{r}_+^4}\right|\nonumber\\
		&+O(r^{-3}).\nonumber\\
		\frac{|h_{vr}|}{|\gb_{vr}|}=&\left|\frac{(392 \xi -87) \left(\hat{L}^2+\hat{r}_+^2\right)^2}{3360 \pi  \hat{L}^4 m^2 \hat{r}_+^4}\right|+O(r^{-6}).
	\end{align}
	From these expressions, it is clear that the perturbation remains bounded for large $r$ but also in the $\hat{L} \to \infty$. In other words, the perturbation induced by a massive field is bounded over the entire exterior spacetime and preserves the asymptotic structure of the background spacetime, even in the Schwarzschild limit. This is in contrast to the unbounded nature of the perturbations induced by a conformal field for large $r$ in the Schwarzschild geometry \cite{York:1985}. 
	
	\subsection{Backreaction for Conformal Fields}
	We now turn to solving for the backreaction for massless, conformally coupled fields by integrating Eqs.~(\ref{eq:ODEs}) with the source-term given by (\ref{eq:T_P_ren}). Integrating these equations is not as elementary as the massive case, indeed it is not even obvious that the Page approximation is regular at the event horizon. It is useful to first rewrite the components in (\ref{eq:T_P_ren}) in terms of $(r_{+},\hat{L})$ rather than $(M, \hat{L})$. Then factorizing the metric function as
	\begin{align}
		f(r)=\frac{(r-r_{+})}{r}h(r),\qquad h(r)=1+\frac{1}{\hat{L}^{2}}(r^{2}+r \,r_{+}+r_{+}^{2}),
	\end{align}
	gives the following form for the source:
	\begin{widetext}
		\begin{align}
			\langle \hat{T}^{t}{}_{t}\rangle=&\frac{(\hat{L}^{2}+r_{+}^{2})^{2}}{7680\pi^{2}\hat{L}^{8}r_{+}^{4}r^{6}h^{2}(r)}\Big\{\frac{32 r_{+}^{5}r^{7}}{(\hat{L}^{2}+r_{+}^{2})}-(\hat{L}^{4}+10 \hat{L}^{2}r_{+}^{2}+17 r_{+}^{4})r^{6}-2 r_{+}(\hat{L}^{4}+10 \hat{L}^{2}r_{+}^{2}+33 r_{+}^{4})r^{5}\nonumber\\
			&-r_{+}^{2}(3 \hat{L}^{4}+30\hat{L}^{2}r_{+}^{2}+11 r_{+}^{4})r^{4}-4 r_{+}^{3}(\hat{L}^{2}-r_{+}^{2})(\hat{L}^{2}+11r_{+}^{2})r^{3}-r_{+}^{4}(5 \hat{L}^{4}-54 \hat{L}^{2}r_{+}^{2}-99 r_{+}^{4})r^{2}\nonumber\\
			&-6 r_{+}^{5}(\hat{L}^{2}-11 r_{+}^{2})(\hat{L}^{2}+r_{+}^{2})r+33r_{+}^{6}(\hat{L}^{2}+r_{+}^{2})^{2}\Big\}\nonumber\\
			\langle \hat{T}^{r}{}_{r}\rangle=&\frac{(\hat{L}^{2}+r_{+}^{2})^{2}}{23040\pi^{2}L^{8}r_{+}^{4}r^{6}h^{2}(r)}\Big\{\frac{(\hat{L}^{2}+9r_{+}^{2})}{(\hat{L}^{2}+r_{+}^{2})}(\hat{L}^{4}+2\hat{L}^{2} r_{+}^{2}+9 r_{+}^{4})r^{6}+\frac{2 r_{+}}{(\hat{L}^{2}+r_{+}^{2})}(\hat{L}^{6}+11 \hat{L}^{4}r_{+}^{2}+75\hat{L}^{2} r_{+}^{4}+81r_{+}^{6})r^{5}\nonumber\\
			&+3r_{+}^{2}(\hat{L}^{4}+10\hat{L}^{2}r_{+}^{2}+41 r_{+}^{4})r^{4}+4r_{+}^{3}(\hat{L}^{4}+18 \hat{L}^{2}r_{+}^{2}+21r_{+}^{4})r^{3}+r_{+}^{4}(5 \hat{L}^{4}+42 \hat{L}^{2}r_{+}^{2}+45 r_{+}^{4})r^{2}\nonumber\\
			&+6 r_{+}^{5}(\hat{L}^{2}+5 r_{+}^{2})(\hat{L}^{2}+r_{+}^{2})\,r+15r_{+}^{6}(\hat{L}^{2}+r_{+}^{2})^{2}\Big\}.
		\end{align}
		
		These particular expressions are explicitly regular on the horizon and are more convenient to integrate. Employing these expressions in (\ref{eq:ODEs}) and as before, writing $\mu(r)=\mu_{0}(r)+C_{0}$ and $\rho(r)=\rho_{0}(r)+K_{0}$ with $\mu_{0}(r_{+})=0=\rho_{0}(r_{+})$ gives expressions for the perturbations $\mu_{0}(r)$ and $\rho_{0}(r)$ in terms of the background horizon radius $r_{+}$. Following the same procedure as in the massive case, we absorb the integration constants $C_{0}$ and $K_{0}$ into a dressed mass $\hat{M}$ and a time-rescaling, respectively, and we replace $r_{+}$ with $\hat{r}_{+}$ in all $\mathcal{O}(\hbar)$ terms. The result is Eq.~(\ref{eq:dressedmetric}) with $\hat{\mu}_{0}(r)$ and $\hat{\rho}_{0}(r)$ given by
		\begin{align}
			\label{eq:solnmuPage}
			\hat{\mu}_{0}(r)&=\frac{1}{960 \pi \hat{L}^{2}\hat{r}_{+}^{5}}\Bigg\{3 \hat{L}^{2}\hat{r}_{+}^{3}\left(1-\frac{\hat{r}_{+}^{2}}{r^{2}}\right)+\frac{5 \hat{L}^{4}\hat{r}_{+}^{3}}{(\hat{L}^{2}+\hat{r}_{+}^{2})}\left(1-\frac{\hat{r}_{+}}{r}\right)-11 \hat{r}_{+}^{3}(\hat{L}^{2}+\hat{r}_{+}^{2})\left(1-\frac{\hat{r}_{+}^{3}}{r^{3}}\right)\nonumber\\
			& -\frac{(r-\hat{r}_{+})}{\hat{L}^{2}(\hat{L}^{2}+\hat{r}_{+}^{2})(4\hat{L}^{2}+3 \hat{r}_{+}^{2})\hat{h}(r)}\Big((2\hat{L}^{2}+3 r_{+}^{2})(\hat{L}^{8}+4 \hat{L}^{6}\hat{r}_{+}^{2}-13 \hat{L}^{4}\hat{r}_{+}^{4}-24 \hat{L}^{2}\hat{r}_{+}^{6}-9 \hat{r}_{+}^{8})\nonumber\\
			& -r\,\hat{r}_{+}(5 \hat{L}^{8}+34 \hat{L}^{6}\hat{r}_{+}^{2}+129 \hat{L}^{4}\hat{r}_{+}^{4}+153 \hat{L}^{2} \hat{r}_{+}^{6}+54 \hat{r}_{+}^{8})\Big)\nonumber\\
			&  +\frac{2(\hat{L}^{4}+12 \hat{L}^{2}\hat{r}_{+}^{2}+9 \hat{r}_{+}^{4})(2 \hat{L}^{8}+8 \hat{L}^{6} \hat{r}_{+}^{2}+22 \hat{L}^{4}\hat{r}_{+}^{4}+24 \hat{L}^{2}\hat{r}_{+}^{6}+9 \hat{r}_{+}^{8})}{(\hat{L}^{2}+\hat{r}_{+}^{2})^{2}(4 \hat{L}^{2}+3 \hat{r}_{+}^{2})^{3/2}}\arctan\left(\frac{(r-\hat{r}_{+})\sqrt{4 \hat{L}^{2}+3 \hat{r}_{+}^{2}}}{2 \hat{L}^{2}+3 r\,\hat{r}_{+}+3 \hat{r}_{+}^{2}}\right)\nonumber\\
			&  +\frac{(2 \hat{L}^{6}\hat{r}_{+}^{3}+27 \hat{L}^{4}\hat{r}_{+}^{5}+54 \hat{L}^{2}\hat{r}_{+}^{7}+27 \hat{r}_{+}^{9})}{(\hat{L}^{2}+\hat{r}_{+}^{2})^{2}}\log\left(\frac{r^{2}\hat{h}(\hat{r}_{+})}{\hat{r}_{+}^{2}\hat{h}(r)}\right)-32 \hat{r}_{+}^{5}\log\left(\frac{r}{\hat{r}_{+}}\right)\Bigg\},
		\end{align}
		and
		\begin{align}
			\label{eq:solnrhoPage}
			\hat{\rho}_{0}(r)&=\frac{1}{2880 \pi \hat{L}^{2}\hat{r}_{+}^{4}}\Bigg\{14 \hat{r}_{+}^{2}(\hat{L}^{2}+\hat{r}_{+}^{2})\left(1-\frac{\hat{r}_{+}^{3}}{r^{3}}\right)+\frac{20 \hat{L}^{4}\hat{r}_{+}^{2}}{ (\hat{L}^{2}+\hat{r}_{+}^{2})}\left(1-\frac{\hat{r}_{+}}{r}\right)+15 \hat{L}^{2}\hat{r}_{+}^{2}\left(1-\frac{\hat{r}_{+}^{2}}{r^{2}}\right)\nonumber\\
			&	    +\frac{(r-\hat{r}_{+})}{\hat{L}^{4}\hat{h}(r)^{2}(4\hat{L}^{2}+3 \hat{r}_{+}^{2})^{2}}\Big(2 \hat{r}_{+}
			\left(\hat{L}^2+3 \hat{r}_{+}^2\right) \left(49 \hat{L}^{10}-173 \hat{L}^8
			\hat{r}_{+}^2-487 \hat{L}^6 \hat{r}_{+}^4-318 \hat{L}^4 \hat{r}_{+}^6-18 \hat{L}^2
			\hat{r}_{+}^8+27 \hat{r}_{+}^{10}\right)\nonumber\\
			&  +\left(16 \hat{L}^{12}+49 \hat{L}^{10}
			\hat{r}_{+}^2-2092 \hat{L}^8 \hat{r}_{+}^4-3825 \hat{L}^6 \hat{r}_{+}^6-1521 \hat{L}^4
			\hat{r}_{+}^8+648 \hat{L}^2 \hat{r}_{+}^{10}+405 \hat{r}_{+}^{12}\right)
			(r-\hat{r}_{+})\nonumber\\
			&   +2 \hat{r}_{+} \left(39 \hat{L}^{10}-440 \hat{L}^8
			\hat{r}_{+}^2-729 \hat{L}^6 \hat{r}_{+}^4-27 \hat{L}^4 \hat{r}_{+}^6+405 \hat{L}^2
			\hat{r}_{+}^8+162 \hat{r}_{+}^{10}\right) (r-\hat{r}_{+})^2\nonumber\\
			&   +\left(8 \hat{L}^{10}-151 \hat{L}^8 \hat{r}_{+}^2-172 \hat{L}^6 \hat{r}_{+}^4+141 \hat{L}^4
			\hat{r}_{+}^6+243 \hat{L}^2 \hat{r}_{+}^8+81 \hat{r}_{+}^{10}\right)
			(r-\hat{r}_{+})^3\Big)\nonumber\\
			&	    +\frac{6 \hat{r}_{+}(10 \hat{L}^{12}-120 \hat{L}^{10}\hat{r}_{+}^{2}-400 \hat{L}^{8}\hat{r}_{+}^{4}-512 \hat{L}^{6}\hat{r}_{+}^{6}-357 \hat{L}^{4}\hat{r}_{+}^{8}-144 \hat{L}^{2}\hat{r}_{+}^{10}-27 \hat{r}_{+}^{12})}{(\hat{L}^{2}+\hat{r}_{+}^{2})^{2}(4 \hat{L}^{2}+3 \hat{r}_{+}^{2})^{5/2}}\arctan\left(\frac{(r-\hat{r}_{+})\sqrt{4 \hat{L}^{2}+3 \hat{r}_{+}^{2}}}{2 \hat{L}^{2}+3 r\,\hat{r}_{+}+3 \hat{r}_{+}^{2}}\right)\nonumber\\
			&	    -\frac{3 \hat{r}_{+}^{2}(-2 \hat{L}^{6}+3 \hat{L}^{4}\hat{r}_{+}^{2}+6 \hat{L}^{2}\hat{r}_{+}^{4}+3 \hat{r}_{+}^{6})}{(\hat{L}^{2}+\hat{r}_{+}^{2})^{2}}\log\left(\frac{r^{2}\hat{h}(\hat{r}_{+})}{\hat{r}_{+}^{2}\hat{h}(r)}\right)\Bigg\},
		\end{align}
	\end{widetext}
	where $\hat{h}(r)=1+\hat{L}^{-2}(r^{2}+r\,\hat{r}_{+}+\hat{r}_{+}^{2})$. One can verify that the $\hat{L}\to\infty$ limit of these solutions agree with the solutions obtained by York \cite{York:1985} for the quantum backreaction of a conformal scalar field on the Schwarzschild geometry.
	
	Hence the solution of the semi-classical equations perturbed about Schwarzschild-AdS with the perturbation induced by a conformally invariant quantum field is given by Eq.~(\ref{eq:dressedmetric}) and (\ref{eq:psihat}) with $\hat{\mu}_{0}(r)$ and $\hat{\rho}_{0}(r)$ given by the expressions above. This perturbation is bounded on the entire exterior region of the corrected geometry but unlike the massive case, does not remain bounded for large $r$ in the Schwarzchild limit in accordance with York's results \cite{York:1985}.
	
	\subsection{Surface Gravity and Minimum Temperature}
	
	Let $\lambda^{a}=(\lambda,0,0,0)$ be parallel to the timelike Killing vector associated with the time-translation invariance, then we can define the surface gravity by
	\begin{align}
		\lambda^{a}\nabla_{b}\lambda_{a}\stackrel{\tiny{\textrm{H}}}{=}-\kappa\,\lambda_{b}
	\end{align}
	where the `$\textrm{H}$' above the equals indicates that the equality holds at the event horizon. Of course the coordinates of the line-element (\ref{eq:dressedmetric}) are singular at the event horizon so we instead employ the Eddington-Finkelstein-type coordinates in which the metric is given by Eq.~(\ref{eq:dressedmetricEF}). In these coordinates, the defining equation reduces to 
	\begin{eqnarray}
		\frac{\lambda}{2}\frac{\partial}{\partial r}g_{vv}\stackrel{\tiny{\textrm{H}}}{=}-\hat{\kappa}(1+\hbar\,\hat{\rho}_{0}).
	\end{eqnarray}
	Evaluating at the horizon and keeping terms only up linear order in $\hbar$ gives
	\begin{eqnarray}
		\hat{\kappa}=\frac{\lambda}{2}\hat{f}'(\hat{r}_{+})\Big(1+\hbar\,\hat{\rho}_{0}(\hat{r}_{+})-2\hbar\,\hat{\psi}(\hat{r}_{+})\Big).
	\end{eqnarray}
	We know that, by construction, $\hat{\rho}_{0}(\hat{r}_{+})=0=\hat{\mu}_{0}(\hat{r}_{+})$, and hence from Eq.~(\ref{eq:psihat}), we obtain
	\begin{align}
		\label{eq:kappadressed}
		\hat{\kappa}=\lambda\left(\kappa_{0}[\hat{r}_{+}]-\hbar\,\frac{\hat{M}\,\hat{\mu}'_{0}(\hat{r}_{+})}{\hat{r}_{+}}\right),
	\end{align}
	where in arriving at this form, we applied l'Hopital's rule and denoted by $\kappa_{0}[\hat{r}_{+}]$ the background surface gravity evaluated at $\hat{r}_{+}$. The normalization $\lambda$ is fixed in asymptotically flat spacetimes by assuming that $\lambda^{a}$ is normalized to unity in the limit $r\to\infty$. This would imply $\lambda=1$. For asymptotically AdS spacetimes, the magnitude of this vector diverges in this limit. In the classical context, one can simply insist on retrieving the asymptotically flat result in the $L\to\infty$ limit which again yields $\lambda=1$. This limit is not appropriate in this context however, since at least for conformal fields, the perturbed spacetime is not asymptotically flat in this limit. It was precisely for this reason that York \cite{York:1985} matches the semi-classical spacetime derived as a solution to the semi-classical equations to an asymptotically flat solution at some finite radius. Another tempting approach to fix the normalization is to insist that we retrieve the correct classical solution in the $\hbar\to 0$ limit but this only fixes the zeroth order term in a small $\hbar$ expansion of the normalization constant, i.e., $\lambda=1+\mathcal{O}(\hbar)$. One can choose the $\mathcal{O}(\hbar)$ term so that $\hat{\kappa}=\kappa_{0}[\hat{r}_{+}]$ to linear order, for example. It is clear that the quantity of physical interest is $\tilde{\kappa}:=\hat{\kappa}/\lambda$ since this is invariant under different choices of normalization.
	
	It is clear from (\ref{eq:kappadressed}) that the surface gravity is independent of $\hat{\rho}_{0}$ regardless of the field mass and coupling. We can also obtain simple explicit expressions for the specific field configurations under consideration. For massive fields where the RSET is given by the DeWitt-Schwinger approximation, we have
	\begin{align}
		\tilde{\kappa}=\frac{3\hat{r}_{+}^{2}+\hat{L}^{2}}{2\,\hat{r}_{+}\hat{L}^{2}}-\hbar\Bigg(\frac{(\xi-\tfrac{2}{9})(\hat{r}_{+}^{2}+\hat{L}^{2})^{2}(4\hat{r}_{+}^{2}+\hat{L}^{2})}{20\,\pi\, m^{2}\hat{L}^{6}\hat{r}_{+}^{5}}\Bigg).
	\end{align}
	For conformally invariant scalar fields whose RSET is approximated by Page's approximation, we obtain
	\begin{align}
		\tilde{\kappa}=\frac{3\hat{r}_{+}^{2}+\hat{L}^{2}}{2\,\hat{r}_{+}\hat{L}^{2}}+\hbar\frac{(\hat{r}_{+}^{2}+\hat{L}^{2})(5 \hat{r}_{+}^{2}+3\hat{L}^{2})}{480\,\pi\,\hat{L}^{4}\hat{r}_{+}^{3}}.
	\end{align}
	For large varying $\hat{r}_{+}$, all other parameters held fixed, the corrections to the surface gravity (and hence temperature) scale linearly with increasing $\hat{r}_{+}$ for both massive and conformally invariant scalar fields. However, for small varying $\hat{r}_{+}$, say $\hat{r}_{+}\sim \ell_{\textrm{pl}}$ with $\ell_{\textrm{pl}}$ the Planck length, all other parameters being fixed, then the corrections scale very differently for varying $\hat{r}_{+}$ for massive and conformally invariant fields. Of course, this is also precisely where the semi-classical approximation breaks down but ostensibly the phenomenology of the corrections for the different scalar fields may diverge before this limit is reached.
	
	Unlike asymptotically flat black holes, for $\Lambda<0$ the temperature as a function of horizon radius has a stationary point \cite{HawkingPage:1983}. To see this for Schwarzschild-AdS, we parametrize the surface gravity of the black hole in terms of $r_{+}$ and $L$ whence $\kappa=(3 r_{+}^{2}+L^{2})/(2 r_{+}L^{2})$. In the Schwarzschild limit $L\to \infty$, the surface gravity, and hence the temperature, is a monotonically decreasing function of horizon radius, that is, smaller black holes are hotter than larger ones. On the other hand, for asymptotically AdS black holes, solving $\partial \kappa/\partial r_{+}=0$ yields a stationary point at $r_{+}=r_{0}=L/\sqrt{3}$. This turns out to be a minimum of the surface gravity or equivalently of the temperature. We get a very different thermodynamical picture whereby, for an event horizon with $r_{+}>r_{0}$, the black hole gets cooler as it evaporates reaching a minimum temperature of $T_{\textrm{min}}=1/(2\pi \,r_{0})$ at $r_{+}=r_{0}$. Since the stationary point is a minimum, these black holes are thermodynamically stable, unlike the Schwarzschild black hole. In general, for any fixed temperature $T> T_{\textrm{min}}$, there are two Schwarzschild-AdS black holes at that temperature, one smaller black hole for which $r_{+}<r_{0}$ and one larger black hole for which $r_{+}>r_{0}$.
	
	Let us compute the quantum correction to the horizon radius at which this stationary point occurs, to first order in $\hbar$. We require a particular horizon radius $\hat{r}_{+}=\hat{r}_{0}$ say, which is a stationary point of the surface gravity, or equivalently of the temperature, i.e., we seek $\hat{r}_{0}$ satisfying
	\begin{align}
		\frac{\partial \tilde{\kappa}}{\partial \hat{r}_{+}}\bigg|_{\hat{r}_{+}=\hat{r}_{0}}=0.
	\end{align}
	We require a solution only to linear order in $\hbar$ so we can look for a solution of the form $\hat{r}_{0}=r_{0}+\hbar\,\delta r_{0}$ with $r_{0}=\hat{L}/\sqrt{3}$ the stationary point in the space of classical Schwarzschild-AdS black holes. For massive fields whose RSET has been approximated by the DeWitt-Schwinger model, the corrected stationary point occurs at
	\begin{align}
		\hat{r}_{0}=r_{0}-\hbar\left(\frac{16(\xi-\tfrac{2}{9})}{27\pi r_{0}^{3}m^{2}}\right).
	\end{align}
	For the conformally-invariant scalar field where we have approximated the RSET by Page's approximation, the corrected stationary point occurs at
	\begin{align}
		\hat{r}_{0}=r_{0}+\hbar\left(\frac{5}{216\pi r_{0}}\right).
	\end{align}
	
	The corresponding correction to the minimum temperature which occurs for black holes with horizon radius $\hat{r}_{0}$ is easily shown to be
	\begin{align}
		\hat{T}_{\textrm{min}}=\frac{1}{2\pi \,r_{0}}-\frac{\hbar}{4\pi}\left(1+\frac{r_{0}^{2}}{L^{2}}\right)\hat{\mu}_{0}'(r_{0}).
	\end{align}
	For massive fields whose RSET is given by the DeWitt-Schwinger approximation, the explicit semi-classical minimum temperature is
	\begin{align}
		\hat{T}_{\textrm{min}}=\frac{1}{2\pi\,r_{0}}-\hbar\left(\frac{14(\xi-\frac{2}{9})}{135\pi^{2}r_{0}^{5}m^{2}}\right),
	\end{align}
	while for conformally-invariant scalar fields whose RSET is given by the Page approximation, we have
	\begin{align}
		\hat{T}_{\textrm{min}}=\frac{1}{2\pi \,r_{0}}+\frac{7\hbar}{1080\pi^{2}r_{0}^{3}}.
	\end{align}
	
	\subsection{Plots of Temperature Profiles for Semi-Classical Black Holes}
	
	In this section we present some plots in order to gain insight into the effect of the backreaction on the geometry of the background spacetime obtained via the reduced-order semi-classical field equations. We choose to focus on the effect to the thermodynamics of the quantum corrected black hole and we examine the correction to the temperature for a massive field only, since in the conformal case the magnitude of backreaction effects on the temperature are too small to be discernible on any plot. 
	To this end, working again in units where $G=c=\hbar=1$ in order to exaggerate the quantum effects, we plot the temperature as a function of $\hat{r}_+$ for various parameters in our model. In these units, the Planck length is unity and so we restrict our attention to values $\hat{r}_+>1$, since below the Planck scale one needs a full quantum theory of gravity. Indeed, our results are dubious for $\hat{r}_+\sim 1$ even when the horizon radius is larger than the Planck scale since this is where the semi-classical approximation completely breaks down. On the other hand, the effects of the backreaction are indiscernible on astrophysical scales, thus we allow $\hat{r}_+$ to take on values of the order of the Planck length in order that the effect may be examined at all, but also we point out that perturbative schemes can sometimes work reasonably well even outside the domain of their formal validity.

	In the plots that follow, the blue and orange curves represents the temperature as a function of the black hole radius, $\hat{r}_+$, for a family of background Schwarzschild-AdS spacetimes and their semi-classical counterparts respectively. As already mentioned above, for a given temperature above the minimum temperature $T_{\textrm{min}}$, there are two classical black holes at that temperature. In the plots, this is represented by the horizontal dashed line which can be seen to intersect the curves at two distinct $\hat{r}_+$ values.
	
	\begin{figure*}
		\begin{subfigure}{.5\textwidth}
			\includegraphics[width=.8\linewidth]{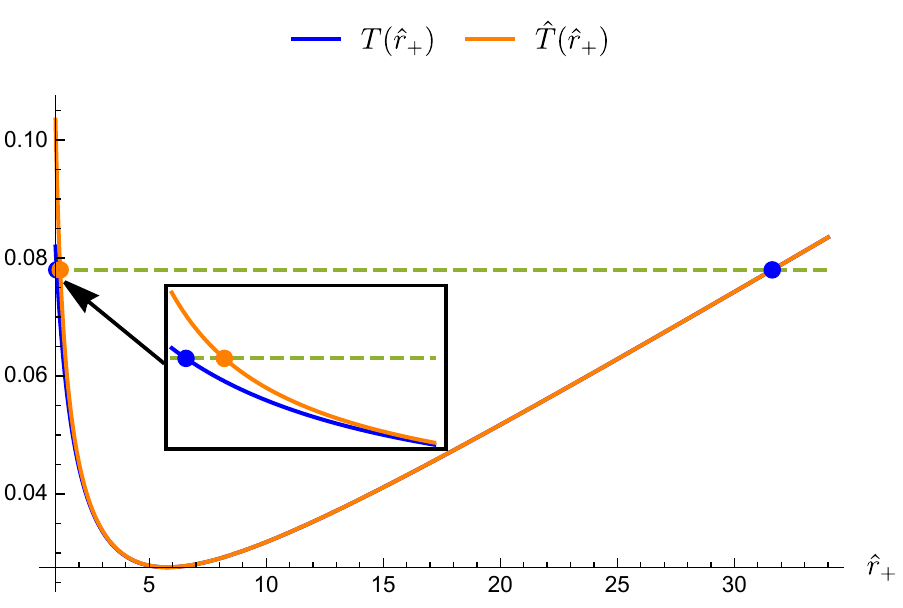}
			\caption{$\hat{L}=10,m=10,\xi=-800$}
			\label{fig:tempm800}
		\end{subfigure}%
		\begin{subfigure}{.5\textwidth}
			\includegraphics[width=.8\linewidth]{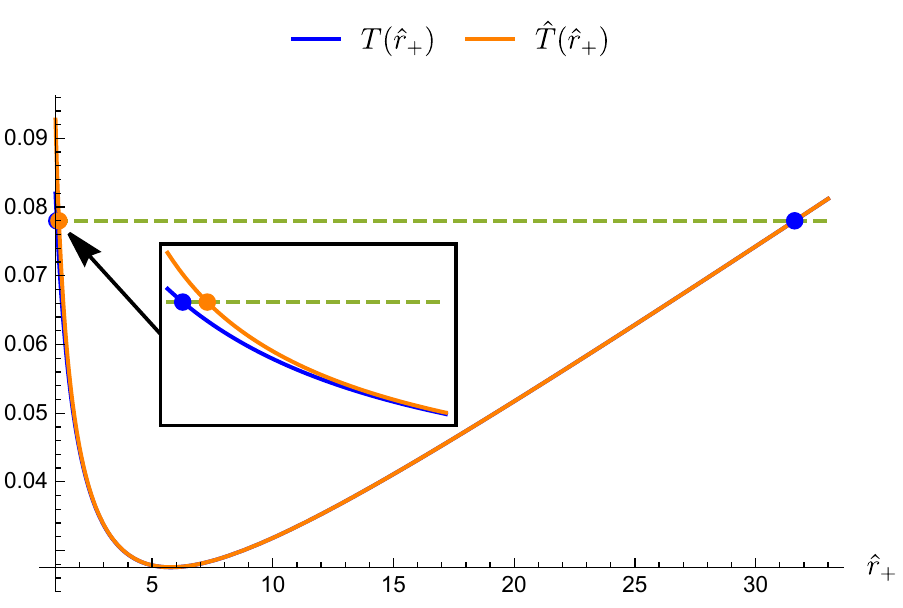}
			\caption{$\hat{L}=10,m=10,\xi=-400$}
			\label{fig:tempm400}
		\end{subfigure}%
		\\
		\begin{subfigure}{.5\textwidth}
			\includegraphics[width=.8\linewidth]{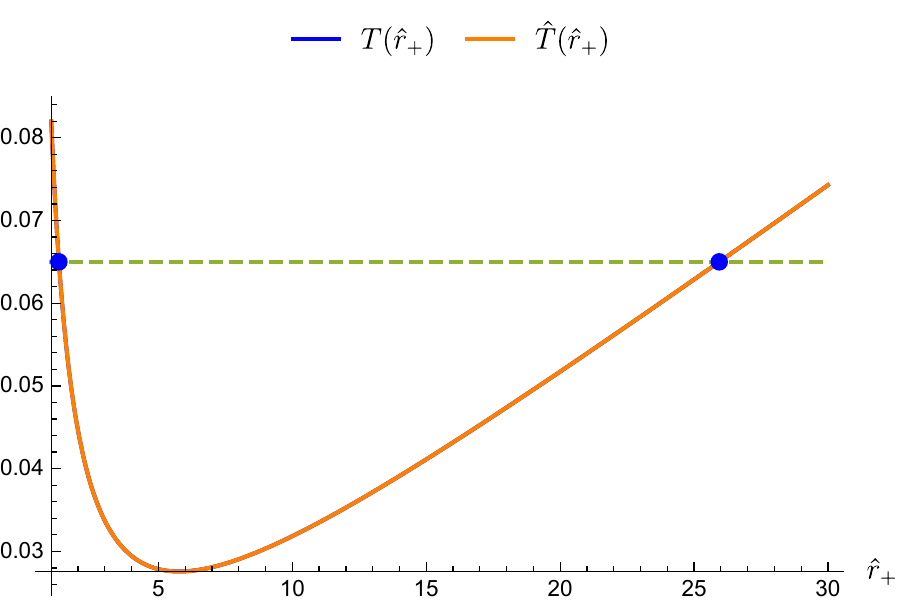}
			\caption{$\hat{L}=10,m=10,\xi=0$}
			\label{fig:temp0}
		\end{subfigure}%
		\begin{subfigure}{.5\textwidth}
			\includegraphics[width=.8\linewidth]{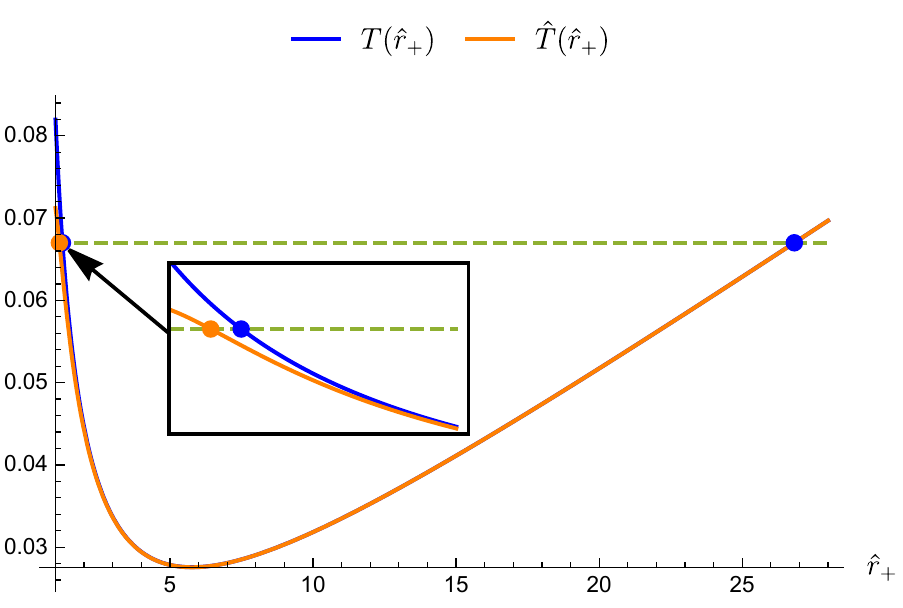}
			\caption{$\hat{L}=10,m=10,\xi=400$}
			\label{fig:temp500}
		\end{subfigure}
		\begin{subfigure}{.5\textwidth}
			\includegraphics[width=.8\linewidth]{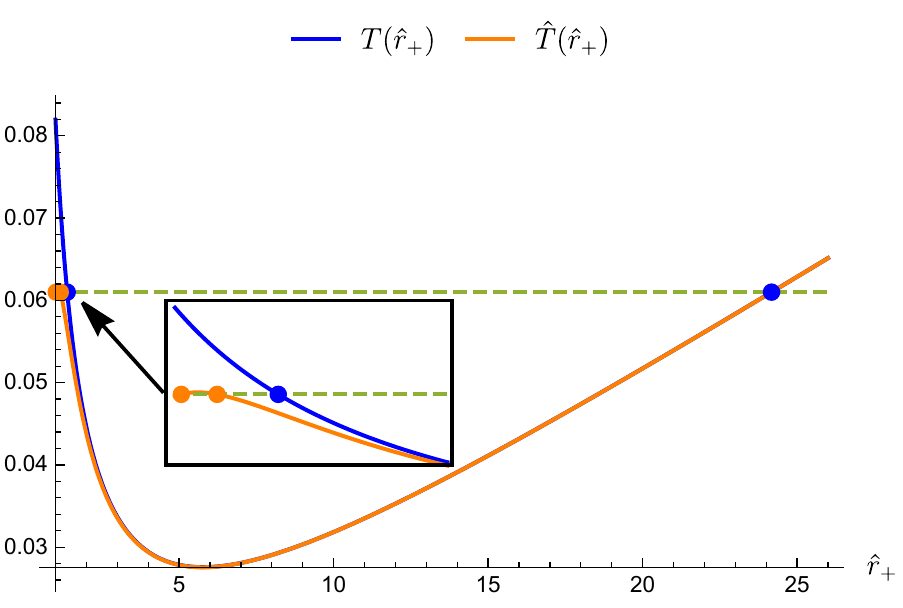}
			\caption{$\hat{L}=10,m=10,\xi=800$}
			\label{fig:temp800}
		\end{subfigure}
		\caption{\label{fig:Tmassive} Plots of temperature as a function of the corrected horizon radius for the family of classical Schwarzschild-AdS black holes (blue curves) and the temperature as a function of the corrected horizon radius for the family of semi-classical solutions perturbed about the Schwarzschild-AdS solutions (orange curves). The perturbation here is induced by a massive quantum scalar field. All parameters are held fixed except the coupling of the quantum field to the background curvature $\xi$ which is increased from $\xi=-800$ in (a) to $\xi=800$ in (e). The inserted plots represent the temperature functions of the classical and semi-classical black holes when $\hat{r}_+\sim 1$. }
	\end{figure*}

	In Fig.~\ref{fig:Tmassive}, we fix the corrected AdS lengthscale $\hat{L}$, the field mass then we vary the parameter $\xi$ which controls the strength of the quantum field coupling to the background curvature.  We note that the allowable range of $\xi$ is bounded above by the Breitenlohner-Freedman bound $m^{2}\hat{L}^2-12(\xi-1/6))\ge -1/4$, for the values of $m$ and $\hat{L}$ chosen in Fig.~\ref{fig:Tmassive} this upper bound is $\xi \approx 833$, while there is no lower bound on $\xi$.  In Fig.~\ref{fig:Tmassive} we examine the effect of allowing $\xi$ to vary from large negative values to large permissible  positive values. For all values considered, we find that for large $\hat{r}_+$ there is no discernible difference between the temperature of the classical and semi-classical black hole. However when $\hat{r}_+ \sim 1$ quantum effects can play a noticeable role, with this phenomenon becoming more pronounced as the magnitude of $\xi$ increases. In Fig. \ref{fig:temp0} we see that for $\xi=0$ there is no discernible difference between the two graphs for all $\hat{r}_+$ but as $\xi$ increases in Figs. \ref{fig:temp500}, \ref{fig:temp800}, for $\hat{r}_+ \sim 1$ the graph of $\hat{T}$ moves slightly to the left. In other words, considering black holes with radii approaching the Planck length for a fixed temperature and sufficiently large coupling $\xi$, the semi-classical black holes will have a smaller horizon radius at that temperature than their classical counterpart. We also see when $\hat{r}_+ \sim 1$, there are classically allowable black hole temperatures which are ruled out for semi-classical black holes of similar size, with the range of these forbidden temperatures increasing with $\xi$. In Figs. \ref{fig:tempm800}, \ref{fig:tempm400}, we see that as $\xi$ becomes more negative, for $\hat{r}_+ \sim 1$ the graph of $\hat{T}$ moves slightly to the right, leading to semi-classical black holes that now have a larger horizon radius at that temperature than their classical counterpart and semi-classical black holes at certain temperatures with no classical counterpart of similar size. Since there is no lower bound on $\xi$ and as the magnitude of the quantum correction grows unbounded with negative $\xi$ (holding all other parameters fixed), there must come a point where the perturbative approach of the semi-classical approximation breaks down for quantum fields with sufficiently negative values of the coupling constant $\xi$, at least in the case of black holes with $\hat{r}_+ \sim 1$.   \\
	
	In Fig. \ref{fig:temp800} we find for values of $\xi$ near, but still below, the upper limit set by the Breitenlohner-Freedman bound, there is a drastic deviation from the typical shape of the temperature curve at small horizon radius, the curve for $\hat{T}$ acquiring an additional stationary point near $\hat{r}_+=1$. The temperature decreases to negative values for $\hat{r}_{+}<1$ (not shown in the graph) signalling a complete breakdown of the semi-classical approximation. It would be interesting to know when exactly the approximation breaks down and more specifically, if the additional turning point is a genuine prediction. If so, then for certain values of the parameters, for a fixed temperature, one may have three semi-classical black hole solutions (as in Fig. \ref{fig:temp800}). Moreover, there could be regions in the space of solutions where these black holes are thermodynamically stable and regions where they are unstable. Again it is worth stressing that all of the interesting results described above occur when $\hat{r}_+ \sim 1$, at the limit of validity of the semi-classical approximation.

	As a final note, we point out that the backreaction seems not to depend strongly on the field mass within this large mass approximation and also, for the case of a massive and conformal field, the effect of the backreaction on the temperature is suppressed for increasing $\hat{L}$  assuming all other quantities fixed.

	\subsection{Photon Sphere in the Semi-Classical Black Hole}
	In the classical Schwarzschild-AdS geometry there exists a photon sphere (a hypersurface on which a massless particle can orbit the black hole on unstable circular null geodesics) at $r_\textrm{u}=3 M$ \cite{Cruz_2005}, in this section we seek the equivalent photon sphere in the corrected geometry. To achieve this we first obtain the radial null geodesic equation for the metric Eq. (\ref{eq:dressedmetric}) by solving the Euler-Lagrange equations associated to this metric in the usual way. Adopting this procedure leads to the following radial equation for a photon in the plane $\theta =\pi/2$:
	\begin{align}
		-\frac{E^2}{\hat{f}(r)(1+2 \hbar \hat{\rho}_0(r) -2 \hbar \hat{\psi}(r))} + \frac{(1+2 \hbar \hat{\psi}(r))}{\hat{f}(r)} \dot{r}^2 +\frac{A^2}{r^2} =0
	\end{align}
	where $E$ and $A$ are constants of motion corresponding to the symmetries of the space-time and $\dot{r}=dr/d\tau$, $\tau$ being an affine parameter along the geodesic. Working to $O(\hbar)$ we may re-express this equation of motion in the form
	\begin{align}
		(1+2 \hbar \hat{\rho}_0(r)) \dot{r}^2= E^2-\frac{A^2 \hat{f}(r) (1+2 \hbar \hat{\rho}_0(r) -2 \hbar \hat{\psi}(r))}{r^2}.
	\end{align}
	The right hand side of this equation is now in the standard form $E^2 - V_{\textrm{eff}}(r)$ where $V_{\textrm{eff}}(r)$ denotes an effective potential. The photon sphere occurs at the value of $r$ for which $V_{\textrm{eff}}(r)$ reaches its maximum value. In the spirit of earlier calculations we seek a solution of the form
	\begin{align}
		\hat{r}_\textrm{u}= 3 \hat{M} + \hbar~\delta r_\textrm{u},
	\end{align}
	leading to the corrected photon sphere occurring at
	\begin{align}
		\hat{r}_\textrm{u}=3\hat{M}\left(1 + \hbar \frac{ \hat{M} \left(\hat{L}^2+27 \hat{M}^2\right) \left(\hat{\rho}_0'(3 \hat{M})-\hat{\psi} '(3 \hat{M})\right)}{\hat{L}^2}\right),
	\end{align}
 or in terms of $\hat{\mu}_{0}(r)$ and $\hat{\rho}_{0}(r)$,
		\begin{align}
			\hat{r}_\textrm{u}=3 \hat{M}\bigg(1 + \hbar\bigg[\hat{\mu}_0(3 \hat{M}) -\hat{M} \hat{\mu}'_0(3 \hat{M}) 
			\nonumber\\
			+\frac{\hat{M}}{\hat{L}^2}(\hat{L}^2+27 \hat{M}^{2})\hat{\rho}'_0(3 \hat{M})\bigg]\bigg).
	\end{align}
	As $\hat{\mu}_0(r)$ and $\hat{\rho}_0(r)$ satisfy the ODEs arising from the reduced order field equations as given in Eq.(\ref{eq:ODEs}), we may simplify the above expression further to give:
	\begin{align}
	\hat{r}_\textrm{u}=3 \hat{M}\bigg(1 + \hbar\bigg[\hat{\mu}_0(3 \hat{M}) + 36\pi \hat{M}^2 \langle \hat{T}^r_{~r} \rangle |_{r=3 \hat{M}} \bigg]\bigg).
	\end{align}
	\begin{widetext}
		For the case of a massive field where the RSET is approximated by the DeWitt-Schwinger model this takes the form
		\begin{align}
			\hat{r}_\textrm{u}=3 \hat{M}\bigg(1- \frac{\hbar}{7348320 \pi  \hat{L}^4 m^2 \hat{r}_+^4 \left(\hat{L}^2+\hat{r}_+^2\right)^4} \bigg[5 \hat{L}^{12} (22344 \xi -4967)+54 \hat{L}^{10} (18144 \xi -4045) \hat{r}_+^2\nonumber\\
			+27 \hat{L}^8 (113400 \xi -25387) \hat{r}_+^4+756 \hat{L}^6 (6480
			\xi -1457) \hat{r}_+^6+3645 \hat{L}^4 (1176 \xi -265) \hat{r}_+^8\nonumber\\
			+4374 \hat{L}^2 (448 \xi -101) \hat{r}_+^{10}+243 (1512 \xi -341)
			\hat{r}_+^{12}\bigg] \bigg),
		\end{align}
		where we have parametrised $\hat{r}_\textrm{u}$ in terms of $(\hat{r}_+,\hat{L})$.
		It is clear from the above expression that provided $\xi>341/1512$ then $\hat{r}_{\textrm{u}} < 3 \hat{M}$ for all $\hat{r}_+$, $\hat{L}$ and $m$.\\
		For a conformal scalar field where the RSET is given by Page's approximation  $\hat{r}_\textrm{u}$ is given by:
		\begin{align}
			\hat{r}_\textrm{u}=3 \hat{M}\bigg(1+\frac{\hbar}{77760 \pi  \hat{L}^2 \hat{r}_+^5 \left(\hat{L}^2+\hat{r}_+^2\right)^2 \left(4 \hat{L}^2+3 \hat{r}_+^2\right)^{2}} \bigg[-2592 \hat{r}_+^5 \left(7 \hat{L}^2 \hat{r}_+^2+4 \hat{L}^4+3 \hat{r}_+^4\right)^2 \log \left(\frac{3}{2} \left(\frac{\hat{r}_+^2}{\hat{L}^2}+1\right)\right)\nonumber\\
			+162 \sqrt{4 \hat{L}^2+3 \hat{r}_+^2} \left(12 \hat{L}^2 \hat{r}_+^2+\hat{L}^4+9 \hat{r}_+^4\right) \left(8 \hat{L}^6 \hat{r}_+^2+22 \hat{L}^4 \hat{r}_+^4+24 \hat{L}^2 \hat{r}_+^6+2 \hat{L}^8+9 \hat{r}_+^8\right) \tan
			^{-1}\left(\frac{\hat{r}_+}{\sqrt{4 \hat{L}^2+3 \hat{r}_+^2}}\right) \nonumber\\
			+81 \hat{r}_+^3 \left(2 \hat{L}^2+3 \hat{r}_+^2\right) \left(4 \hat{L}^2+3 \hat{r}_+^2\right)^2 \left(12 \hat{L}^2 \hat{r}_+^2+\hat{L}^4+9 \hat{r}_+^4\right) \log \left(3-\frac{3 \hat{L}^2}{4 \hat{L}^2+3 \hat{r}_+^2}\right)\nonumber\\
			+ 15795 \hat{L}^2 \hat{r}_+^{11}+11988 \hat{L}^4 \hat{r}_+^9-5202 \hat{L}^6 \hat{r}_+^7-8607 \hat{L}^8 \hat{r}_+^5-2417 \hat{L}^{10} \hat{r}_+^3-324 \hat{L}^{12} \hat{r}_++5103 \hat{r}_+^{13}\bigg]
			\bigg).
		\end{align}
	\end{widetext}
	In this case $\hat{r}_\textrm{u} > 3 \hat{M}$ for small values of $\hat{r}_+$, however as $\hat{r}_+$ increases, holding $\hat{L}$ fixed, the correction decreases to zero before becoming increasingly negative. The value of $\hat{r}_+$ for which the correction goes to zero increases with $\hat{L}$ and in the limit $\hat{L} \to \infty$ we recover York's result that $\hat{r}_\textrm{u} > 3 \hat{M}$ for all $\hat{r}_+$. 
	
	While the location of the photon sphere in the background geometry depends only on the black hole mass, in the corrected metric we see that this orbit also depends on the AdS length scale $\hat{L}$, as well as the scalar hair in the case of a massive field. As was observed for the surface gravity in the corrected geometry, for small varying $\hat{r}_+$ and all other parameters held fixed, the corrections scale very differently for massive and conformally invariant fields. However the behaviour of these corrections also differs for large $\hat{r}_+$, all other parameters held fixed, with $\delta r_\textrm{u}$ scaling like $\hat{r}_+^3$ and $\log(\hat{r}_+)$ for the massive and conformally invariant field, respectively.
	
	The location of the unstable photon orbit has implications for the quasi-normal modes (QNMs) of the semi-classical black hole since this is associated with the peak of the potential in the radial perturbation equation. Computing the QNM spectrum could ostensibly distinguish between a classical and semi-classical black hole by, for example, matching the ringdown from the gravitational waveform to the ringdown predicted by the QNM spectrum of both the classical and semi-classical black hole. For astrophysical black holes, the difference is presumably too small to measure. In fact, using the most straightforward technique for computing the QNMs in asymptotically AdS spacetimes, we found the difference even numerically indistinguishable. In particular, if one adapts the Frobenius method of Refs.~\cite{HorowitzHubeny2000, CardosoLemos2001} to compute the QNMs in our semi-classical spacetime, then this method amounts to solving the equation
	\begin{align}
		\sum_{k=0}^{\infty}a_{k}(-1/\hat{r}_{+})^{k}=0,
	\end{align}
	where the $a_{k}$ are frequency dependent coefficients from a Frobenius expansion about the horizon. The point is that there are only a discrete set of complex frequencies that satisfy this equation and these are precisely the QNMs we are trying to compute. Obviously this infinite series must be truncated to solve in practice but the problem is that, while for larger black holes the series converges sufficiently fast that only a modest number of terms are needed, the modes are indistinguishable from the classical modes up to 4-5 decimal places. One would expect a deviation from the classical QNM spectrum for smaller black holes, but this is the regime where the convergence of the series above becomes very slow and it becomes impractical to adopt this method. It has been shown \cite{BertiCardosoPani} that the theory of Breit-Wigner resonances is an efficient numerical tool for computing the QNMs for small Schwarzschild-AdS black holes, thus circumventing the problems of slow convergence suffered by the Frobenius method. This resonance method ought to be adaptable to compute the QNM spectrum for the semi-classical black hole spacetimes described in this paper too. We hope to return to this in future work.

	\subsection{Correction to the Curvature Invariants}	
	In this section we calculate the backreaction correction to select curvature invariants of dimension $[\mbox{length}^{-4}]$. Such quantities play an important role in quantum field theory in curved spacetimes, for instance the trace anomaly for a conformal scalar field is given by \cite{DecaniniFolacci:2008}:
	\begin{align}
		\label{eq:Trace}
		\langle \hat{T}^{a}{}_{a}\rangle=\frac{1}{2880\pi^2}\left(R_{abcd}R^{abcd} - R_{ab}R^{ab}+\Box R\right),
	\end{align}
	where $R_{abcd}$ denotes the Riemann tensor. Using a symbolic computational package such as Mathematica it is straightforward to calculate the invariants contained in the above expression for the corrected geometry, although the resulting expressions can be quite lengthy. Therefore for the sake of brevity we evaluate each of the curvature invariants on the event horizon of the corrected geometry $\hat{r}_+$. Adopting this procedure for the case of a massive field with a RSET given by the DeWitt-Schwinger approximation yields to $O(\hbar)$:
	\begin{widetext}
		\begin{align}
			\hat{R}_{abcd}\hat{R}^{abcd}[\hat{r}_+]&= R_{abcd}R^{abcd}[\hat{r}_+]  +\hbar \frac{2  \left(\hat{L}^2+3 \hat{r}_+^2\right)
				\left(\hat{L}^2+\hat{r}_+^2\right)^2 \left(\hat{L}^2+56 (2-9 \xi )
				\hat{r}_+^2\right)}{315 \pi  \hat{L}^8 m^2 \hat{r}_+^8}\nonumber\\
			\hat{R}_{ab}\hat{R}^{ab}[\hat{r_+}]&=      R_{ab}R^{ab}[\hat{r}_+]-\hbar \frac{
				\left(\hat{L}^2+\hat{r}_+^2\right)^2 \left(\hat{L}^2 (378 \xi
				-83)+(1386 \xi -305) \hat{r}_+^2\right)}{105 \pi  \hat{L}^8 m^2
				\hat{r}_+^6}   \nonumber\\
			\hat{\Box} \hat{R}[\hat{r}_+]&=  -\hbar \frac{ \left(\hat{L}^2+\hat{r}_+^2\right)^2 \left(\hat{L}^2+3
				\hat{r}_+^2\right) \left(\hat{L}^2 (2352 \xi -517)+3 (2240 \xi
				-493) \hat{r}_+^2\right)}{280 \pi  \hat{L}^8 m^2 \hat{r}_+^8},               \nonumber\\   
		\end{align}
	\end{widetext}
	where $ R_{abcd}R^{abcd}[\hat{r}_+] $ and $ R_{ab}R^{ab}[\hat{r}_+] $ denote the square of the Riemann and Ricci tensors for the background geometry evaluated at $r=\hat{r}_+$.
	
	For a conformal field with Page's approximation to the RSET we obtain:
	\begin{align}
		\hat{R}_{abcd}\hat{R}^{abcd}[\hat{r}_+]&= R_{abcd}R^{abcd}[\hat{r}_+]  \nonumber\\	&+\hbar \frac{\left(\hat{L}^2+\hat{r}_+^2\right)^2 \left(\hat{L}^2+3
			\hat{r}_+^2\right)}{6 \pi  \hat{L}^6 \hat{r}_+^6}\nonumber\\
		\hat{R}_{ab}\hat{R}^{ab}[\hat{r}_+]&=      R_{ab}R^{ab}[\hat{r}_+]+\hbar \frac{\left(\hat{L}^2+\hat{r}_+^2\right)^2}{5 \pi  \hat{L}^6
			\hat{r}_+^4}  \nonumber\\
		\hat{\Box} \hat{R}[\hat{r}_+]&=  \hbar\frac{\left(\hat{L}^2+\hat{r}_+^2\right)^2 \left(\hat{L}^2+3
			\hat{r}_+^2\right)}{5 \pi  \hat{L}^6 \hat{r}_+^6}           \nonumber\\   
	\end{align}
	Taking the Schwarzschild limit of the above invariants for a conformal field gives agreement with Page's results.
	
	For large $\hat{r}_+$ and all other parameters fixed, the corrections to the curvature invariants considered here scale the same for both the conformal and massive case, however the behaviour of these corrections for small $\hat{r}_+$, holding other parameters fixed, differs between the conformal and massive case. 
	
	Finally for the case of a conformal field, by evaluating Eq. (\ref{eq:Trace}) in the corrected geometry we find that the trace anomaly evaluated at $\hat{r}_+$ is given by
	\begin{align}
		\langle \hat{T}^{a}{}_{a}\rangle=\frac{\hat{L}^2+2 \hat{r}_+^2}{240 \pi ^2 \hat{L}^2
			\hat{r}_+^4}+\hbar\frac{  \left(11 \hat{L}^2+27 \hat{r}_+^2\right)
			\left(\hat{L}^2+\hat{r}_+^2\right)^2}{86400 \pi ^3 \hat{L}^6
			\hat{r}_+^6},
	\end{align}
	where it is understood that this is now the trace of $\langle \hat{T}^{a}{}_{b}\rangle$ for a conformal scalar thermal field propagating on the corrected geometry with metric Eq. (\ref{eq:dressedmetric}). Given that the corrected geometry possess the same symmetries as the background geometry, then once armed with the above expression for the trace all that one would require in order to determine $\langle \hat{T}^{a}{}_{b}\rangle$ on the horizon of this corrected geometry is one of the diagonal components, $\langle \hat{T}^{\theta}{}_{\theta}\rangle$ say, evaluated there.

	\section{Summary and Outlook}
	\label{sec:conclusions}
	In this paper we studied the backreaction of a quantum scalar field on the geometry of an asymptotically AdS black hole. Rather than attempt to compute the expectation value of the quantum stress-energy tensor exactly and then try to numerically integrate, we employ analytical approximations for the stress-energy of the scalar field which acts as the source term in the semi-classical field equations. We explore the phenomenology of the backreaction for both massive and massless scalar fields. For massless scalar fields, the only known approximations typically rely on setting the coupling to $\xi=1/6$ and exploiting the conformal invariance of the theory. To this end, we employ Page's approximation for the stress-energy tensor. For black hole spacetimes, Page's prescription naturally approximates a conformal field in the Hartle-Hawking state. For massive fields, we adopt the local DeWitt-Schwinger approximation to the stress-energy tensor. Since this is a local approximation, it is agnostic to the quantum state. However, this approximation is regular on the event horizon in a freely falling frame and satisfies the same symmetry properties as the spacetime. Since these are criteria satisfied only by the Hartle-Hawking state, there is still a sense in which we associate this approximation with an approximation to the exact RSET in the Hartle-Hawking state.

	In order to assess the validity of the approximations employed in this study, a comparison with the exact numerical results was undertaken. The exact numerical results were generated using the extended Green-Liouville method of Breen and Ottewill \cite{breenottewill:2012b}, but computed only with sufficient accuracy to faithfully generate a plot of the components of the RSET, not with the refinement needed to be employed in the exact numerical integration of the semi-classical equations. We found that for conformal scalar fields, the Page approximation for the stress-energy tensor on Schwarzschild-AdS spacetimes fares worse than the same approximation applied to the asymptotically flat Schwarzschild spacetime. We suggest that the comparative difference is due to the timelike infinity in the Schwarzschild-AdS spacetime which endows the geometry with a richer geodesic structure. This matters in the approximation since the underlying approximate propagator out of which the stress-energy tensor is constructed assumes contributions only from the shortest geodesic between two points. It is likely that this approximation ignores more geodesics in the asymptotically AdS case; hence the greater error in the overall approximation. Nevertheless, the approximation is a reasonable one and in the absence of a better analytical approximation, we adopt the Page stress-energy tensor. Assessing the approximation for massive fields is also not without issue. In this case, the problem is somewhat more subtle in that it is not clear what two objects one ought to compare in the first place. The choice is tantamount to the renormalization ambiguity again. We have a one-parameter family of numerical RSETs, parametrized by some lengthscale $\ell_{\textrm{numeric}}$ say, and a one-parameter family of approximate stress-energy tensors parametrized by $\ell$. Which particular members of these one-parameter families should be compared? This raises a question about the very meaning of the accuracy of an approximation in the context of the renormalization ambiguity. While the implications of the renormalization freedom have been discussed in detail in the literature, we are not aware of any authors that discuss the implications in the context of assessing an approximation. On the contrary, it is often claimed that the DeWitt-Schwinger approximation is a good approximation for sufficiently large field mass, but this begs the question: ``Accurate compared to the exact numerical RSET for what choice of lengthscale?''. Notwithstanding this pedantry, we show that there exists a choice of renormalization lengthscales for which the modified DeWitt-Schwinger is a reasonable approximation to the exact RSET. The particular choices were made by insisting that both the approximation and exact RSET had the same asymptotic form near spatial infinity. While it seems a moot point in this context since the renormalization ambiguity gets absorbed into a renormalization of the cosmological constant and hence plays no role in sourcing the semi-classical equations, we stress that when talking about approximations of RSETs, one is comparing one-parameter families of tensors.
	
	Furnished with these analytical approximations, we showed that the reduced-order semi-classical field equations reduced to two simple ODEs for two unknown functions, provided the perturbed spacetime is static and spherically symmetric. These functions could in turn be expanded in $\hbar$ and solved to first order in a straightforward way. The method here generalizes York's calculation \cite{York:1985} for the backreaction on Schwarzschild spacetime. It turns out there is a distinct advantage to working in the asymptotically AdS spacetime since the confining AdS potential provides a barrier which prevents radiation from reaching infinity and effectively amounts to ensuring the perturbed spacetime preserves the asymptotic structure of the background. The equivalent calculation in Schwarzschild necessitates that the perturbation be matched to an asymptotically flat metric at finite radius, a complication which is naturally avoided here. Having obtained the first order correction to the metric components, we were then able to calculate the corrections to the event horizon, the surface gravity and the minimum temperature in the space of solutions. We further calculate corrections to the photon sphere and to quadratic curvature invariants.
	
	The result of solving the backreaction is a multi-dimensional space of black hole solutions, parametrized by quantum dressed black hole parameters $\hat{r}_{+}$ and $\hat{L}$ as well as the quantum field parameters such as the mass and coupling. It was shown that the two integration constants that arise from solving the ODEs for the static perturbation could be absorbed into a dressed black hole mass and a renormalization of the time in static coordinates. Similarly, we found that the renormalization ambiguity could be absorbed into a quantum dressed cosmological lengthscale. For the case of a perturbation induced by a conformal field, our solution breaks down in the limit where the cosmological lengthscale becomes unbounded. This is because in this limit, our background is Schwarzschild and in this asymptotically flat spacetime, the semi-classical perturbation induced by a conformal field grows without bound as $r$ increases and must be matched to an asymptotically flat spacetime, as shown by York \cite{York:1985}. This issue does not arise for the case of perturbations induced by a massive quantum field, where our solution remains bounded in the Schwarzschild limit.

	Having obtained the first order corrections to the background geometry, it was then of interest to plot these corrections in order to discern the qualitative effects of the backreaction for the various parameters. We chose to focus on the correction to the black hole temperature profile, that is, in the space of solutions parametrized by the horizon radius (for fixed $\hat{L}$), how does the temperature depend on the horizon radius. To see how the backreaction affects these temperature profiles, we plotted both the classical and semi-classical temperature as a function of the corrected black hole radius. We focused solely on the  case of a massive field, as effects were negligible for the conformal case. By increasing $\xi$ with all other parameters held fixed, we saw that for a fixed temperature above the classical minimum temperature and considering black hole radii near the Planck length, the quantum-corrected black holes will have a smaller horizon radius at that temperature then their classical counterparts and there is a non-negligible temperature range at which classical black holes exist but semi-classical black holes do not. Considering values of $\xi$ near the maximum allowable value as set by the Breitenlohner-Freedman bound, we find the temperature profiles become very different to their classical counterparts for smaller black holes, the curves develop a local maximum near the Planck length. This presumably is pointing to the breakdown of the semi-classical approximation. Allowing $\xi$ to take on increasingly negative values and again considering black hole radii near the Planck length we saw that the semi-classical black holes will now have a larger horizon radius at a fixed temperature than their classical counterpart and there exists semi-classical black holes at certain temperatures with no classical counterpart of similar size.

	A further numerical investigation would reveal how robust these features are as a prediction of the semi-classical equations. It is likely they are a result of pushing the semi-classical approximation beyond its domain of validity. While effective field theory may be employed to give an order-of-magnitude estimate for when the semi-classical approximation breaks down in terms of natural lengthscales in the problem, perhaps a more pragmatic perspective here is that the semi-classical approximation breaks down when the solutions deviate strongly from the classical background, notwithstanding the possibility of large secular effects such as the Hawking process when we relax the static assumption.
	
	Finally, we note that the work herein may be extended in many directions. First and most obvious, an exact computation of the backreaction with the exact numerical RSET sourcing the semi-classical equations. This requires very efficient and accurate mode-sum prescriptions for the RSET for a range of field parameters. Since the calculation is on a static background, the extended-coordinate method of Refs.~\cite{taylorbreen:2016, taylorbreen:2017, BreenTaylor:18} ought to provide such a prescription. In this work, our approximations pertained to a scalar field satisfying Dirichlet boundary conditions at the timelike boundary of spatial infinity, but other boundary conditions are of course possible. In fact, the asymptotic values of the expectation values are generically not those produced by the Dirichlet boundary conditions \cite{MTW2, MTW3}, but instead all other Robin boundary conditions \textit{except the Dirichlet case} asymptote to the same value. It would be interesting therefore to examine how strongly dependent is the backreaction on the choice of boundary condition. A somewhat different direction is to address how one might distinguish observationally between the classical and semi-classical black holes. As mentioned above, the QNMs for astrophysical black holes are indistinguishable for Schwarzschild-AdS and the semi-classical counterpart, however, the difference ought to become significant for smaller black holes. Computing the QNMs in this regime is technically challenging but the theory of Breit-Wigner resonances applied to Schwarzschild-AdS black holes provides a promising template for asymptotically AdS semi-classical black holes. The last direction these results might be extended is for topological black holes. For $\Lambda<0$, black holes with other horizon topologies are permissible solutions to the vacuum field equations. In fact recently, techniques have been developed \cite{MTW1} for numerically computing expectation values for the square of a quantum scalar field in these topological black hole spacetimes and these techniques could be extended to the calculation of the RSET and used to compute the backreaction. 
	
	\section*{Acknowledgements}
	We wish to thank an anonymous referee for their astute comments on the first draft of this manuscript.


\begin{thebibliography}{70}%
		\makeatletter
		\providecommand \@ifxundefined [1]{%
			\@ifx{#1\undefined}
		}%
		\providecommand \@ifnum [1]{%
			\ifnum #1\expandafter \@firstoftwo
			\else \expandafter \@secondoftwo
			\fi
		}%
		\providecommand \@ifx [1]{%
			\ifx #1\expandafter \@firstoftwo
			\else \expandafter \@secondoftwo
			\fi
		}%
		\providecommand \natexlab [1]{#1}%
		\providecommand \enquote  [1]{``#1''}%
		\providecommand \bibnamefont  [1]{#1}%
		\providecommand \bibfnamefont [1]{#1}%
		\providecommand \citenamefont [1]{#1}%
		\providecommand \href@noop [0]{\@secondoftwo}%
		\providecommand \href [0]{\begingroup \@sanitize@url \@href}%
		\providecommand \@href[1]{\@@startlink{#1}\@@href}%
		\providecommand \@@href[1]{\endgroup#1\@@endlink}%
		\providecommand \@sanitize@url [0]{\catcode `\\12\catcode `\$12\catcode
			`\&12\catcode `\#12\catcode `\^12\catcode `\_12\catcode `\%12\relax}%
		\providecommand \@@startlink[1]{}%
		\providecommand \@@endlink[0]{}%
		\providecommand \url  [0]{\begingroup\@sanitize@url \@url }%
		\providecommand \@url [1]{\endgroup\@href {#1}{\urlprefix }}%
		\providecommand \urlprefix  [0]{URL }%
		\providecommand \Eprint [0]{\href }%
		\providecommand \doibase [0]{http://dx.doi.org/}%
		\providecommand \selectlanguage [0]{\@gobble}%
		\providecommand \bibinfo  [0]{\@secondoftwo}%
		\providecommand \bibfield  [0]{\@secondoftwo}%
		\providecommand \translation [1]{[#1]}%
		\providecommand \BibitemOpen [0]{}%
		\providecommand \bibitemStop [0]{}%
		\providecommand \bibitemNoStop [0]{.\EOS\space}%
		\providecommand \EOS [0]{\spacefactor3000\relax}%
		\providecommand \BibitemShut  [1]{\csname bibitem#1\endcsname}%
		\let\auto@bib@innerbib\@empty
		\bibitem [{\citenamefont {Hawking}(1975)}]{Hawking:1975}%
		\BibitemOpen
		\bibfield  {author} {\bibinfo {author} {\bibfnamefont {S.~W.}\ \bibnamefont
				{Hawking}},\ }\href {\doibase 10.1007/BF02345020} {\bibfield  {journal}
			{\bibinfo  {journal} {Commun. Math. Phys.}\ }\textbf {\bibinfo {volume}
				{43}},\ \bibinfo {pages} {199} (\bibinfo {year} {1975})}\BibitemShut
		{NoStop}%
		\bibitem [{\citenamefont {Hawking}(1976)}]{Hawking:1976}%
		\BibitemOpen
		\bibfield  {author} {\bibinfo {author} {\bibfnamefont {S.~W.}\ \bibnamefont
				{Hawking}},\ }\href {\doibase 10.1103/PhysRevD.13.191} {\bibfield  {journal}
			{\bibinfo  {journal} {Phys. Rev. D}\ }\textbf {\bibinfo {volume} {13}},\
			\bibinfo {pages} {191} (\bibinfo {year} {1976})}\BibitemShut {NoStop}%
		\bibitem [{\citenamefont {Horowitz}(1980)}]{Horowitz:1980}%
		\BibitemOpen
		\bibfield  {author} {\bibinfo {author} {\bibfnamefont {G.~T.}\ \bibnamefont
				{Horowitz}},\ }\href {\doibase 10.1103/PhysRevD.21.1445} {\bibfield
			{journal} {\bibinfo  {journal} {Phys. Rev. D}\ }\textbf {\bibinfo {volume}
				{21}},\ \bibinfo {pages} {1445} (\bibinfo {year} {1980})}\BibitemShut
		{NoStop}%
		\bibitem [{\citenamefont {Horowitz}\ and\ \citenamefont
			{Wald}(1978)}]{HorowitzWald:1978}%
		\BibitemOpen
		\bibfield  {author} {\bibinfo {author} {\bibfnamefont {G.~T.}\ \bibnamefont
				{Horowitz}}\ and\ \bibinfo {author} {\bibfnamefont {R.~M.}\ \bibnamefont
				{Wald}},\ }\href {\doibase 10.1103/PhysRevD.17.414} {\bibfield  {journal}
			{\bibinfo  {journal} {Phys. Rev. D}\ }\textbf {\bibinfo {volume} {17}},\
			\bibinfo {pages} {414} (\bibinfo {year} {1978})}\BibitemShut {NoStop}%
		\bibitem [{\citenamefont {Singh}\ and\ \citenamefont
			{Padmanabhan}(1989)}]{Singh:1989}%
		\BibitemOpen
		\bibfield  {author} {\bibinfo {author} {\bibfnamefont {T.}~\bibnamefont
				{Singh}}\ and\ \bibinfo {author} {\bibfnamefont {T.}~\bibnamefont
				{Padmanabhan}},\ }\href {\doibase
			https://doi.org/10.1016/0003-4916(89)90180-2} {\bibfield  {journal} {\bibinfo
				{journal} {Annals of Physics}\ }\textbf {\bibinfo {volume} {196}},\ \bibinfo
			{pages} {296 } (\bibinfo {year} {1989})}\BibitemShut {NoStop}%
		\bibitem [{\citenamefont {Paz}\ and\ \citenamefont
			{Sinha}(1991)}]{PazSinha:1991}%
		\BibitemOpen
		\bibfield  {author} {\bibinfo {author} {\bibfnamefont {J.~P.}\ \bibnamefont
				{Paz}}\ and\ \bibinfo {author} {\bibfnamefont {S.}~\bibnamefont {Sinha}},\
		}\href {\doibase 10.1103/PhysRevD.44.1038} {\bibfield  {journal} {\bibinfo
				{journal} {Phys. Rev. D}\ }\textbf {\bibinfo {volume} {44}},\ \bibinfo
			{pages} {1038} (\bibinfo {year} {1991})}\BibitemShut {NoStop}%
		\bibitem [{\citenamefont {Padmanabhan}\ and\ \citenamefont
			{Singh}(1990)}]{Padmanabhan:1990}%
		\BibitemOpen
		\bibfield  {author} {\bibinfo {author} {\bibfnamefont {T.}~\bibnamefont
				{Padmanabhan}}\ and\ \bibinfo {author} {\bibfnamefont {T.~P.}\ \bibnamefont
				{Singh}},\ }\href {\doibase 10.1088/0264-9381/7/3/015} {\bibfield  {journal}
			{\bibinfo  {journal} {Classical and Quantum Gravity}\ }\textbf {\bibinfo
				{volume} {7}},\ \bibinfo {pages} {411} (\bibinfo {year} {1990})}\BibitemShut
		{NoStop}%
		\bibitem [{\citenamefont {Simon}(1990)}]{Simon:1990}%
		\BibitemOpen
		\bibfield  {author} {\bibinfo {author} {\bibfnamefont {J.~Z.}\ \bibnamefont
				{Simon}},\ }\href {\doibase 10.1103/PhysRevD.41.3720} {\bibfield  {journal}
			{\bibinfo  {journal} {Phys. Rev. D}\ }\textbf {\bibinfo {volume} {41}},\
			\bibinfo {pages} {3720} (\bibinfo {year} {1990})}\BibitemShut {NoStop}%
		\bibitem [{\citenamefont {Parker}\ and\ \citenamefont
			{Simon}(1993)}]{ParkerSimon:1993}%
		\BibitemOpen
		\bibfield  {author} {\bibinfo {author} {\bibfnamefont {L.}~\bibnamefont
				{Parker}}\ and\ \bibinfo {author} {\bibfnamefont {J.~Z.}\ \bibnamefont
				{Simon}},\ }\href {\doibase 10.1103/PhysRevD.47.1339} {\bibfield  {journal}
			{\bibinfo  {journal} {Phys. Rev. D}\ }\textbf {\bibinfo {volume} {47}},\
			\bibinfo {pages} {1339} (\bibinfo {year} {1993})}\BibitemShut {NoStop}%
		\bibitem [{\citenamefont {Flanagan}\ and\ \citenamefont
			{Wald}(1996)}]{FlanaganWald:1996}%
		\BibitemOpen
		\bibfield  {author} {\bibinfo {author} {\bibfnamefont {E.~E.}\ \bibnamefont
				{Flanagan}}\ and\ \bibinfo {author} {\bibfnamefont {R.~M.}\ \bibnamefont
				{Wald}},\ }\href {\doibase 10.1103/PhysRevD.54.6233} {\bibfield  {journal}
			{\bibinfo  {journal} {Phys. Rev. D}\ }\textbf {\bibinfo {volume} {54}},\
			\bibinfo {pages} {6233} (\bibinfo {year} {1996})}\BibitemShut {NoStop}%
		\bibitem [{\citenamefont {Candelas}\ and\ \citenamefont
			{Howard}(1984)}]{CandelasHoward:84}%
		\BibitemOpen
		\bibfield  {author} {\bibinfo {author} {\bibfnamefont {P.}~\bibnamefont
				{Candelas}}\ and\ \bibinfo {author} {\bibfnamefont {K.~W.}\ \bibnamefont
				{Howard}},\ }\href {\doibase 10.1103/PhysRevD.29.1618} {\bibfield  {journal}
			{\bibinfo  {journal} {Phys. Rev. D}\ }\textbf {\bibinfo {volume} {29}},\
			\bibinfo {pages} {1618} (\bibinfo {year} {1984})}\BibitemShut {NoStop}%
		\bibitem [{\citenamefont {Howard}\ and\ \citenamefont
			{Candelas}(1984)}]{CandelasHoward:1984}%
		\BibitemOpen
		\bibfield  {author} {\bibinfo {author} {\bibfnamefont {K.~W.}\ \bibnamefont
				{Howard}}\ and\ \bibinfo {author} {\bibfnamefont {P.}~\bibnamefont
				{Candelas}},\ }\href {\doibase 10.1103/PhysRevLett.53.403} {\bibfield
			{journal} {\bibinfo  {journal} {Phys. Rev. Lett.}\ }\textbf {\bibinfo
				{volume} {53}},\ \bibinfo {pages} {403} (\bibinfo {year} {1984})}\BibitemShut
		{NoStop}%
		\bibitem [{\citenamefont {Anderson}(1990)}]{Anderson:1990}%
		\BibitemOpen
		\bibfield  {author} {\bibinfo {author} {\bibfnamefont {P.~R.}\ \bibnamefont
				{Anderson}},\ }\href {\doibase 10.1103/PhysRevD.41.1152} {\bibfield
			{journal} {\bibinfo  {journal} {Phys. Rev. D}\ }\textbf {\bibinfo {volume}
				{41}},\ \bibinfo {pages} {1152} (\bibinfo {year} {1990})}\BibitemShut
		{NoStop}%
		\bibitem [{\citenamefont {Anderson}\ \emph {et~al.}(1995)\citenamefont
			{Anderson}, \citenamefont {Hiscock},\ and\ \citenamefont
			{Samuel}}]{AHS:1995}%
		\BibitemOpen
		\bibfield  {author} {\bibinfo {author} {\bibfnamefont {P.~R.}\ \bibnamefont
				{Anderson}}, \bibinfo {author} {\bibfnamefont {W.~A.}\ \bibnamefont
				{Hiscock}}, \ and\ \bibinfo {author} {\bibfnamefont {D.~A.}\ \bibnamefont
				{Samuel}},\ }\href {\doibase 10.1103/PhysRevD.51.4337} {\bibfield  {journal}
			{\bibinfo  {journal} {Phys. Rev. D}\ }\textbf {\bibinfo {volume} {51}},\
			\bibinfo {pages} {4337} (\bibinfo {year} {1995})}\BibitemShut {NoStop}%
		\bibitem [{\citenamefont {Flachi}\ and\ \citenamefont
			{Tanaka}(2008)}]{FlachiTanaka:08}%
		\BibitemOpen
		\bibfield  {author} {\bibinfo {author} {\bibfnamefont {A.}~\bibnamefont
				{Flachi}}\ and\ \bibinfo {author} {\bibfnamefont {T.}~\bibnamefont
				{Tanaka}},\ }\href {\doibase 10.1103/PhysRevD.78.064011} {\bibfield
			{journal} {\bibinfo  {journal} {Phys. Rev. D}\ }\textbf {\bibinfo {volume}
				{78}},\ \bibinfo {pages} {064011} (\bibinfo {year} {2008})}\BibitemShut
		{NoStop}%
		\bibitem [{\citenamefont {Ottewill}\ and\ \citenamefont
			{Taylor}(2010)}]{OttewillTaylor:10}%
		\BibitemOpen
		\bibfield  {author} {\bibinfo {author} {\bibfnamefont {A.~C.}\ \bibnamefont
				{Ottewill}}\ and\ \bibinfo {author} {\bibfnamefont {P.}~\bibnamefont
				{Taylor}},\ }\href {\doibase 10.1103/PhysRevD.82.104013} {\bibfield
			{journal} {\bibinfo  {journal} {Phys. Rev. D}\ }\textbf {\bibinfo {volume}
				{82}},\ \bibinfo {pages} {104013} (\bibinfo {year} {2010})}\BibitemShut
		{NoStop}%
		\bibitem [{\citenamefont {Breen}\ and\ \citenamefont
			{Ottewill}(2012{\natexlab{a}})}]{breenottewill:2012a}%
		\BibitemOpen
		\bibfield  {author} {\bibinfo {author} {\bibfnamefont {C.}~\bibnamefont
				{Breen}}\ and\ \bibinfo {author} {\bibfnamefont {A.~C.}\ \bibnamefont
				{Ottewill}},\ }\href {\doibase 10.1103/PhysRevD.85.064026} {\bibfield
			{journal} {\bibinfo  {journal} {Phys. Rev. D}\ }\textbf {\bibinfo {volume}
				{85}},\ \bibinfo {pages} {064026} (\bibinfo {year}
			{2012}{\natexlab{a}})}\BibitemShut {NoStop}%
		\bibitem [{\citenamefont {Breen}\ and\ \citenamefont
			{Ottewill}(2012{\natexlab{b}})}]{breenottewill:2012b}%
		\BibitemOpen
		\bibfield  {author} {\bibinfo {author} {\bibfnamefont {C.}~\bibnamefont
				{Breen}}\ and\ \bibinfo {author} {\bibfnamefont {A.~C.}\ \bibnamefont
				{Ottewill}},\ }\href {\doibase 10.1103/PhysRevD.85.084029} {\bibfield
			{journal} {\bibinfo  {journal} {Phys. Rev. D}\ }\textbf {\bibinfo {volume}
				{85}},\ \bibinfo {pages} {084029} (\bibinfo {year}
			{2012}{\natexlab{b}})}\BibitemShut {NoStop}%
		\bibitem [{\citenamefont {Taylor}\ and\ \citenamefont
			{Breen}(2016)}]{taylorbreen:2016}%
		\BibitemOpen
		\bibfield  {author} {\bibinfo {author} {\bibfnamefont {P.}~\bibnamefont
				{Taylor}}\ and\ \bibinfo {author} {\bibfnamefont {C.}~\bibnamefont {Breen}},\
		}\href {\doibase 10.1103/PhysRevD.94.125024} {\bibfield  {journal} {\bibinfo
				{journal} {Phys. Rev. D}\ }\textbf {\bibinfo {volume} {94}},\ \bibinfo
			{pages} {125024} (\bibinfo {year} {2016})}\BibitemShut {NoStop}%
		\bibitem [{\citenamefont {Taylor}\ and\ \citenamefont
			{Breen}(2017)}]{taylorbreen:2017}%
		\BibitemOpen
		\bibfield  {author} {\bibinfo {author} {\bibfnamefont {P.}~\bibnamefont
				{Taylor}}\ and\ \bibinfo {author} {\bibfnamefont {C.}~\bibnamefont {Breen}},\
		}\href {\doibase 10.1103/PhysRevD.96.105020} {\bibfield  {journal} {\bibinfo
				{journal} {Phys. Rev. D}\ }\textbf {\bibinfo {volume} {96}},\ \bibinfo
			{pages} {105020} (\bibinfo {year} {2017})}\BibitemShut {NoStop}%
		\bibitem [{\citenamefont {Levi}\ and\ \citenamefont
			{Ori}(2016{\natexlab{a}})}]{amoslevi:2016}%
		\BibitemOpen
		\bibfield  {author} {\bibinfo {author} {\bibfnamefont {A.}~\bibnamefont
				{Levi}}\ and\ \bibinfo {author} {\bibfnamefont {A.}~\bibnamefont {Ori}},\
		}\href {\doibase 10.1103/PhysRevLett.117.231101} {\bibfield  {journal}
			{\bibinfo  {journal} {Phys. Rev. Lett.}\ }\textbf {\bibinfo {volume} {117}},\
			\bibinfo {pages} {231101} (\bibinfo {year} {2016}{\natexlab{a}})}\BibitemShut
		{NoStop}%
		\bibitem [{\citenamefont {Levi}\ \emph {et~al.}(2017)\citenamefont {Levi},
			\citenamefont {Eilon}, \citenamefont {Ori},\ and\ \citenamefont {van~de
				Meent}}]{LeviOriKerr:2017}%
		\BibitemOpen
		\bibfield  {author} {\bibinfo {author} {\bibfnamefont {A.}~\bibnamefont
				{Levi}}, \bibinfo {author} {\bibfnamefont {E.}~\bibnamefont {Eilon}},
			\bibinfo {author} {\bibfnamefont {A.}~\bibnamefont {Ori}}, \ and\ \bibinfo
			{author} {\bibfnamefont {M.}~\bibnamefont {van~de Meent}},\ }\href {\doibase
			10.1103/PhysRevLett.118.141102} {\bibfield  {journal} {\bibinfo  {journal}
				{Phys. Rev. Lett.}\ }\textbf {\bibinfo {volume} {118}},\ \bibinfo {pages}
			{141102} (\bibinfo {year} {2017})}\BibitemShut {NoStop}%
		\bibitem [{\citenamefont {DeWitt}(1975)}]{DeWitt:1975}%
		\BibitemOpen
		\bibfield  {author} {\bibinfo {author} {\bibfnamefont {B.~S.}\ \bibnamefont
				{DeWitt}},\ }\href {\doibase https://doi.org/10.1016/0370-1573(75)90051-4}
		{\bibfield  {journal} {\bibinfo  {journal} {Physics Reports}\ }\textbf
			{\bibinfo {volume} {19}},\ \bibinfo {pages} {295 } (\bibinfo {year}
			{1975})}\BibitemShut {NoStop}%
		\bibitem [{\citenamefont {Christensen}(1976)}]{Christensen:1976}%
		\BibitemOpen
		\bibfield  {author} {\bibinfo {author} {\bibfnamefont {S.~M.}\ \bibnamefont
				{Christensen}},\ }\href {\doibase 10.1103/PhysRevD.14.2490} {\bibfield
			{journal} {\bibinfo  {journal} {Phys. Rev. D}\ }\textbf {\bibinfo {volume}
				{14}},\ \bibinfo {pages} {2490} (\bibinfo {year} {1976})}\BibitemShut
		{NoStop}%
		\bibitem [{\citenamefont {Christensen}(1978)}]{Christensen:1978}%
		\BibitemOpen
		\bibfield  {author} {\bibinfo {author} {\bibfnamefont {S.~M.}\ \bibnamefont
				{Christensen}},\ }\href {\doibase 10.1103/PhysRevD.17.946} {\bibfield
			{journal} {\bibinfo  {journal} {Phys. Rev. D}\ }\textbf {\bibinfo {volume}
				{17}},\ \bibinfo {pages} {946} (\bibinfo {year} {1978})}\BibitemShut
		{NoStop}%
		\bibitem [{\citenamefont {Fulling}\ \emph {et~al.}(1978)\citenamefont
			{Fulling}, \citenamefont {Sweeny},\ and\ \citenamefont
			{Wald}}]{FullingWald:1978}%
		\BibitemOpen
		\bibfield  {author} {\bibinfo {author} {\bibfnamefont {S.~A.}\ \bibnamefont
				{Fulling}}, \bibinfo {author} {\bibfnamefont {M.}~\bibnamefont {Sweeny}}, \
			and\ \bibinfo {author} {\bibfnamefont {R.~M.}\ \bibnamefont {Wald}},\ }\href
		{\doibase 10.1007/BF01196934} {\bibfield  {journal} {\bibinfo  {journal}
				{Commun. Math. Phys.}\ }\textbf {\bibinfo {volume} {63}},\ \bibinfo {pages}
			{257} (\bibinfo {year} {1978})}\BibitemShut {NoStop}%
		\bibitem [{\citenamefont {Fulling}\ \emph {et~al.}(1981)\citenamefont
			{Fulling}, \citenamefont {Narcowich},\ and\ \citenamefont
			{Wald}}]{FullingWald:1981}%
		\BibitemOpen
		\bibfield  {author} {\bibinfo {author} {\bibfnamefont {S.}~\bibnamefont
				{Fulling}}, \bibinfo {author} {\bibfnamefont {F.}~\bibnamefont {Narcowich}},
			\ and\ \bibinfo {author} {\bibfnamefont {R.~M.}\ \bibnamefont {Wald}},\
		}\href {\doibase https://doi.org/10.1016/0003-4916(81)90098-1} {\bibfield
			{journal} {\bibinfo  {journal} {Annals of Physics}\ }\textbf {\bibinfo
				{volume} {136}},\ \bibinfo {pages} {243 } (\bibinfo {year}
			{1981})}\BibitemShut {NoStop}%
		\bibitem [{\citenamefont {Brown}\ and\ \citenamefont
			{Ottewill}(1986)}]{BrownOttewill:1986}%
		\BibitemOpen
		\bibfield  {author} {\bibinfo {author} {\bibfnamefont {M.~R.}\ \bibnamefont
				{Brown}}\ and\ \bibinfo {author} {\bibfnamefont {A.~C.}\ \bibnamefont
				{Ottewill}},\ }\href {\doibase 10.1103/PhysRevD.34.1776} {\bibfield
			{journal} {\bibinfo  {journal} {Phys. Rev. D}\ }\textbf {\bibinfo {volume}
				{34}},\ \bibinfo {pages} {1776} (\bibinfo {year} {1986})}\BibitemShut
		{NoStop}%
		\bibitem [{\citenamefont {Klemm}\ and\ \citenamefont
			{Vanzo}(1998)}]{KlemmVanzo98}%
		\BibitemOpen
		\bibfield  {author} {\bibinfo {author} {\bibfnamefont {D.}~\bibnamefont
				{Klemm}}\ and\ \bibinfo {author} {\bibfnamefont {L.}~\bibnamefont {Vanzo}},\
		}\href {\doibase 10.1103/PhysRevD.58.104025} {\bibfield  {journal} {\bibinfo
				{journal} {Phys. Rev. D}\ }\textbf {\bibinfo {volume} {58}},\ \bibinfo
			{pages} {104025} (\bibinfo {year} {1998})}\BibitemShut {NoStop}%
		\bibitem [{\citenamefont {York}(1985)}]{York:1985}%
		\BibitemOpen
		\bibfield  {author} {\bibinfo {author} {\bibfnamefont {J.~W.}\ \bibnamefont
				{York}},\ }\href {\doibase 10.1103/PhysRevD.31.775} {\bibfield  {journal}
			{\bibinfo  {journal} {Phys. Rev. D}\ }\textbf {\bibinfo {volume} {31}},\
			\bibinfo {pages} {775} (\bibinfo {year} {1985})}\BibitemShut {NoStop}%
		\bibitem [{\citenamefont {Hochberg}\ \emph {et~al.}(1993)\citenamefont
			{Hochberg}, \citenamefont {Kephart},\ and\ \citenamefont {York}}]{York:1993}%
		\BibitemOpen
		\bibfield  {author} {\bibinfo {author} {\bibfnamefont {D.}~\bibnamefont
				{Hochberg}}, \bibinfo {author} {\bibfnamefont {T.~W.}\ \bibnamefont
				{Kephart}}, \ and\ \bibinfo {author} {\bibfnamefont {J.~W.}\ \bibnamefont
				{York}},\ }\href {\doibase 10.1103/PhysRevD.48.479} {\bibfield  {journal}
			{\bibinfo  {journal} {Phys. Rev. D}\ }\textbf {\bibinfo {volume} {48}},\
			\bibinfo {pages} {479} (\bibinfo {year} {1993})}\BibitemShut {NoStop}%
		\bibitem [{\citenamefont {Anderson}\ \emph {et~al.}(1994)\citenamefont
			{Anderson}, \citenamefont {Hiscock}, \citenamefont {Whitesell},\ and\
			\citenamefont {York}}]{AndersonYork:1994}%
		\BibitemOpen
		\bibfield  {author} {\bibinfo {author} {\bibfnamefont {P.~R.}\ \bibnamefont
				{Anderson}}, \bibinfo {author} {\bibfnamefont {W.~A.}\ \bibnamefont
				{Hiscock}}, \bibinfo {author} {\bibfnamefont {J.}~\bibnamefont {Whitesell}},
			\ and\ \bibinfo {author} {\bibfnamefont {J.~W.}\ \bibnamefont {York}},\
		}\href {\doibase 10.1103/PhysRevD.50.6427} {\bibfield  {journal} {\bibinfo
				{journal} {Phys. Rev. D}\ }\textbf {\bibinfo {volume} {50}},\ \bibinfo
			{pages} {6427} (\bibinfo {year} {1994})}\BibitemShut {NoStop}%
		\bibitem [{\citenamefont {Massar}(1995)}]{Massar:1995}%
		\BibitemOpen
		\bibfield  {author} {\bibinfo {author} {\bibfnamefont {S.}~\bibnamefont
				{Massar}},\ }\href {\doibase 10.1103/PhysRevD.52.5857} {\bibfield  {journal}
			{\bibinfo  {journal} {Phys. Rev. D}\ }\textbf {\bibinfo {volume} {52}},\
			\bibinfo {pages} {5857} (\bibinfo {year} {1995})}\BibitemShut {NoStop}%
		\bibitem [{\citenamefont {Loustó}\ and\ \citenamefont
			{Sánchez}(1988)}]{LoustoSanchez:1988}%
		\BibitemOpen
		\bibfield  {author} {\bibinfo {author} {\bibfnamefont {C.}~\bibnamefont
				{Loustó}}\ and\ \bibinfo {author} {\bibfnamefont {N.}~\bibnamefont
				{Sánchez}},\ }\href {\doibase https://doi.org/10.1016/0370-2693(88)91789-3}
		{\bibfield  {journal} {\bibinfo  {journal} {Physics Letters B}\ }\textbf
			{\bibinfo {volume} {212}},\ \bibinfo {pages} {411 } (\bibinfo {year}
			{1988})}\BibitemShut {NoStop}%
		\bibitem [{\citenamefont {Taylor}\ \emph {et~al.}(2000)\citenamefont {Taylor},
			\citenamefont {Hiscock},\ and\ \citenamefont {Anderson}}]{THA:2000}%
		\BibitemOpen
		\bibfield  {author} {\bibinfo {author} {\bibfnamefont {B.~E.}\ \bibnamefont
				{Taylor}}, \bibinfo {author} {\bibfnamefont {W.~A.}\ \bibnamefont {Hiscock}},
			\ and\ \bibinfo {author} {\bibfnamefont {P.~R.}\ \bibnamefont {Anderson}},\
		}\href {\doibase 10.1103/PhysRevD.61.084021} {\bibfield  {journal} {\bibinfo
				{journal} {Phys. Rev. D}\ }\textbf {\bibinfo {volume} {61}},\ \bibinfo
			{pages} {084021} (\bibinfo {year} {2000})}\BibitemShut {NoStop}%
		\bibitem [{\citenamefont {Fabbri}\ \emph {et~al.}(2006)\citenamefont {Fabbri},
			\citenamefont {Farese}, \citenamefont {Navarro-Salas}, \citenamefont {Olmo},\
			and\ \citenamefont {Sanchis-Alepuz}}]{Fabbri:2006}%
		\BibitemOpen
		\bibfield  {author} {\bibinfo {author} {\bibfnamefont {A.}~\bibnamefont
				{Fabbri}}, \bibinfo {author} {\bibfnamefont {S.}~\bibnamefont {Farese}},
			\bibinfo {author} {\bibfnamefont {J.}~\bibnamefont {Navarro-Salas}}, \bibinfo
			{author} {\bibfnamefont {G.~J.}\ \bibnamefont {Olmo}}, \ and\ \bibinfo
			{author} {\bibfnamefont {H.}~\bibnamefont {Sanchis-Alepuz}},\ }\href
		{\doibase 10.1103/PhysRevD.73.104023} {\bibfield  {journal} {\bibinfo
				{journal} {Phys. Rev. D}\ }\textbf {\bibinfo {volume} {73}},\ \bibinfo
			{pages} {104023} (\bibinfo {year} {2006})}\BibitemShut {NoStop}%
		\bibitem [{\citenamefont {Page}(1982)}]{Page1982}%
		\BibitemOpen
		\bibfield  {author} {\bibinfo {author} {\bibfnamefont {D.~N.}\ \bibnamefont
				{Page}},\ }\href {\doibase 10.1103/PhysRevD.25.1499} {\bibfield  {journal}
			{\bibinfo  {journal} {Phys. Rev. D}\ }\textbf {\bibinfo {volume} {25}},\
			\bibinfo {pages} {1499} (\bibinfo {year} {1982})}\BibitemShut {NoStop}%
		\bibitem [{\citenamefont {{Wald}}(1994)}]{WaldBook:1994}%
		\BibitemOpen
		\bibfield  {author} {\bibinfo {author} {\bibfnamefont {R.~M.}\ \bibnamefont
				{{Wald}}},\ }\href@noop {} {\emph {\bibinfo {title} {{Quantum field theory in
						curved spacetime and black hole thermodynamics.}}}}\ (\bibinfo  {publisher}
		{The University of Chicago Press},\ \bibinfo {year} {1994})\BibitemShut
		{NoStop}%
		\bibitem [{\citenamefont {D\'ecanini}\ and\ \citenamefont
			{Folacci}(2008)}]{DecaniniFolacci:2008}%
		\BibitemOpen
		\bibfield  {author} {\bibinfo {author} {\bibfnamefont {Y.}~\bibnamefont
				{D\'ecanini}}\ and\ \bibinfo {author} {\bibfnamefont {A.}~\bibnamefont
				{Folacci}},\ }\href {\doibase 10.1103/PhysRevD.78.044025} {\bibfield
			{journal} {\bibinfo  {journal} {Phys. Rev. D}\ }\textbf {\bibinfo {volume}
				{78}},\ \bibinfo {pages} {044025} (\bibinfo {year} {2008})}\BibitemShut
		{NoStop}%
		\bibitem [{\citenamefont {Harte}\ \emph {et~al.}(2018)\citenamefont {Harte},
			\citenamefont {Taylor},\ and\ \citenamefont
			{Flanagan}}]{HarteFlanaganTaylor}%
		\BibitemOpen
		\bibfield  {author} {\bibinfo {author} {\bibfnamefont {A.~I.}\ \bibnamefont
				{Harte}}, \bibinfo {author} {\bibfnamefont {P.}~\bibnamefont {Taylor}}, \
			and\ \bibinfo {author} {\bibfnamefont {E.~E.}\ \bibnamefont {Flanagan}},\
		}\href {\doibase 10.1103/PhysRevD.97.124053} {\bibfield  {journal} {\bibinfo
				{journal} {Phys. Rev. D}\ }\textbf {\bibinfo {volume} {97}},\ \bibinfo
			{pages} {124053} (\bibinfo {year} {2018})}\BibitemShut {NoStop}%
		\bibitem [{\citenamefont {Brown}\ and\ \citenamefont
			{Ottewill}(1985)}]{BrownOttewill1985}%
		\BibitemOpen
		\bibfield  {author} {\bibinfo {author} {\bibfnamefont {M.~R.}\ \bibnamefont
				{Brown}}\ and\ \bibinfo {author} {\bibfnamefont {A.~C.}\ \bibnamefont
				{Ottewill}},\ }\href {\doibase 10.1103/PhysRevD.31.2514} {\bibfield
			{journal} {\bibinfo  {journal} {Phys. Rev. D}\ }\textbf {\bibinfo {volume}
				{31}},\ \bibinfo {pages} {2514} (\bibinfo {year} {1985})}\BibitemShut
		{NoStop}%
		\bibitem [{\citenamefont {Landau}\ and\ \citenamefont
			{Lifshitz}(1962)}]{LandauLifschitz}%
		\BibitemOpen
		\bibfield  {author} {\bibinfo {author} {\bibfnamefont {L.~D.}\ \bibnamefont
				{Landau}}\ and\ \bibinfo {author} {\bibfnamefont {E.~M.}\ \bibnamefont
				{Lifshitz}},\ }\href@noop {} {\emph {\bibinfo {title} {{The Classical Theory
						of Fields }}}}\ (\bibinfo  {publisher} {Pergamon, Oxford},\ \bibinfo {year}
		{1962})\BibitemShut {NoStop}%
		\bibitem [{\citenamefont {Levi}\ and\ \citenamefont
			{Ori}(2015)}]{LeviOri:2015}%
		\BibitemOpen
		\bibfield  {author} {\bibinfo {author} {\bibfnamefont {A.}~\bibnamefont
				{Levi}}\ and\ \bibinfo {author} {\bibfnamefont {A.}~\bibnamefont {Ori}},\
		}\href {\doibase 10.1103/PhysRevD.91.104028} {\bibfield  {journal} {\bibinfo
				{journal} {Phys. Rev. D}\ }\textbf {\bibinfo {volume} {91}},\ \bibinfo
			{pages} {104028} (\bibinfo {year} {2015})}\BibitemShut {NoStop}%
		\bibitem [{\citenamefont {Levi}\ and\ \citenamefont
			{Ori}(2016{\natexlab{b}})}]{LeviOri:2016}%
		\BibitemOpen
		\bibfield  {author} {\bibinfo {author} {\bibfnamefont {A.}~\bibnamefont
				{Levi}}\ and\ \bibinfo {author} {\bibfnamefont {A.}~\bibnamefont {Ori}},\
		}\href {\doibase 10.1103/PhysRevD.94.044054} {\bibfield  {journal} {\bibinfo
				{journal} {Phys. Rev. D}\ }\textbf {\bibinfo {volume} {94}},\ \bibinfo
			{pages} {044054} (\bibinfo {year} {2016}{\natexlab{b}})}\BibitemShut
		{NoStop}%
		\bibitem [{\citenamefont {Frolov}\ and\ \citenamefont
			{Zel'nikov}(1985)}]{FZ1985a}%
		\BibitemOpen
		\bibfield  {author} {\bibinfo {author} {\bibfnamefont {V.~P.}\ \bibnamefont
				{Frolov}}\ and\ \bibinfo {author} {\bibfnamefont {A.~I.}\ \bibnamefont
				{Zel'nikov}},\ }\href {\doibase 10.1103/PhysRevD.32.3150} {\bibfield
			{journal} {\bibinfo  {journal} {Phys. Rev. D}\ }\textbf {\bibinfo {volume}
				{32}},\ \bibinfo {pages} {3150} (\bibinfo {year} {1985})}\BibitemShut
		{NoStop}%
		\bibitem [{\citenamefont {Jensen}\ and\ \citenamefont
			{Ottewill}(1989)}]{JensenOttewill1989}%
		\BibitemOpen
		\bibfield  {author} {\bibinfo {author} {\bibfnamefont {B.~P.}\ \bibnamefont
				{Jensen}}\ and\ \bibinfo {author} {\bibfnamefont {A.}~\bibnamefont
				{Ottewill}},\ }\href {\doibase 10.1103/PhysRevD.39.1130} {\bibfield
			{journal} {\bibinfo  {journal} {Phys. Rev. D}\ }\textbf {\bibinfo {volume}
				{39}},\ \bibinfo {pages} {1130} (\bibinfo {year} {1989})}\BibitemShut
		{NoStop}%
		\bibitem [{\citenamefont {D{\'{e}}canini}\ and\ \citenamefont
			{Folacci}(2007)}]{Decanini_2007}%
		\BibitemOpen
		\bibfield  {author} {\bibinfo {author} {\bibfnamefont {Y.}~\bibnamefont
				{D{\'{e}}canini}}\ and\ \bibinfo {author} {\bibfnamefont {A.}~\bibnamefont
				{Folacci}},\ }\href {\doibase 10.1088/0264-9381/24/18/014} {\bibfield
			{journal} {\bibinfo  {journal} {Classical and Quantum Gravity}\ }\textbf
			{\bibinfo {volume} {24}},\ \bibinfo {pages} {4777} (\bibinfo {year}
			{2007})}\BibitemShut {NoStop}%
		\bibitem [{\citenamefont {Wald}(1978)}]{Wald1978}%
		\BibitemOpen
		\bibfield  {author} {\bibinfo {author} {\bibfnamefont {R.~M.}\ \bibnamefont
				{Wald}},\ }\href {\doibase 10.1103/PhysRevD.17.1477} {\bibfield  {journal}
			{\bibinfo  {journal} {Phys. Rev. D}\ }\textbf {\bibinfo {volume} {17}},\
			\bibinfo {pages} {1477} (\bibinfo {year} {1978})}\BibitemShut {NoStop}%
		\bibitem [{\citenamefont {Holzegel}(2011)}]{Holzegel2011}%
		\BibitemOpen
		\bibfield  {author} {\bibinfo {author} {\bibfnamefont {G.}~\bibnamefont
				{Holzegel}},\ }\href@noop {} {\  (\bibinfo {year} {2011})},\ \Eprint
		{http://arxiv.org/abs/1103.0710} {arXiv:1103.0710 [gr-qc]} \BibitemShut
		{NoStop}%
		\bibitem [{\citenamefont {Vasy}(2012)}]{Vasy}%
		\BibitemOpen
		\bibfield  {author} {\bibinfo {author} {\bibfnamefont {A.}~\bibnamefont
				{Vasy}},\ }\href {\doibase 10.2140/apde.2012.5.81} {\bibfield  {journal}
			{\bibinfo  {journal} {Analysis and PDE}\ }\textbf {\bibinfo {volume} {5}},\
			\bibinfo {pages} {81} (\bibinfo {year} {2012})},\ \Eprint
		{http://arxiv.org/abs/0911.5440} {arXiv:0911.5440 [Math.AP]} \BibitemShut
		{NoStop}%
		\bibitem [{\citenamefont {Kent}\ and\ \citenamefont
			{Winstanley}(2015)}]{winstanleykent:2014}%
		\BibitemOpen
		\bibfield  {author} {\bibinfo {author} {\bibfnamefont {C.}~\bibnamefont
				{Kent}}\ and\ \bibinfo {author} {\bibfnamefont {E.}~\bibnamefont
				{Winstanley}},\ }\href {\doibase 10.1103/PhysRevD.91.044044} {\bibfield
			{journal} {\bibinfo  {journal} {Phys. Rev. D}\ }\textbf {\bibinfo {volume}
				{91}},\ \bibinfo {pages} {044044} (\bibinfo {year} {2015})}\BibitemShut
		{NoStop}%
		\bibitem [{\citenamefont {Bekenstein}\ and\ \citenamefont
			{Parker}(1981)}]{BekensteinParker1981}%
		\BibitemOpen
		\bibfield  {author} {\bibinfo {author} {\bibfnamefont {J.~D.}\ \bibnamefont
				{Bekenstein}}\ and\ \bibinfo {author} {\bibfnamefont {L.}~\bibnamefont
				{Parker}},\ }\href {\doibase 10.1103/PhysRevD.23.2850} {\bibfield  {journal}
			{\bibinfo  {journal} {Phys. Rev. D}\ }\textbf {\bibinfo {volume} {23}},\
			\bibinfo {pages} {2850} (\bibinfo {year} {1981})}\BibitemShut {NoStop}%
		\bibitem [{\citenamefont {Cruz}\ \emph {et~al.}(2005)\citenamefont {Cruz},
			\citenamefont {Olivares},\ and\ \citenamefont {Villanueva}}]{Cruz_2005}%
		\BibitemOpen
		\bibfield  {author} {\bibinfo {author} {\bibfnamefont {N.}~\bibnamefont
				{Cruz}}, \bibinfo {author} {\bibfnamefont {M.}~\bibnamefont {Olivares}}, \
			and\ \bibinfo {author} {\bibfnamefont {J.~R.}\ \bibnamefont {Villanueva}},\
		}\href {\doibase 10.1088/0264-9381/22/6/016} {\bibfield  {journal} {\bibinfo
				{journal} {Classical and Quantum Gravity}\ }\textbf {\bibinfo {volume}
				{22}},\ \bibinfo {pages} {1167} (\bibinfo {year} {2005})}\BibitemShut
		{NoStop}%
		\bibitem [{\citenamefont {Hayward}(2006)}]{Hayward:2006}%
		\BibitemOpen
		\bibfield  {author} {\bibinfo {author} {\bibfnamefont {S.~A.}\ \bibnamefont
				{Hayward}},\ }\href {\doibase 10.1103/PhysRevLett.96.031103} {\bibfield
			{journal} {\bibinfo  {journal} {Phys. Rev. Lett.}\ }\textbf {\bibinfo
				{volume} {96}},\ \bibinfo {pages} {031103} (\bibinfo {year}
			{2006})}\BibitemShut {NoStop}%
		\bibitem [{\citenamefont {{Frolov}}\ and\ \citenamefont
			{{Vilkovisky}}(1981)}]{Frolov:1981}%
		\BibitemOpen
		\bibfield  {author} {\bibinfo {author} {\bibfnamefont {V.~P.}\ \bibnamefont
				{{Frolov}}}\ and\ \bibinfo {author} {\bibfnamefont {G.~A.}\ \bibnamefont
				{{Vilkovisky}}},\ }\href {\doibase 10.1016/0370-2693(81)90542-6} {\bibfield
			{journal} {\bibinfo  {journal} {Physics Letters B}\ }\textbf {\bibinfo
				{volume} {106}},\ \bibinfo {pages} {307} (\bibinfo {year}
			{1981})}\BibitemShut {NoStop}%
		\bibitem [{\citenamefont {Stephens}\ \emph {et~al.}(1994)\citenamefont
			{Stephens}, \citenamefont {t'Hooft},\ and\ \citenamefont
			{Whiting}}]{Stephens:1994}%
		\BibitemOpen
		\bibfield  {author} {\bibinfo {author} {\bibfnamefont {C.~R.}\ \bibnamefont
				{Stephens}}, \bibinfo {author} {\bibfnamefont {G.}~\bibnamefont {t'Hooft}}, \
			and\ \bibinfo {author} {\bibfnamefont {B.~F.}\ \bibnamefont {Whiting}},\
		}\href {\doibase 10.1088/0264-9381/11/3/014} {\bibfield  {journal} {\bibinfo
				{journal} {Classical and Quantum Gravity}\ }\textbf {\bibinfo {volume}
				{11}},\ \bibinfo {pages} {621} (\bibinfo {year} {1994})}\BibitemShut
		{NoStop}%
		\bibitem [{\citenamefont {Ashtekar}\ and\ \citenamefont
			{Bojowald}(2005)}]{Ashtekar:2005}%
		\BibitemOpen
		\bibfield  {author} {\bibinfo {author} {\bibfnamefont {A.}~\bibnamefont
				{Ashtekar}}\ and\ \bibinfo {author} {\bibfnamefont {M.}~\bibnamefont
				{Bojowald}},\ }\href {\doibase 10.1088/0264-9381/22/16/014} {\bibfield
			{journal} {\bibinfo  {journal} {Classical and Quantum Gravity}\ }\textbf
			{\bibinfo {volume} {22}},\ \bibinfo {pages} {3349} (\bibinfo {year}
			{2005})}\BibitemShut {NoStop}%
		\bibitem [{\citenamefont {Haggard}\ and\ \citenamefont
			{Rovelli}(2015)}]{Haggard:2015}%
		\BibitemOpen
		\bibfield  {author} {\bibinfo {author} {\bibfnamefont {H.~M.}\ \bibnamefont
				{Haggard}}\ and\ \bibinfo {author} {\bibfnamefont {C.}~\bibnamefont
				{Rovelli}},\ }\href {\doibase 10.1103/PhysRevD.92.104020} {\bibfield
			{journal} {\bibinfo  {journal} {Phys. Rev. D}\ }\textbf {\bibinfo {volume}
				{92}},\ \bibinfo {pages} {104020} (\bibinfo {year} {2015})}\BibitemShut
		{NoStop}%
		\bibitem [{\citenamefont {KAWAI}\ \emph {et~al.}(2013)\citenamefont {KAWAI},
			\citenamefont {MATSUO},\ and\ \citenamefont {YOKOKURA}}]{Yokokura:2013}%
		\BibitemOpen
		\bibfield  {author} {\bibinfo {author} {\bibfnamefont {H.}~\bibnamefont
				{KAWAI}}, \bibinfo {author} {\bibfnamefont {Y.}~\bibnamefont {MATSUO}}, \
			and\ \bibinfo {author} {\bibfnamefont {Y.}~\bibnamefont {YOKOKURA}},\ }\href
		{\doibase 10.1142/S0217751X13500504} {\bibfield  {journal} {\bibinfo
				{journal} {International Journal of Modern Physics A}\ }\textbf {\bibinfo
				{volume} {28}},\ \bibinfo {pages} {1350050} (\bibinfo {year} {2013})},\
		\Eprint {http://arxiv.org/abs/https://doi.org/10.1142/S0217751X13500504}
		{https://doi.org/10.1142/S0217751X13500504} \BibitemShut {NoStop}%
		\bibitem [{\citenamefont {Kawai}\ and\ \citenamefont
			{Yokokura}(2016)}]{Yokokura:2016}%
		\BibitemOpen
		\bibfield  {author} {\bibinfo {author} {\bibfnamefont {H.}~\bibnamefont
				{Kawai}}\ and\ \bibinfo {author} {\bibfnamefont {Y.}~\bibnamefont
				{Yokokura}},\ }\href {\doibase 10.1103/PhysRevD.93.044011} {\bibfield
			{journal} {\bibinfo  {journal} {Phys. Rev. D}\ }\textbf {\bibinfo {volume}
				{93}},\ \bibinfo {pages} {044011} (\bibinfo {year} {2016})}\BibitemShut
		{NoStop}%
		\bibitem [{\citenamefont {Kawai}\ and\ \citenamefont
			{Yokokura}(2017)}]{Yokokura:2017}%
		\BibitemOpen
		\bibfield  {author} {\bibinfo {author} {\bibfnamefont {H.}~\bibnamefont
				{Kawai}}\ and\ \bibinfo {author} {\bibfnamefont {Y.}~\bibnamefont
				{Yokokura}},\ }\href {\doibase 10.3390/universe3020051} {\bibfield  {journal}
			{\bibinfo  {journal} {Universe}\ }\textbf {\bibinfo {volume} {3}},\ \bibinfo
			{pages} {51} (\bibinfo {year} {2017})}\BibitemShut {NoStop}%
		\bibitem [{\citenamefont {Kawai}\ and\ \citenamefont
			{Yokokura}(2020)}]{Yokokura:2020}%
		\BibitemOpen
		\bibfield  {author} {\bibinfo {author} {\bibfnamefont {H.}~\bibnamefont
				{Kawai}}\ and\ \bibinfo {author} {\bibfnamefont {Y.}~\bibnamefont
				{Yokokura}},\ }\href {\doibase 10.3390/universe6060077} {\bibfield  {journal}
			{\bibinfo  {journal} {Universe}\ }\textbf {\bibinfo {volume} {6}},\ \bibinfo
			{pages} {77} (\bibinfo {year} {2020})}\BibitemShut {NoStop}%
		\bibitem [{\citenamefont {Ho}(2016)}]{Ho:2016}%
		\BibitemOpen
		\bibfield  {author} {\bibinfo {author} {\bibfnamefont {P.-M.}\ \bibnamefont
				{Ho}},\ }\href {\doibase https://doi.org/10.1016/j.nuclphysb.2016.05.016}
		{\bibfield  {journal} {\bibinfo  {journal} {Nuclear Physics B}\ }\textbf
			{\bibinfo {volume} {909}},\ \bibinfo {pages} {394 } (\bibinfo {year}
			{2016})}\BibitemShut {NoStop}%
		\bibitem [{\citenamefont {Hawking}\ and\ \citenamefont
			{Page}(1983)}]{HawkingPage:1983}%
		\BibitemOpen
		\bibfield  {author} {\bibinfo {author} {\bibfnamefont {S.~W.}\ \bibnamefont
				{Hawking}}\ and\ \bibinfo {author} {\bibfnamefont {D.~N.}\ \bibnamefont
				{Page}},\ }\href {\doibase 10.1007/BF01208266} {\bibfield  {journal}
			{\bibinfo  {journal} {Commun. Math. Phys.}\ }\textbf {\bibinfo {volume}
				{87}},\ \bibinfo {pages} {577} (\bibinfo {year} {1983})}\BibitemShut
		{NoStop}%
		\bibitem [{\citenamefont {Horowitz}\ and\ \citenamefont
			{Hubeny}(2000)}]{HorowitzHubeny2000}%
		\BibitemOpen
		\bibfield  {author} {\bibinfo {author} {\bibfnamefont {G.~T.}\ \bibnamefont
				{Horowitz}}\ and\ \bibinfo {author} {\bibfnamefont {V.~E.}\ \bibnamefont
				{Hubeny}},\ }\href {\doibase 10.1103/PhysRevD.62.024027} {\bibfield
			{journal} {\bibinfo  {journal} {Phys. Rev. D}\ }\textbf {\bibinfo {volume}
				{62}},\ \bibinfo {pages} {024027} (\bibinfo {year} {2000})}\BibitemShut
		{NoStop}%
		\bibitem [{\citenamefont {Cardoso}\ and\ \citenamefont
			{Lemos}(2001)}]{CardosoLemos2001}%
		\BibitemOpen
		\bibfield  {author} {\bibinfo {author} {\bibfnamefont {V.}~\bibnamefont
				{Cardoso}}\ and\ \bibinfo {author} {\bibfnamefont {J.~P.~S.}\ \bibnamefont
				{Lemos}},\ }\href {\doibase 10.1103/PhysRevD.64.084017} {\bibfield  {journal}
			{\bibinfo  {journal} {Phys. Rev. D}\ }\textbf {\bibinfo {volume} {64}},\
			\bibinfo {pages} {084017} (\bibinfo {year} {2001})}\BibitemShut {NoStop}%
		\bibitem [{\citenamefont {Berti}\ \emph {et~al.}(2009)\citenamefont {Berti},
			\citenamefont {Cardoso},\ and\ \citenamefont {Pani}}]{BertiCardosoPani}%
		\BibitemOpen
		\bibfield  {author} {\bibinfo {author} {\bibfnamefont {E.}~\bibnamefont
				{Berti}}, \bibinfo {author} {\bibfnamefont {V.}~\bibnamefont {Cardoso}}, \
			and\ \bibinfo {author} {\bibfnamefont {P.}~\bibnamefont {Pani}},\ }\href
		{\doibase 10.1103/PhysRevD.79.101501} {\bibfield  {journal} {\bibinfo
				{journal} {Phys. Rev. D}\ }\textbf {\bibinfo {volume} {79}},\ \bibinfo
			{pages} {101501} (\bibinfo {year} {2009})}\BibitemShut {NoStop}%
		\bibitem [{\citenamefont {Breen}\ and\ \citenamefont
			{Taylor}(2018)}]{BreenTaylor:18}%
		\BibitemOpen
		\bibfield  {author} {\bibinfo {author} {\bibfnamefont {C.}~\bibnamefont
				{Breen}}\ and\ \bibinfo {author} {\bibfnamefont {P.}~\bibnamefont {Taylor}},\
		}\href {\doibase 10.1103/PhysRevD.98.105006} {\bibfield  {journal} {\bibinfo
				{journal} {Phys. Rev. D}\ }\textbf {\bibinfo {volume} {98}},\ \bibinfo
			{pages} {105006} (\bibinfo {year} {2018})}\BibitemShut {NoStop}%
		\bibitem [{\citenamefont {Morley}\ \emph {et~al.}(2020)\citenamefont {Morley},
			\citenamefont {Taylor},\ and\ \citenamefont {Winstanley}}]{MTW2}%
		\BibitemOpen
		\bibfield  {author} {\bibinfo {author} {\bibfnamefont {T.}~\bibnamefont
				{Morley}}, \bibinfo {author} {\bibfnamefont {P.}~\bibnamefont {Taylor}}, \
			and\ \bibinfo {author} {\bibfnamefont {E.}~\bibnamefont {Winstanley}},\
		}\href {} {\bibfield
			{journal} {\bibinfo  {journal} {accepted to CQG}\ } (\bibinfo
			{year} {2020})}\BibitemShut {NoStop}%
		\bibitem [{\citenamefont {Morley}\ \emph {et~al.}(2020)\citenamefont {Morley},
			\citenamefont {Taylor},\ and\ \citenamefont {Winstanley}}]{MTW3}%
		\BibitemOpen
		\bibfield  {author} {\bibinfo {author} {\bibfnamefont {T.}~\bibnamefont
				{Morley}}, \bibinfo {author} {\bibfnamefont {E.}~\bibnamefont {Winstanley}}, \
			and\ \bibinfo {author} {\bibfnamefont {P.}~\bibnamefont {Taylor}},\
		}\href {} {\bibfield
			{journal} {\bibinfo  {journal} {accepted to PRD}\ } (\bibinfo
			{year} {2020})}\BibitemShut {NoStop}%
		\bibitem [{\citenamefont {Morley}\ \emph {et~al.}(2018)\citenamefont {Morley},
			\citenamefont {Taylor},\ and\ \citenamefont {Winstanley}}]{MTW1}%
		\BibitemOpen
		\bibfield  {author} {\bibinfo {author} {\bibfnamefont {T.}~\bibnamefont
				{Morley}}, \bibinfo {author} {\bibfnamefont {P.}~\bibnamefont {Taylor}}, \
			and\ \bibinfo {author} {\bibfnamefont {E.}~\bibnamefont {Winstanley}},\
		}\href {\doibase 10.1088/1361-6382/aae45b} {\bibfield  {journal} {\bibinfo
				{journal} {Classical and Quantum Gravity}\ }\textbf {\bibinfo {volume}
				{35}},\ \bibinfo {pages} {235010} (\bibinfo {year} {2018})}\BibitemShut
		{NoStop}%
	\end{thebibliography}
\end{document}